\documentclass[12pt]{article}
\usepackage{makeidx}
\usepackage{amsfonts}
\usepackage{amssymb}
\usepackage{amsmath}
\usepackage{sw20elba}
\usepackage{hyperref}

\newtheorem{theorem}{Theorem}

\newtheorem{remark}[theorem]{Remark}

\input{tcilatex}
\begin{document}

\title{Electrodynamics of balanced charges}
\author{Anatoli Babin and Alexander Figotin \\
%EndAName
University of California at Irvine}
\maketitle

\begin{abstract}
In this work we modify the wave-corpuscle mechanics for elementary charges
introduced by us recently. This modification is designed to better describe
electromagnetic (EM) phenomena at atomic scales. It includes a modification
of the concept of the classical EM field and a new model for the elementary
charge which we call a balanced charge (b-charge). A b-charge does not
interact with itself electromagnetically, and every b-charge possesses its
own elementary EM field. The EM energy is naturally partitioned as the
interaction energy of pairs of different b-charges. We construct EM theory
of b-charges (BEM) based on a relativistic Lagrangian with the following
properties: (i) b-charges interact only through their elementary EM
potentials and fields; (ii) the field equations for the elementary EM fields
are exactly the Maxwell equations with proper currents; (iii) a free charge
moves uniformly preserving up to the Lorentz contraction its shape; (iv) the
Newton equations with the Lorentz forces hold approximately when charges are
well separated and move with non-relativistic velocities. The BEM theory can
be characterized as neoclassical one which covers the macroscopic as well as
the atomic spatial scales, it describes EM phenomena at atomic scale
differently than the classical EM theory. It yields in macroscopic regimes
the Newton equations with Lorentz forces for centers of well separated
charges moving with nonrelativistic velocities. Applied to atomic scales it
yields a hydrogen atom model with a frequency spectrum matching the same for
the Schrodinger model with any desired accuracy.
\end{abstract}

\tableofcontents

\section{Introduction}

It is well recognized that the classical electromagnetic (CEM) theory
formulated in the form of Maxwell-Lorentz equations provides an excellent
description of electromagnetic phenomena at the macroscopic length scales.
It also well known that CEM theory is inadequate in explaining
electromagnetic (EM) phenomena at the atomic scales including spectroscopic
data of the hydrogen atom (HA).

We develop here an EM theory which accounts for all classical EM phenomena
at the macroscopic scales as well at least some EM phenomena at the atomic
scale including the HA spectral lines. Our theory is classical, though we
can apply it to some phenomena (HA energy spectrum in particular) at spatial
scales compared with Bohr radius, that is spatial scales of order $\ 0.1$
nm, and our theory produces the same type of results as the quantum
mechanics which is commonly used at such scales. We think that expanding the
classical theory down to smallest possible spatial scales is important
because the classical description of physical systems allows (at least in
principle) more details compared with the probabilistic quantum-mechanical
description. Since our neoclassical description applied to HA does not
generate contradictions at atomic scales of order $0.1$ nm (as CEM theory
did), we expect it to be applicable to describe details of electromagnetic
processes at nanometer scales. Note that at such scales the \ electron
cannot be considered a point or a charged ball and we have to use a complete
description of an electron which is presented in our model as a
wave-corpuscle.

When attempting to change the CEM theory we want: (i) to stay on solid
ground of the Lagrangian mechanics and the relativity principle; (ii) to
recover in this new EM theory all well established experimental facts
described by the CEM theory. The foundational pillars of the CEM theory -
the Maxwell equations and the Lorentz force expression - remain to be key
elements in proposed here new EM theory, and before we proceed with the new
developments let us briefly recall the CEM theory fundamentals. First of
all, the EM fields driven by prescribed currents in vacuum are described by
the Maxwell equations%
\begin{equation}
\frac{1}{\mathrm{c}}\frac{\partial \mathbf{B}}{\partial t}+\nabla \times 
\mathbf{E}=\boldsymbol{0},\ \nabla \cdot \mathbf{B}=0,  \label{psys1}
\end{equation}%
\begin{equation}
\frac{1}{\mathrm{c}}\frac{\partial \mathbf{E}}{\partial t}-\nabla \times 
\mathbf{B}=-\frac{4\pi }{\mathrm{c}}\mathbf{J},\ \nabla \cdot \mathbf{E}%
=4\pi \rho ,  \label{psys2}
\end{equation}%
where $\mathbf{E}$ and $\mathbf{B}$ are respectively electric field and the
magnetic induction, and $\rho \left( t,x\right) $, $\mathbf{J}\left(
t,x\right) $ are respectively prescribed charge and current densities. In
particular, when the fields are generated by point charges, the sources $%
\rho $, $\mathbf{J}$ are written in the form%
\begin{equation}
\rho =\sum_{\ell }q^{\ell }\delta \left( \mathbf{x}-\mathbf{r}^{\ell }\left(
t\right) \right) ,\quad \mathbf{J}=\sum_{\ell }q^{\ell }\delta \left( 
\mathbf{x}-\mathbf{r}^{\ell }\left( t\right) \right) \mathbf{v}^{\ell
}\left( t\right) ,  \label{rosumdel}
\end{equation}%
where $q^{\ell }$ is charge value of the $\ell $-th point charge, $\mathbf{r}%
^{\ell }$ and is $\mathbf{v}^{\ell }\left( t\right) =\frac{\mathrm{d}\mathbf{%
r}^{\ell }}{\mathrm{d}t}$ are respectively its position and the velocity,
and $\delta $ is the Dirac delta-function. Second, for a given EM field the
motion of every point charge in the field is determined from the equation 
\begin{equation}
\frac{\mathrm{d}}{\mathrm{d}t}\left[ m^{\ell }\mathbf{v}^{\ell }\left(
t\right) \right] =q^{\ell }\left[ \mathbf{E}\left( t,\mathbf{r}^{\ell
}\left( t\right) \right) +\frac{1}{\mathrm{c}}\mathbf{v}^{\ell }\left(
t\right) \times \mathbf{B}\left( t,\mathbf{r}^{\ell }\left( t\right) \right) %
\right] ,  \label{pchar1}
\end{equation}%
where $m^{\ell }$ is the $\ell $-th point charge mass and the right hand
side of (\ref{pchar1}) is the \emph{Lorentz force}. The CEM theory
essentially treats three types of problems: (i) studies of EM fields for
prescribed charge and current densities described by the Maxwell equations (%
\ref{psys1}), (\ref{psys2}); (ii) the motion of charges in a prescribed
external field; (iii) interaction of charges and their EM fields. The
classical Maxwell-Lorentz system though very successful in describing many
EM phenomena has well known problems including infinite self-energy which
are discussed widely in the literature. As to important for the CEM theory
concept of point charges or particles F. Rohrlich writes, \cite[Section 2.1,
p.9; Section 6.1, p.123]{Rohrlich}: "... the classical theory of charged
particles as first conceived by Lorentz emerged as a hybrid theory of
particles and fields: charged particles are interacting via an
electromagnetic field", "Macroscopic Maxwell electrodynamics knows only
charge distributions. In this theory electrostatic charge is derived from a
charge density (linear, surface, or volume density). The concept of charge
as an aggregate of elementary charged particles is foreign to it." There is
also a fundamental thermodynamical problem for the classical Maxwell theory
at atomic scale namely the lack of an \emph{"elementary process of
absorbtion"} as it was formulated by A. Einsteins in his seminal paper \cite%
{Einstein 1909b}. He wrote there: "\emph{The fundamental property of the
oscillation theory that engenders these difficulties seems to me the
following. In the kinetic theory of molecules, for every process in which
only a few elementary particles participate (e.g., molecular collisions),
the inverse process also exists. But that is not the case for the elementary
processes of radiation. ... The elementary process of emission is not
invertible. In this, I believe, our oscillation theory does not hit the
mark. Newton's emission theory of light seems to contain more truth with
respect to this point than the oscillation theory since, first of all, the
energy given to a light particle is not scattered over infinite space, but
remains available for an elementary process of absorption.}" \ In
particular, the proposed here theory does have an elementary process of
asbroption as A. Einstein suggested. We call this process \emph{"negative
radiation"} and its essense is that the EM energy is moving toward two
closely located elementary charges if they oscillate with the same frequency
but oposite phases.

The proposed here EM is based on a new concept for elementary charge which
we call \emph{balanced charge or b-charge} for short, and we refer the
theory itself as balanced electromagnetic (BEM) theory. A key element of the
BEM theory is a concept of an \emph{elementary EM field} assigned to every
single \emph{b-charge}. A single b-charge is described by a pair $\left(
\psi ,A^{\mu }\right) $, where $\psi $ is its wave function and $A^{\mu
}=\left( \varphi ,\mathbf{A}\right) $ is its 4-vector \emph{elementary
potential} with the corresponding \emph{elementary EM field} defined by the
familiar formula $F^{\mu \nu }=\partial ^{\mu }A^{\nu }-\partial ^{\nu
}A^{\mu }$. So, \emph{b-charge is a field over 4-dimensional space-time
continuum for which the wave function }$\psi $\emph{\ represents its matter
properties and the elementary potential }$A^{\mu }$\emph{\ mediates its EM
interactions with all other b-charges. Importantly,\ (i) all internal forces
of a b-charge are exclusively of non-electromagnetic origin; (ii) every
b-charge is a source of its elementary EM field which represents force
exerted by this charge on any other b-charge but not upon itself. The later
allows to view a single b-charge as truly elementary one with respect to the
electromagnetic interactions.}

An idea to introduce an extended charge instead of the point one is not new,
and the most known models for it are the Abraham rigid charge model and the
Lorentz relativistically covariant model. These models are studied and
advanced in many papers, see \cite[Sections 16]{Jackson}, \cite{Kiessling2}, 
\cite{Pearle1}, \cite[Sections 2, 6]{Rohrlich}, \cite{Schwinger}, \cite%
{Spohn}, \cite{Yaghjian}. In contrast to those models, here and in \cite%
{BabFig1}, \cite{BabFig2} we do not prescribe to an elementary charge a
certain geometry, but instead the elementary charge has a wave function
governed by a nonlinear Klein-Gordon or a nonlinear Schr\"{o}dinger equation
in the relativistic and nonrelativistic cases respectively. An idea to
eliminate self-interaction is also, of course, not new. The latest to our
best knowledge attempt to have this feature in the electrodynamics is due J.
Wheeler and R. Feynman, \cite{Wheeler Feynmann 1}, \cite{Wheeler Feynmann 2}%
, but the EM theory proposed here is very different from it.

The BEM theory is constructed based on a relativistic Lagrangian with the
following properties: (i) b-charges interact only through their elementary
EM potentials and fields; (ii) the field equations for the elementary EM
fields are exactly the Maxwell equations with proper conserved currents;
(iii) a free charge moves uniformly preserving up to the Lorentz contraction
its shape; (iv) the Newton equations with the Lorentz forces hold
approximately when charges are well separated and move with non-relativistic
velocities. Since an overwhelming number of EM phenomena are explained
within the CEM theory by the Maxwell equations and the Lorentz forces the
BEM theory is equally successful in explaining the same phenomena.

A system of $N$ elementary charges in the BEM theory is modeled by $N$ pairs 
$\left( \psi ^{\ell },A^{\ell \mu }\right) $, $1\leq \ell \leq N$, and every
b-charge $\left( \psi ^{\ell },A^{\ell \mu }\right) $ is naturally assigned
via the Lagrangian its \emph{elementary conserved 4-current} $J^{\ell \nu }$%
. The later property provides additional justification for calling the
elementary field $F^{\ell \mu \nu }$ electromagnetic. The classical (total)
EM field $A^{\mu }=\left( \varphi ,\mathbf{A}\right) $ is recovered in this
theory as the sum of all elementary EM fields, namely%
\begin{equation}
\varphi =\sum_{1\leq \ell \leq N}\varphi ^{\ell },\quad \mathbf{A}%
=\sum_{1\leq \ell \leq N}\mathbf{A}^{\ell },  \label{fiasum}
\end{equation}%
but, importantly, this \emph{total field is not an independent entity with
its degrees of freedom.} Notice then that since in the BEM theory there is
no EM self-interaction the action on $\ell $-th charge by EM fields of other
charges is described by a field $A_{\neq \ell }^{\mu }=\left( \varphi _{\neq
\ell },\mathbf{A}_{\neq \ell }\right) $ which is the total field $\left(
\varphi ,\mathbf{A}\right) $ "balanced" by the removal from it the
self-interaction, namely 
\begin{equation}
\varphi _{\neq \ell }=\sum_{\ell ^{\prime }\neq \ell }\varphi ^{\ell
^{\prime }},\quad \mathbf{A}_{\neq \ell }=\sum_{\ell ^{\prime }\neq \ell }%
\mathbf{A}^{\ell }.  \label{fineqL}
\end{equation}%
Use of EM fields similar to ones in (\ref{fineqL}) is, of course, not a
discovery and they can be found in many textbooks, but the \emph{BEM theory
goes further than that and removes the elementary EM self-actions for the
CEM Lagrangian and consequently the elementary self-energies from the
classical EM energy-momentum tensor}. Only with this removal of the
elementary self-actions from the classical EM Lagrangian one gets the field
equations in the form of the \emph{elementary Maxwell equations} $\partial
_{\mu }F^{\ell \mu \nu }=\frac{4\pi }{\mathrm{c}}J^{\ell \nu }$ for the
elementary EM $F^{\ell \mu \nu }$ fields with the elementary conserved
4-currents $J^{\ell \nu }$.

In Section \ref{srbem} we study differences and similarities between the CEM
and BEM theories. The differences and similarities stem from the differences
and similarities between their Lagrangians and with consequent differences
and similarities in the energy-momentum tensor definition. Since the both
theories are based on the Maxwell equations they have exactly the same EM
fields for prescribed currents but their total actions and the
energy-momentum tensors differ with consequent differences in radiation
phenomena. We show below that the CEM theory is a limit of the BEM theory.
That can be already seen by comparing formulas (\ref{fiasum}) and (\ref%
{fineqL}) for a large $N$ since these large sums differ just by one term.
The classical charge can be treated within the BEM theory as a cluster of
many b-charges and it is shown, in particular, that such a classical charge
does interact with itself electromagnetically and its interaction with other
classical charges can be described approximately through the interaction
with the single effective EM field. Notice also that the classical single EM
field is defined by (\ref{fiasum} as the sum of all elementary EM fields,
but since the energy is a quadratic function of fields, \emph{\ the total
single classical EM field has energy equal to the combined energy of
interaction between all the pairs of elementary EM fields only approximately}%
. Most of EM phenomena at the macroscopic scale can be described based on
the single EM field with astounding precision and the relative difference
with the BEM theory since the inverse of the Avogadro constant, that is $%
10^{-23},$ is very small. But the differences between the CEM and BEM
theories become more pronounced for smaller systems with fewer b-charges.
The BEM theory predictions can significantly deviate from those of the CEM
theory in the following situations: (i) there are just a few b-charges which
are in close proximity; (ii) there is a large but highly coherent system of
b-charges similar to those \emph{collective, coherent systems}
(superconducting ring, laser and more) described by C. Mead \cite[p.5]{Mead}
in his "collective electrodynamics". In the BEM theory the EM energy is the
energy of EM interaction of pairs of b-charges and consequently it is
naturally partioned between the pairs of charges. We show that for every
pair of b-charges their EM interaction energy satisfies elementary
energy-momentum conservations governed by the relevant Lorentz force
densities.

In Section \ref{snonrbem} we study particle-like properties of b-charges $%
\left( \psi ^{\ell },A^{\ell \mu }\right) $ which are captured by the
concept of wave-corpuscle similarly to the wave-corpuscle mechanics (WCM), 
\cite{BabFig1}, \cite{BabFig2}. A key ingredient providing for particle-like
behavior of a b-charge are self-interaction nonlinearities $G_{a,\ell }$
where size parameter $a=a^{\ell }$ determines the size of a free particle
which though small is non-zero. The shape of the wave-corpuscle $\psi ^{\ell
}$ is intimately related to the the nonlinearity $G_{a,\ell }$ via the rest
charge equation - a nonlinear Klein-Gordon equation- which in
non-relativistic case turns into a nonlinear Schrodinger equation. We also
derive in this section the Newton equations with the Lorentz forces as an
approximation when charges are well separated and move with non-relativistic
velocities. \ 

In Section \ref{sbemha} we provide a detailed sketch of the hydrogen atom
model. At atomic scales when there are just a few elementary charges in a
close proximity the BEM theory differs significantly from the CEM theory. 
\emph{It yields, in particular, the BEM hydrogen atom model with a frequency
spectrum matching the same for the Schrodinger hydrogen atom with any
desired accuracy.}\ The difference between the two HA models depends on on
the size of the free electron as a parameter in the self-interaction
nonlinearity $G_{a}$, and an analysis suggests the size of a free electron
to be of about 100 times of the Bohr radius.

\section{Relativistic theory\label{srbem}}

In this section we introduce BEM Lagrangian for a system of bounced charges
and derive the corresponding field equations, currents, energy-momentum
tensors (EnMT) for system components and more. An important objective of
this section is to study the EM interactions in the BEM theory and find out
common and different features of the BEM theory and the CEM theory.

\subsection{Lagrangian, field equations and currents\label{slagbem}}

Let us consider a system of elementary charges $\left( \psi ^{\ell },A^{\ell
\mu }\right) $, $1\leq \ell \leq N$. In what follows the $\ell $-th charge
potential $A^{\ell \mu }$ and its EM field $F^{\ell \mu \nu }=\partial ^{\mu
}A^{\ell \nu }-\partial ^{\nu }A^{\ell \mu }$ completely account for its
action upon all other charges $\ell ^{\prime }\neq \ell $. Consequently, the
action upon $\ell $-th charge by all other charges is described by $\ell $%
-th exterior potential $A_{\neq }^{\ell \mu }$ and its EM field $F_{\neq
}^{\ell \mu \nu }$ defined by%
\begin{equation}
A_{\neq }^{\ell \mu }=\sum\limits_{\ell ^{\prime }\neq \ell }A^{\ell
^{\prime }\mu },\quad A_{\neq }^{\ell \mu }=\left( \varphi _{\neq }^{\ell },%
\mathbf{A}_{\neq }^{\ell }\right) ,\quad F_{\neq }^{\ell \mu \nu
}=\sum\limits_{\ell ^{\prime }\neq \ell }F^{\ell ^{\prime }\mu \nu }.
\label{ilag2}
\end{equation}%
We also introduce for the total potential $\mathcal{A}^{\mu }$ and the
corresponding total EM field $\mathcal{F}^{\mu \nu }$ by the following
formulas%
\begin{gather}
\mathcal{A}^{\mu }=\sum\limits_{1\leq \ell \leq N}A^{\ell \mu },\qquad 
\mathcal{F}^{\mu \nu }=\sum\limits_{1\leq \ell \leq N}F^{\ell \mu \nu },
\label{ilag1} \\
F^{\ell \mu \nu }=\partial ^{\mu }A^{\ell \nu }-\partial ^{\nu }A^{\ell \mu
}.  \notag
\end{gather}%
We furnish now the system of $N$ b-charges with the following Lagrangian%
\begin{gather}
\mathcal{L}\left( \left\{ \psi ^{\ell },\psi _{;\mu }^{\ell }\right\}
,\left\{ \psi ^{\ell \ast },\psi _{;\mu }^{\ell \ast }\right\} ,A^{\ell \mu
}\right) =\sum\limits_{1\leq \ell \leq N}L^{\ell }\left( \psi ^{\ell },\psi
_{;\mu }^{\ell },\psi ^{\ell \ast },\psi _{;\mu }^{\ell \ast }\right) +%
\mathcal{L}_{\mathrm{BEM}},  \label{ilag3} \\
\mathcal{L}_{\mathrm{BEM}}=\mathcal{L}_{\mathrm{CEM}}-\mathcal{L}_{\mathrm{e}%
},\ \mathcal{L}_{\mathrm{CEM}}=-\frac{\mathcal{F}^{\mu \nu }\mathcal{F}_{\mu
\nu }}{16\pi },\ \mathcal{L}_{\mathrm{e}}=-\sum\limits_{1\leq \ell \leq N}%
\frac{F^{\ell \mu \nu }F_{\mu \nu }^{\ell }}{16\pi },  \label{ilag3a}
\end{gather}%
where $L^{\ell }$ is the Lagrangian of the $\ell $-th bare charge, and the 
\emph{covariant derivatives} are defined by the following formulas%
\begin{gather}
\psi _{;\mu }^{\ell }=\tilde{\partial}^{\ell \mu }\psi ^{\ell },\ \psi
_{;\mu }^{\ell \ast }=\tilde{\partial}^{\ell \mu \ast }\psi ^{\ell \ast },
\label{ilag4} \\
\tilde{\partial}^{\ell \mu }=\partial ^{\mu }+\frac{\mathrm{i}q^{\ell
}A_{\neq }^{\ell \mu }}{\chi \mathrm{c}},\ \tilde{\partial}^{\ell \mu \ast
}=\partial ^{\mu }-\frac{\mathrm{i}q^{\ell }A_{\neq }^{\ell \mu }}{\chi 
\mathrm{c}}.  \notag
\end{gather}%
where $\psi ^{\ast }$ is complex conjugate to $\psi $

Observe that EM part $\mathcal{L}_{\mathrm{BEM}}$ of the Lagrangian $%
\mathcal{L}$ according to (\ref{ilag3a}) is obtained by the removal from the
classical EM Lagrangian $\mathcal{L}_{\mathrm{CEM}}$ all self-interaction
contributions $\mathcal{L}_{\mathrm{e}}$ of the elementary EM fields and it
can be recast as%
\begin{equation}
\mathcal{L}_{\mathrm{BEM}}=-\sum\limits_{\left\{ \ell ,\ell ^{\prime
}\right\} :\ell ^{\prime }\neq \ell }\frac{F^{\ell \mu \nu }F_{\mu \nu
}^{\ell ^{\prime }}}{16\pi }=-\sum\limits_{1\leq \ell \leq N}\frac{F^{\ell
\mu \nu }F_{\neq \mu \nu }^{\ell }}{16\pi }.  \label{ilag5}
\end{equation}%
The "bare" charge Lagrangians $L^{\ell }$ are defined by exactly same
expressions as in \cite{BabFig1}, \cite{BabFig2}, namely 
\begin{equation}
L^{\ell }\left( \psi ^{\ell },\psi _{;\mu }^{\ell },\psi ^{\ell \ast },\psi
_{;\mu }^{\ell \ast }\right) =\frac{\chi ^{2}}{2m^{\ell }}\left\{ \psi
_{;\mu }^{\ell \ast }\psi ^{\ell ;\mu }-\kappa ^{\ell 2}\psi ^{\ell \ast
}\psi ^{\ell }-G^{\ell }\left( \psi ^{\ell \ast }\psi ^{\ell }\right)
\right\} ,  \label{ilag6}
\end{equation}%
where (i) $G^{\ell }$ is a nonlinear self-interaction function of the $\ell $%
-th charge described below; (ii) $m^{\ell }>0$ is the charge mass; (iii) $%
q^{\ell }$ is the value of the charge; (iv) $\chi >0$ is a constant similar
to the Planck constant $\hbar =\frac{h}{2\pi }$ and 
\begin{equation}
\kappa ^{\ell }=\frac{\omega ^{\ell }}{\mathrm{c}}=\frac{m^{\ell }\mathrm{c}%
}{\chi },\qquad \omega ^{\ell }=\frac{m^{\ell }\mathrm{c}^{2}}{\chi }.
\label{ilag7}
\end{equation}%
The system Lagrangian$\mathcal{\ L}$ defined by (\ref{ilag5})-(\ref{ilag7})
is manifestly Lorentz and gauge invariant with respect to the gauge
transformations of the first kind (\ref{flagr8}). The gauge invariance via
the Noether's theorem allows to introduce elementary conserved currents, 
\cite{BabFig1}, \cite{BabFig2},%
\begin{equation}
J^{\ell \nu }=-\mathrm{i}\frac{q^{\ell }}{\chi }\left( \frac{\partial
L^{\ell }}{\partial \psi _{;\nu }^{\ell }}\psi ^{\ell }-\frac{\partial
L^{\ell }}{\partial \psi _{;\nu }^{\ast \ell }}\psi ^{\ast \ell }\right) =-%
\mathrm{c}\frac{\partial L^{\ell }}{\partial A_{\neq \nu }^{\ell }},
\label{ilag8}
\end{equation}%
with the conservation law%
\begin{equation}
\partial _{\nu }J^{\ell \nu }=0,\ \partial _{t}\rho ^{\ell }+\nabla \cdot 
\mathbf{J}^{\ell }=0,\ J^{\ell \nu }=\left( \rho ^{\ell }\mathrm{c},\mathbf{J%
}^{\ell }\right) .  \label{ilag8a}
\end{equation}%
The Euler-Lagrange field equations for the above Lagrangian $\mathcal{L}$
are (i) \emph{elementary wave equations}%
\begin{equation}
\left[ \tilde{\partial}_{\mu }^{\ell }\tilde{\partial}^{\ell \mu }+\kappa
^{\ell 2}+G^{\ell \prime }\left( \left\vert \psi ^{\ell }\right\vert
^{2}\right) \right] \psi ^{\ell }=0,\ \tilde{\partial}^{\ell \mu }=\partial
^{\mu }+\frac{\mathrm{i}q^{\ell }A_{\neq }^{\ell \mu }}{\chi \mathrm{c}},
\label{ilag9}
\end{equation}%
together with the conjugate equation for $\psi ^{\ast \ell }$%
\begin{equation}
\left[ \tilde{\partial}_{\mu }^{\ell \ast }\tilde{\partial}^{\ell \ast \mu
}+\kappa ^{\ell 2}+G^{\ell \prime }\left( \left\vert \psi ^{\ell
}\right\vert ^{2}\right) \right] \psi ^{\ast \ell }=0,\ \tilde{\partial}%
^{\ell \ast \mu }=\partial ^{\mu }-\frac{\mathrm{i}q^{\ell }A_{\neq }^{\ell
\mu }}{\chi \mathrm{c}},  \label{ilag10}
\end{equation}%
and (ii) the Maxwell equations for the elementary EM fields%
\begin{equation}
\partial _{\mu }F^{\ell \mu \nu }=\frac{4\pi }{\mathrm{c}}J^{\ell \nu },
\label{ilag11}
\end{equation}%
with the familiar vector form%
\begin{equation}
\nabla \cdot \mathbf{E}^{\ell }=4\pi \varrho ^{\ell },\qquad \nabla \cdot 
\mathbf{B}^{\ell }=0,  \label{ilag11a}
\end{equation}%
\begin{equation}
\nabla \times \mathbf{E}^{\ell }+\frac{1}{\mathrm{c}}\partial _{t}\mathbf{B}%
^{\ell }=0,\qquad \nabla \times \mathbf{B}^{\ell }-\frac{1}{\mathrm{c}}%
\partial _{t}\mathbf{E}^{\ell }=\frac{4\pi }{\mathrm{c}}\mathbf{J}^{\ell }.
\label{ilag11b}
\end{equation}%
We will refer to the equations (\ref{ilag11}), (\ref{ilag11a}) as the \emph{%
elementary Maxwell equations} and to the field equations (\ref{ilag9})-(\ref%
{ilag11}) as \emph{field equations for b-charges}%
\index{field equation!balanced charges}. Using (\ref{ilag6})-(\ref{ilag8})
we obtain the following representation for the $\ell $-th \emph{elementary
current} $J^{\ell \nu }=\left( \mathrm{c}\rho ^{\ell },\mathbf{J}^{\ell
}\right) $%
\begin{gather}
J^{\ell \nu }=-\mathrm{i}%
\frac{q^{\ell }\chi \left[ \left( \tilde{\partial}^{\ell \nu \ast }\psi
^{\ell \ast }\right) \psi ^{\ell }-\psi ^{\ell \ast }\tilde{\partial}^{\ell
\nu }\psi ^{\ell }\right] }{2m^{\ell }}=  \label{ilag12} \\
=-\frac{q^{\ell }\chi \left\vert \psi ^{\ell }\right\vert ^{2}}{m^{\ell }}%
\left( \func{Im}\frac{\partial ^{\nu }\psi ^{\ell }}{\psi ^{\ell }}+\frac{%
q^{\ell }A_{\neq }^{\ell \nu }}{\chi \mathrm{c}}\right) ,  \notag
\end{gather}%
or its vector form%
\begin{gather}
\rho ^{\ell }=-\frac{q^{\ell }\left\vert \psi ^{\ell }\right\vert ^{2}}{%
m^{\ell }\mathrm{c}^{2}}\left( \chi \func{Im}\frac{\partial _{t}\psi ^{\ell }%
}{\psi ^{\ell }}+q^{\ell }\varphi _{\neq }^{\ell }\right) ,  \label{mplag10}
\\
\mathbf{J}^{\ell }=\frac{q^{\ell }\left\vert \psi ^{\ell }\right\vert ^{2}}{%
m^{\ell }}\left( \chi \func{Im}\frac{\nabla \psi ^{\ell }}{\psi ^{\ell }}-%
\frac{q^{\ell }\mathbf{A}_{\neq }^{\ell }}{\mathrm{c}}\right) .
\label{mplag11}
\end{gather}%
The Maxwell equations for elementary EM fields (\ref{ilag11}) combined with
the equalities (\ref{ilag1})-(\ref{ilag2}) readily imply that total and
exterior fields also satisfy the Maxwell equations%
\begin{gather}
\partial _{\mu }\mathcal{F}^{\ell \mu \nu }=\frac{4\pi }{\mathrm{c}}\mathcal{%
J}^{\nu },\text{ where }\mathcal{J}^{\nu }=\sum\limits_{1\leq \ell \leq
N}J^{\ell \nu },  \label{ilag13} \\
\partial _{\mu }F_{\neq }^{\ell \mu \nu }=\frac{4\pi }{\mathrm{c}}J_{\neq
}^{\nu },\text{ where }J_{\neq }^{\nu }=\sum\limits_{\ell ^{\prime }\neq
\ell }J^{\ell ^{\prime }\nu }.  \label{ilag14}
\end{gather}

Observe that: (i) the field equations (\ref{ilag11}) for the elementary EM
fields are exactly the Maxwell equations with the corresponding elementary
currents; (ii) every elementary wave equation (\ref{ilag9})-(\ref{ilag11})
indicates that the $\ell $-th charge is driven by its exterior potential $%
A_{\neq }^{\ell \nu }$ indicating that there is no self-interaction. We can
see also from equation (\ref{ilag13}) that the total EM field satisfies the
Maxwell equations for the total currents as in the\ CEM theory.

It is instructive to see how the BEM system Lagrangian $\mathcal{L}$ defined
by (\ref{ilag3})-(\ref{ilag5}) can be obtained by a modification of a
similar WCM Lagrangian involving the classical single EM field introduced in
, \cite{BabFig1}, \cite{BabFig2}. The modification consists of two actions:
(i) alteration of the covariant derivatives (\ref{ilag4}) which removes the
self-action by using exterior potentials $A_{\neq }^{\ell \mu }$ instead of
the same total field potential $\mathcal{A}^{\mu }$; (ii) subtracting from
the classical action of the total EM field the sum $\mathcal{L}_{\mathrm{e}}$
of the classical actions of the individual EM fields as in (\ref{ilag3}). It
is the two described actions combined yield an EM theory with the Maxwell
equations and the Lorentz force densities as its exact components. The
natural alteration of the covariant derivatives alone is not sufficient
since then the field equations will be not exactly the Maxwell equations nor
there will be force densities described exactly by the Lorentz formula.
Importantly, an additional action - the subtraction of the Lagrangian
component $\mathcal{L}_{\mathrm{e}}$ defined in (\ref{ilag3}) - is necessary
to have the Maxwell equations and the Lorentz forces as fundamentally exact
parts of the new EM theory. One can see already significant differences
brought into the EM theory by many elementary EM fields. It is evident from
the formula (\ref{ilag5}) for $\mathcal{L}_{\mathrm{BEM}}$ that the EM
action is the sum of elementary EM actions $\mathcal{L}_{\mathrm{BEM}}^{\ell
\ell ^{\prime }}$ associated with all pairs $\left\{ \ell ,\ell ^{\prime
}\right\} :\ell ^{\prime }\neq \ell $ of elementary charges and every
elementary action $\mathcal{L}_{\mathrm{BEM}}^{\ell \ell ^{\prime }}$
depends on the fields $F^{\ell \mu \nu }$ and $F^{\ell ^{\prime }\mu \nu }$
only. Continue this line we observe that though the EM Lagrangian $\mathcal{L%
}_{\mathrm{BEM}}$ according to formula (\ref{ilag5}) is a simple and natural
summatory function of the elementary EM fields, it can not be reduced
exactly to any function of a single EM field as in the CEM Lagrangian.

In what follows we often use a vector form of the system Lagrangian $%
\mathcal{L}$ defined by (\ref{ilag3}), (\ref{ilag3a}), that\ is 
\begin{gather}
\mathcal{L}\left( \left\{ \psi ^{\ell }\right\} _{\ell =1}^{N},\left\{
\left( \varphi ^{\ell },\mathbf{A}^{\ell }\right) \right\} _{\ell
=1}^{N}\right) =\mathcal{L}_{\mathrm{BEM}}+  \label{relsh} \\
+\sum\limits_{1\leq \ell \leq N}\frac{\chi ^{2}}{2m^{\ell }}\left\{ \frac{%
\left\vert \tilde{\partial}_{t}^{\ell }\psi ^{\ell }\right\vert ^{2}}{%
\mathrm{c}^{2}}-\left\vert \tilde{\nabla}\psi ^{\ell }\right\vert
^{2}-\kappa _{0\ell }^{2}\left\vert \psi ^{\ell }\right\vert ^{2}-G^{\ell
}\left( \psi ^{\ell \ast }\psi ^{\ell }\right) \right\}  \notag
\end{gather}%
where%
\begin{gather}
\mathcal{L}_{\mathrm{BEM}}=\mathcal{L}_{\mathrm{CEM}}-\mathcal{L}_{\mathrm{e}%
},  \label{relsh1} \\
\mathcal{L}_{\mathrm{CEM}}=-\frac{1}{8\pi }\left[ \left( \nabla \varphi +%
\frac{\partial _{t}\mathbf{A}}{\mathrm{c}}\right) ^{2}-\left( \nabla \times 
\mathbf{A}\right) ^{2}\right] ,  \notag \\
\mathcal{L}_{\mathrm{e}}=\frac{1}{8\pi }\sum\limits_{1\leq \ell \leq N}\left[
\left( \nabla \varphi ^{\ell }+\frac{\partial _{t}\mathbf{A}^{\ell }}{%
\mathrm{c}}\right) ^{2}-\left( \nabla \times \mathbf{A}^{\ell }\right) ^{2}%
\right] .  \notag
\end{gather}%
The corresponding field Euler-Lagrange field equations are, first of all,
the Maxwell equations for all elementary EM potentials $\varphi ^{\ell }$, $%
\mathbf{A}^{\ell }$ 
\begin{gather}
\nabla \cdot \left( \frac{1}{\mathrm{c}}\partial _{t}\mathbf{A}^{\ell
}+\nabla \varphi ^{\ell }\right) =-4\pi \rho ^{\ell },  \label{fiash} \\
\ \nabla \times \left( \nabla \times \mathbf{A}^{\ell }\right) +\frac{1}{%
\mathrm{c}}\partial _{t}\left( \frac{1}{\mathrm{c}}\partial _{t}\mathbf{A}%
^{\ell }+\nabla \varphi ^{\ell }\right) =\frac{4\pi }{\mathrm{c}}\mathbf{J}%
^{\ell },\ \ell =1,...,N,  \label{fial}
\end{gather}%
where the charge densities and currents are defined by (\ref{mplag10}), (\ref%
{mplag11}), and, second of all, the equations for the wave functions $\psi
^{\ell }$ in the form of the \emph{nonlinear Klein-Gordon equations}%
\begin{equation}
-\frac{1}{\mathrm{c}^{2}}\tilde{\partial}_{t}^{\ell }\tilde{\partial}%
_{t}^{\ell }\psi ^{\ell }+\tilde{\nabla}^{\ell 2}\psi ^{\ell }-G^{\ell
\prime }\left( \psi ^{\ell \ast }\psi ^{\ell }\right) \psi ^{\ell }-\kappa
_{0}^{2}\psi ^{\ell }=0,\ \ell =1,...,N,  \label{KG}
\end{equation}%
and similar equations for the complex conjugate variables $\psi ^{\ell \ast
} $. Note that equations (\ref{KG}) for $\psi ^{\ell }$ are coupled with the
equations for EM potentials via the covariant derivatives (\ref{ilag3}). If
we choose for all elementary potentials the Lorentz gauge 
\begin{equation}
\frac{1}{\mathrm{c}}\partial _{t}\varphi ^{\ell }+\nabla \cdot \mathbf{A}%
^{\ell }=0,  \label{Lorg}
\end{equation}%
then equations (\ref{fiash})-(\ref{fial}) turn into the wave equations%
\begin{gather}
\nabla ^{2}\varphi ^{\ell }-\frac{1}{\mathrm{c}^{2}}\partial _{t}^{2}\varphi
^{\ell }=-4\pi \rho ^{\ell },  \label{filor} \\
\frac{1}{\mathrm{c}^{2}}\partial _{t}^{2}\mathbf{A}^{\ell }-\nabla ^{2}%
\mathbf{A}^{\ell }=\frac{4\pi }{\mathrm{c}}\mathbf{J}^{\ell },\quad \ell
=1,...,N,  \label{alor}
\end{gather}%
as in the CEM theory.

Importantly, as in the WCM, \cite{BabFig1}, \cite{BabFig2}, the
nonlinearities $G^{\ell }$ are determined based on the single $\ell $-th
charge equations (\ref{fiash})-(\ref{KG}) under the assumptions that it is
isolated and is at rest. Namely, let us consider a single b-charge, set $N=1$
in (\ref{fiash})-(\ref{KG}) and simplify notations $\psi =\psi ^{\ell }$, $%
\varphi =\varphi ^{\ell }$, $\mathbf{A}=\mathbf{A}^{\ell }$, $G^{\ell }=G$.
Then according to (\ref{ilag3}) the covariant derivatives are $\tilde{%
\partial}_{t}^{\ell }=\partial _{t}$, $\tilde{\nabla}^{\ell }=\nabla $ and
the Lagrangian for a single b-charge is 
\begin{equation}
L_{0}=\frac{\chi ^{2}}{2m}\left\{ \frac{\left\vert \tilde{\partial}_{t}\psi
\right\vert ^{2}}{\mathrm{c}^{2}}-\left\vert \tilde{\nabla}\psi \right\vert
^{2}-\kappa _{0}^{2}\left\vert \psi \right\vert ^{2}-G\left( \psi ^{\ast
}\psi \right) \right\} .  \label{blag1}
\end{equation}%
Observe that the Lagrangian $L_{0}$ above does not depend on the potentials $%
\varphi $, $\mathbf{A}$ since there is no self-interaction. Though we can
still find the potentials based on the elementary Maxwell equations (\ref%
{filor}), (\ref{alor}) they have no role to play and carry no energy. Let us
consider now the \emph{rest state of the b-charge} which we set as in the
WCM, \cite{BabFig1}, \cite{BabFig2}, to be of the form 
\begin{gather}
\psi \left( t,\mathbf{x}\right) =\mathrm{e}^{-\mathrm{i}\omega _{0}t}%
\mathring{\psi}\left( \mathbf{x}\right) ,\quad \omega _{0}=\frac{m\mathrm{c}%
^{2}}{\chi }=\mathrm{c}\kappa _{0},  \label{psioml} \\
\varphi \left( t,\mathbf{x}\right) =\mathring{\varphi}\left( \mathbf{x}%
\right) ,\quad \mathbf{A}\left( t,\mathbf{x}\right) =\mathbf{0},  \notag
\end{gather}%
where $\mathring{\psi}\left( \left\vert \mathbf{x}\right\vert \right) $ and $%
\mathring{\varphi}=\mathring{\varphi}\left( \left\vert \mathbf{x}\right\vert
\right) $ are real-valued radial functions. Substituting the $\psi $, $%
\varphi $ and $\mathbf{A}$ defined by the relations (\ref{psioml}) into the
field equations (\ref{KG})(\ref{KG}) we obtain the following \emph{rest
charge equations}:%
\begin{gather}
-\nabla ^{2}\mathring{\psi}+G^{\prime }\left( \left\vert \mathring{\psi}%
\right\vert ^{2}\right) \mathring{\psi}=0,  \label{nop40} \\
-\nabla ^{2}\mathring{\varphi}=4\pi \left\vert \mathring{\psi}\right\vert
^{2}.  \label{stst}
\end{gather}%
The quantities $\mathring{\psi}$ and $\mathring{\varphi}$ are fundamental
for our theory and we refer to them, respectively, as \emph{form factor} and 
\emph{form factor potential}. The equation (\ref{nop40}) signifies a
complete balance of the two forces acting upon the resting charge: (i)
internal elastic deformation force associated with the term $-\Delta 
\mathring{\psi}$; (ii) internal nonlinear self-interaction of the charge
associated with the term $G^{\prime }\left( |\mathring{\psi}|^{2}\right) 
\mathring{\psi}$. We refer to the equation (\ref{nop40}), which establishes
an explicit relation between the form factor $\mathring{\psi}$ and the
self-interaction nonlinearity $G,$ as the \emph{charge equilibrium equation}%
. Hence, if the form factor $\mathring{\psi}$ is given we can find from the
equilibrium equation (\ref{nop40}) the self-interaction nonlinearity $G$
which exactly produces this factor under assumptions that $\mathring{\psi}%
\left( r\right) $ is a nonnegative, monotonically decaying and sufficiently
smooth function of $r\geq 0$. \emph{Now we pick the form factor }$\mathring{%
\psi}$\emph{\ considering it as the model parameter\ and then the nonlinear
self interaction function }$G$\emph{\ is determined based on the charge
equilibrium\ equation (\ref{nop40}). As in the WCM, \cite{BabFig1}, \cite%
{BabFig2}, such a determination of the nonlinearity is a key feature of our
approach: it allows to choose the form factor }$\mathring{\psi}$\emph{\ and
then to determine matching self-interaction nonlinearity }$G$\emph{\ rather
than to deal with solving a nontrivial nonlinear partial differential
equation.}

To explicitly integrate the size of the resting b-charge into its model we
introduce \emph{size parameter} $a>0$ into $G=G_{a}$ through its derivative
as follows 
\begin{equation}
G_{a}^{\prime }\left( s\right) =a^{-2}G_{1}^{\prime }\left( a^{3}s\right) ,%
\text{ where }G^{\prime }\left( s\right) =\partial _{s}G\left( s\right) .
\label{totgkap}
\end{equation}%
In the following Section \ref{snonlin} we give examples of the nonlinearity $%
G$ and discuss its properties.

\subsection{Energy-Momentum tensors and Lorentz forces}

The conservations laws, particularly the energy-momentum conservation, play
an important role in the physics of EM phenomena. A general source of
conservation laws in Lagrangian theories is the Noether's theorem, \cite[%
Section 13.7]{Goldstein}, which yields canonically conservations laws based
on the Lagrangian symmetries, that is its invariance with respect to
continuous groups of transformations. The conservation laws are not
independent equations, and they hold only if the fields satisfy the field
Euler-Lagrange equations. The energy-momentum and charge conservations are
two important laws in any EM theory and they have special significance in
our studies for several reasons. First of all, the conservation of
energy-momentum describes their transport in the space and is directly
related to the point charge approximations. Second of all, the Lorentz force
density expression which the one of the most important components of any EM
theory arises in the energy-momentum conservation equations (\ref{maxw17})
and not in the field Maxwell equations. Third of all, the elementary charge
and energy-momentum conservation equations (\ref{ilag8a}), (\ref{iemt6})
account for charges individuality. In fact, the BEM Lagrangian is invariant
with respect to a wider group of elementary gauge transformations (\ref%
{gaum1}), (\ref{gaum2}), and it is due to this symmetry the "two-way"
representation (\ref{ilag8}) holds for every elementary current: one that
involves the differentiation with respect to the wave functions derivatives
and another one that involves the differentiation with the respect to the
elementary potentials. It is due to this two-way representation the
conserved Noether's elementary current is exactly the source current in the
corresponding elementary Maxwell equations (\ref{ilag11}), (\ref{ilag11a}).

To get further insight into b-charges properties we need to find the
symmetric EnMT $\mathcal{T}^{\mu \nu }$ of the system Lagrangian$\mathcal{\ L%
}$ defined by (\ref{ilag5})-(\ref{ilag7}) making use of a general method
used in \cite{BabFig1}, \cite{BabFig2}. The method yields the following
representation%
\begin{equation}
\mathcal{T}^{\mu \nu }=\sum\limits_{1\leq \ell \leq N}T^{\ell \mu \nu }+\Xi
^{\mu \nu },\ \Xi ^{\mu \nu }=\sum\limits_{\ell ^{\prime }\neq \ell }\Xi
^{\ell \ell ^{\prime }\mu \nu },  \label{iemt1}
\end{equation}%
where the individual EnMT $T^{\ell \mu \nu }$ of the bare $\ell $-th charge
and the EnMT components $\Xi ^{\ell \ell ^{\prime }\mu \nu }$ for the EM
fields are as follows%
\begin{gather}
T^{\ell \mu \nu }=\frac{\partial L^{\ell }}{\partial \psi _{;\mu }^{\ell }}%
\psi ^{\ell ;\nu }+\frac{\partial L^{\ell }}{\partial \psi _{;\mu }^{\ell
\ast }}\psi ^{\ell ;\nu \ast }-g^{\mu \nu }L^{\ell }=  \label{iemt2} \\
=\frac{\chi ^{2}}{2m^{\ell }}\left\{ \left( \psi ^{\ell ;\mu \ast }\psi
^{\ell ;\nu }+\psi ^{\ell ;\mu }\psi ^{\ell ;\nu \ast }\right) -\right. 
\notag \\
-\left. \left[ \psi _{;\mu }^{\ell \ast }\psi ^{\ell ;\mu }-\kappa ^{\ell
2}\psi ^{\ell \ast }\psi ^{\ell }-G^{\ell }\left( \psi ^{\ell \ast }\psi
^{\ell }\right) \right] \delta ^{\mu \nu }\right\} ,  \notag
\end{gather}%
\begin{equation}
\Xi ^{\ell \ell ^{\prime }\mu \nu }=\frac{1}{4\pi }\left( g^{\mu \gamma
}F_{\gamma \xi }^{\ell }F^{\ell ^{\prime }\xi \nu }+\frac{1}{4}g^{\mu \nu
}F_{\gamma \xi }^{\ell }F^{\ell ^{\prime }\gamma \xi }\right) ,\ \ell
^{\prime }\neq \ell .  \label{iemt3}
\end{equation}%
Notice that the expression (\ref{iemt2}) is the same as in \cite{BabFig1}, 
\cite{BabFig2} with the only difference that the covariant derivatives here
are defined by equalities (\ref{ilag4}). To emphasize notationally the new
meaning of the EnMT of the EM fields we use for it the symbol $\Xi ^{\mu \nu
}$ and we continue to use the symbol $\Theta ^{\mu \nu }$ for the classical
EnMT of the the EM field as in formula (\ref{maxw11}). As to the expression (%
\ref{iemt3}) for the individual components of the symmetric EnMT it is
obtained from the representation (\ref{ilag3}) and evidently is similar to
the expression (\ref{maxw11}) for the classical EnMT $\Theta ^{\mu \nu }$.
An alternative and more elementary way to derive the expression (\ref{iemt3}%
) is based on the expression (\ref{flagr7b}) for the canonical EnMT $%
\mathring{\Theta}^{\mu \nu }$ as it is done in \cite[Section 12.10]{Jackson}%
. Indeed, in our case the canonical EnMT takes the form%
\begin{equation}
\mathring{\Xi}^{\ell \ell ^{\prime }\mu \nu }=-\frac{F^{\ell \mu \xi
}\partial ^{\nu }A_{\xi }^{\ell ^{\prime }}}{4\pi }+g^{\mu \nu }\frac{%
F_{\gamma \xi }^{\ell }F^{\ell ^{\prime }\gamma \xi }}{16\pi },\ \ell
^{\prime }\neq \ell .  \label{iemt3a}
\end{equation}%
The identity $\partial ^{\nu }A^{\ell ^{\prime }\xi }=-F^{\xi \nu }+\partial
^{\xi }A^{\ell ^{\prime }\nu }$ allows to recast the expression (\ref{iemt3a}%
) into%
\begin{equation}
\mathring{\Xi}^{\ell \ell ^{\prime }\mu \nu }=\frac{1}{4\pi }\left( g^{\mu
\gamma }F_{\gamma \xi }^{\ell }F^{\ell ^{\prime }\xi \nu }+\frac{1}{4}g^{\mu
\nu }F_{\gamma \xi }^{\ell }F^{\ell ^{\prime }\gamma \xi }\right) -\frac{1}{%
4\pi }g^{\mu \gamma }F_{\gamma \xi }^{\ell }\partial ^{\xi }A^{\ell ^{\prime
}\nu }.  \label{iemt3b}
\end{equation}%
Then using the Maxwell equations (\ref{ilag11}) we can recast the second
term as 
\begin{equation}
-\frac{1}{4\pi }g^{\mu \gamma }F_{\gamma \xi }^{\ell }\partial ^{\xi
}A^{\ell ^{\prime }\nu }=\partial ^{\xi }\left( -\frac{1}{4\pi }g^{\mu
\gamma }F_{\gamma \xi }^{\ell }A^{\ell ^{\prime }\nu }\right) -J_{\gamma \xi
}^{\ell }A^{\ell ^{\prime }\nu }.  \label{iemt3c}
\end{equation}%
The term $-J_{\gamma \xi }^{\ell }A^{\ell ^{\prime }\nu }$ combined with the
canonical EnMT $\mathring{T}^{\ell \mu \nu }$ yields the the symmetric $%
T^{\ell \mu \nu }$ and hence%
\begin{gather}
\mathcal{\mathring{T}}^{\mu \nu }=\sum\limits_{1\leq \ell \leq N}\mathring{T}%
^{\ell \mu \nu }+\sum\limits_{\ell ^{\prime }\neq \ell }\mathring{\Xi}^{\ell
\ell ^{\prime }\mu \nu }=  \label{iemt3d} \\
=\sum\limits_{1\leq \ell \leq N}T^{\ell \mu \nu }+\sum\limits_{\ell ^{\prime
}\neq \ell }\Xi ^{\ell \ell ^{\prime }\mu \nu }-\frac{1}{4\pi }\partial
^{\xi }\left( \sum\limits_{\ell ^{\prime }\neq \ell }g^{\mu \gamma
}F_{\gamma \xi }^{\ell }A^{\ell ^{\prime }\nu }\right) .  \notag
\end{gather}%
The later by the standard argument implies the conservation law%
\begin{equation}
\partial _{\mu }\mathcal{T}^{\mu \nu }=0  \label{iemt4c}
\end{equation}%
for the symmetric EnMT $\mathcal{T}^{\mu \nu }$ defined by the expressions (%
\ref{iemt1})-(\ref{iemt3}).

Observe that using representations (\ref{iemt3}) we can also recast
expression (\ref{iemt1}) for EnMT $\Xi ^{\mu \nu }$ as follows%
\begin{equation}
\Xi ^{\mu \nu }=\sum\limits_{\ell ^{\prime }\neq \ell }\Xi ^{\ell \ell
^{\prime }\mu \nu }=\sum\limits_{1\leq \ell \leq N}\Xi ^{\ell \mu \nu },
\label{iemt4}
\end{equation}%
where%
\begin{gather}
\Xi ^{\ell \ell ^{\prime }\mu \nu }=\frac{1}{4\pi }\left( g^{\mu \gamma
}F_{\gamma \xi }^{\ell }F^{\ell ^{\prime }\xi \nu }+\frac{1}{4}g^{\mu \nu
}F_{\gamma \xi }^{\ell }F^{\ell ^{\prime }\gamma \xi }\right) ,
\label{iemt4b} \\
\Xi ^{\ell \mu \nu }=\frac{1}{4\pi }\left( g^{\mu \gamma }F_{\gamma \xi
}^{\ell }F_{\neq }^{\ell \xi \nu }+\frac{1}{4}g^{\mu \nu }F_{\gamma \xi
}^{\ell }F_{\neq }^{\ell \gamma \xi }\right) .  \notag
\end{gather}%
The relation between the classical and the new EnMTs $\Theta ^{\mu \nu }$
and $\Xi ^{\mu \nu }$ in view of their representations (\ref{iemt1}), (\ref%
{iemt3}) and (\ref{maxw11}) is as follows%
\begin{gather}
\Xi ^{\mu \nu }=\sum\limits_{\ell ^{\prime }\neq \ell }\Xi ^{\ell \ell
^{\prime }\mu \nu }=\Theta ^{\mu \nu }\left( \sum\limits_{1\leq \ell \leq
N}F^{\ell \mu \nu }\right) -\sum\limits_{1\leq \ell \leq N}\Theta ^{\mu \nu
}\left( F^{\ell \mu \nu }\right) =  \label{iemt4d} \\
=\Theta ^{\mu \nu }\left( \mathcal{F}^{\mu \nu }\right) -\sum\limits_{1\leq
\ell \leq N}\Theta ^{\mu \nu }\left( F^{\ell \mu \nu }\right) .  \notag
\end{gather}%
Evidently the formula (\ref{iemt4d}) reads that the new expression of the
EnMT equals to the classical EnMT of the total EM field minus the sum of the
the the classical EnMTs for the elementary EM fields.

Based on the formula (\ref{iemt3}) for the components $\Xi ^{\ell \ell
^{\prime }}$ for $\ell ^{\prime }\neq \ell $ of the symmetric EnMT we can
represent their entries in terms of the fields $\mathbf{E}^{\ell }$ and $%
\mathbf{B}^{\ell }$, namely%
\begin{eqnarray}
w^{\ell \ell ^{\prime }} &=&\Xi ^{\ell \ell ^{\prime }00}=\frac{\mathbf{E}%
^{\ell }\cdot \mathbf{E}^{\ell ^{\prime }}+\mathbf{B}^{\ell }\cdot \mathbf{B}%
^{\ell ^{\prime }}}{8\pi },  \label{iemt4e} \\
\mathrm{c}g_{i}^{\ell \ell ^{\prime }} &=&\Xi ^{\ell \ell ^{\prime }0i}=\Xi
^{\ell \ell ^{\prime }i0}=\frac{\mathbf{E}^{\ell }\times \mathbf{B}^{\ell
^{\prime }}}{4\pi },  \notag
\end{eqnarray}%
\begin{equation}
\Xi ^{\ell \ell ^{\prime }ij}=-\frac{1}{8\pi }\left[ E_{i}^{\ell
}E_{j}^{\ell ^{\prime }}+B_{i}^{\ell }B_{j}^{\ell ^{\prime }}-\frac{1}{2}%
\delta _{ij}\left( \mathbf{E}^{\ell }\cdot \mathbf{E}^{\ell ^{\prime }}+%
\mathbf{B}^{\ell }\cdot \mathbf{B}^{\ell ^{\prime }}\right) \right] ,
\label{iemt4f}
\end{equation}%
\begin{equation}
\Xi ^{\ell \ell ^{\prime }\alpha \beta }=%
\begin{bmatrix}
w^{\ell \ell ^{\prime }} & \mathrm{c}\mathbf{g}^{\ell \ell ^{\prime }} \\ 
\mathrm{c}\mathbf{g}^{\ell \ell ^{\prime }} & -\tau _{ij}^{\ell \ell
^{\prime }}%
\end{bmatrix}%
,  \label{iemt4g}
\end{equation}%
which are evidently similar to the same entries (\ref{maxw12})-(\ref{maxw14}%
) for the classical EM theory. In particular, we have the following
expression for the Poynting vectors similar to (\ref{maxw16}) 
\begin{equation}
\mathbf{S}^{\ell \ell ^{\prime }}=\mathrm{c}^{2}g_{i}^{\ell \ell ^{\prime }}=%
\frac{\mathrm{c}\mathbf{E}^{\ell }\times \mathbf{B}^{\ell ^{\prime }}}{4\pi }%
.  \label{iemt4h}
\end{equation}%
To find expressions for the EM force densities acting upon b-charges let us
examine their conservation laws. Combining the Maxwell field equations (\ref%
{ilag11}), (\ref{ilag14}), an elementary identity%
\begin{equation}
\partial ^{\mu }F^{\xi \nu }-\partial ^{\xi }F^{\mu \nu }=\partial ^{\nu
}F^{\mu \xi },\text{ for }F^{\mu \xi }=\partial ^{\mu }A^{\xi }-\partial
^{\xi }A^{\mu },  \label{iemt4a}
\end{equation}%
and the antisymmetry of the EM field tensors $F^{\ell \mu \nu }$ we obtain
for $\ell ^{\prime }\neq \ell $:%
\index{conservation laws!energy-momentum!detailed} 
\begin{equation}
\partial _{\mu }\Xi ^{\ell \ell ^{\prime }\mu \nu }=%
\frac{1}{\mathrm{c}}J_{\xi }^{\ell }F^{\ell ^{\prime }\xi \nu }+\frac{1}{%
4\pi }\left[ -\frac{1}{2}F_{\mu \xi }^{\ell }\partial ^{\nu }F^{\ell
^{\prime }\mu \xi }+\frac{1}{4}\partial ^{\nu }\left( F_{\gamma \xi }^{\ell
}F^{\ell ^{\prime }\gamma \xi }\right) \right] ,  \label{iemt5}
\end{equation}%
Let us introduce an \emph{elementary EM field interaction energy for the pair%
} $\left\{ \ell ,\ell ^{\prime }\right\} $ of b-charges%
\begin{equation}
\Xi ^{\left\{ \ell ,\ell ^{\prime }\right\} \mu \nu }=\Xi ^{\ell \ell
^{\prime }\mu \nu }+\Xi ^{\ell ^{\prime }\ell \mu \nu },  \label{iemt5b}
\end{equation}%
which in view of (\ref{iemt5}) satisfies the following \emph{elementary
energy-momentum conservation laws} 
\begin{equation}
\partial _{\mu }\Xi ^{\left\{ \ell ,\ell ^{\prime }\right\} \mu \nu }=-\frac{%
1}{\mathrm{c}}\left( J_{\xi }^{\ell }F^{\ell ^{\prime }\nu \xi }+J_{\xi
}^{\ell ^{\prime }}F^{\ell \nu \xi }\right)  \label{iemt6}
\end{equation}%
with the right-hand side being the negative of the sum of the corresponding
Lorentz force density.

It readily follows from relations (\ref{iemt5b}) and (\ref{iemt4b}) that the
interaction EnMT $\Xi ^{\left\{ \ell ,\ell ^{\prime }\right\} \mu \nu }$ has
the following representation in terms of the classical EnMT $\Theta ^{\mu
\nu }$:%
\begin{equation}
\Xi ^{\left\{ \ell ,\ell ^{\prime }\right\} \mu \nu }=\Theta ^{\mu \nu
}\left( F^{\ell \mu \nu }+F^{\ell ^{\prime }\mu \nu }\right) -\Theta ^{\mu
\nu }\left( F^{\ell \mu \nu }\right) -\Theta ^{\mu \nu }\left( F^{\ell
^{\prime }\mu \nu }\right) .  \label{iemt5a}
\end{equation}%
The representation (\ref{iemt5a}) is evidently similar to (\ref{iemt4d}).

Notice that in view of (\ref{iemt4e}), (\ref{iemt4f}) and (\ref{iemt4h}) the
entries of the tensor $\Xi ^{\left\{ \ell ,\ell ^{\prime }\right\} \mu \nu }$
are as follows%
\begin{gather}
w^{\left\{ \ell ,\ell ^{\prime }\right\} }=\Xi ^{\left\{ \ell ,\ell ^{\prime
}\right\} 00}=\frac{\mathbf{E}^{\ell }\cdot \mathbf{E}^{\ell ^{\prime }}+%
\mathbf{B}^{\ell }\cdot \mathbf{B}^{\ell ^{\prime }}}{4\pi },  \label{iemt6b}
\\
\mathrm{c}g_{i}^{\left\{ \ell ,\ell ^{\prime }\right\} }=\Xi ^{\left\{ \ell
,\ell ^{\prime }\right\} 0i}=\Xi ^{\left\{ \ell ,\ell ^{\prime }\right\} i0}=%
\frac{\mathbf{E}^{\ell }\times \mathbf{B}^{\ell ^{\prime }}+\mathbf{E}^{\ell
^{\prime }}\times \mathbf{B}^{\ell }}{4\pi },  \notag \\
\mathbf{S}^{\left\{ \ell ,\ell ^{\prime }\right\} }=\mathrm{c}\frac{\mathbf{E%
}^{\ell }\times \mathbf{B}^{\ell ^{\prime }}+\mathbf{E}^{\ell ^{\prime
}}\times \mathbf{B}^{\ell }}{4\pi },  \notag
\end{gather}%
\begin{gather}
\tau _{ij}^{\left\{ \ell ,\ell ^{\prime }\right\} }=\Xi ^{\left\{ \ell ,\ell
^{\prime }\right\} ij}=  \label{iemt6c} \\
=-\frac{1}{4\pi }\left[ E_{i}^{\ell }E_{j}^{\ell ^{\prime }}+B_{i}^{\ell
}B_{j}^{\ell ^{\prime }}-\frac{1}{2}\delta _{ij}\left( \mathbf{E}^{\ell
}\cdot \mathbf{E}^{\ell ^{\prime }}+\mathbf{B}^{\ell }\cdot \mathbf{B}^{\ell
^{\prime }}\right) \right] .  \notag
\end{gather}%
The expressions (\ref{iemt6b}) and (\ref{iemt6c}) can be alternatively
derived from their classical counterparts (\ref{maxw12}), (\ref{maxw13}) and
(\ref{maxw16}) based on relation (\ref{iemt5a})

The relations (\ref{iemt1}) together with (\ref{iemt6}) readily imply the
total energy-momentum conservation law%
\begin{equation}
\partial _{\mu }\Xi ^{\mu \nu }=\frac{1}{2}\sum\limits_{\ell ^{\prime }\neq
\ell }\partial _{\mu }\left( \Xi ^{\ell \ell ^{\prime }\mu \nu }+\Xi ^{\ell
^{\prime }\ell \mu \nu }\right) =-\frac{1}{\mathrm{c}}\sum\limits_{1\leq
\ell \leq N}J_{\xi }^{\ell }F_{\neq }^{\ell \nu \xi },  \label{iemt7}
\end{equation}%
with the sum of the negative of the Lorentz forces in the right-hand side.

Using expression (\ref{iemt2}) for $T^{\ell \mu \nu }$ and the same
transformations as in the similar case considered in \cite{BabFig2} we
obtain the following elementary conservation laws%
\index{conservation laws!energy-momentum!detailed}%
\begin{equation}
\partial _{\mu }T^{\ell \mu \nu }=%
\frac{1}{\mathrm{c}}J_{\xi }^{\ell }\sum\limits_{\ell ^{\prime }\neq \ell
}F^{\ell ^{\prime }\nu \xi }=\frac{1}{\mathrm{c}}J_{\xi }^{\ell }F_{\neq
}^{\ell \nu \xi }.  \label{iemt8}
\end{equation}%
Observe that the expression on the right-hand side of the above equality is
the Lorentz force density 4-vector acting upon the 4-vector current $J_{\xi
}^{\ell }$ by its exterior field $F_{\neq }^{\ell \nu \xi }$, and the same
vectors with the minus sign arise in the conservations laws for the EM
fields (\ref{iemt6}) and (\ref{iemt7}). Notice also that the natural
partition (\ref{iemt1})-(\ref{iemt3}) of the system EnMT into charges EnMTs $%
T^{\ell \mu \nu }$ and the EnMTs $\Xi ^{\left\{ \ell ,\ell ^{\prime
}\right\} \mu \nu }$ for pairs $\left\{ \ell ,\ell ^{\prime }\right\} $ of
interacting elementary EM fields produces exactly Lorentz force densities in
the elementary conservation laws (\ref{iemt6}), (\ref{iemt7}) and (\ref%
{iemt8}). This provides a solid justification for the energy-momentum
partition (\ref{iemt1})-(\ref{iemt3}), and it seems there is no simple way
to alter it when preserving the Lorentz forces expression.

The vector form of the elementary conservation law (\ref{iemt6}) is similar
to the CEM theory conservation law \ref{maxw18}, \cite[Section 6.8]{Jackson1}%
, namely 
\begin{equation}
\frac{1}{\mathrm{c}}\left[ \partial _{t}w^{\left\{ \ell ,\ell ^{\prime
}\right\} }+\nabla \cdot \mathbf{S}^{\left\{ \ell ,\ell ^{\prime }\right\} }%
\right] =-\frac{1}{\mathrm{c}}\left[ \mathbf{J}^{\ell }\cdot \mathbf{E}%
^{\ell ^{\prime }}+\mathbf{J}^{\ell ^{\prime }}\cdot \mathbf{E}^{\ell }%
\right] ,  \label{iemt9}
\end{equation}%
\begin{equation}
\frac{\partial g_{i}^{\left\{ \ell ,\ell ^{\prime }\right\} }}{\partial t}%
-\sum\limits_{j=1}^{3}\frac{\partial }{\partial x^{j}}\tau _{ij}^{\left\{
\ell ,\ell ^{\prime }\right\} }=f_{i}^{\left\{ \ell ,\ell ^{\prime }\right\}
},\quad i=1,2,3,  \label{iemt10}
\end{equation}%
where $f_{i}^{\left\{ \ell ,\ell ^{\prime }\right\} }$ is the Lorentz force
density satisfying 
\begin{equation}
f_{i}^{\left\{ \ell ,\ell ^{\prime }\right\} }=f_{i}^{\ell \ell ^{\prime
}}+f_{i}^{\ell ^{\prime }\ell },\quad f_{i}^{\ell \ell ^{\prime }}=\rho
^{\ell }E_{i}^{\ell ^{\prime }}+\frac{1}{\mathrm{c}}\left( \mathbf{J}^{\ell
}\times \mathbf{B}^{\ell ^{\prime }}\right) _{i}.  \label{iemt10a}
\end{equation}

\begin{remark}
The conservation law (\ref{iemt9}) can be easily derived from the classical
energy conservation law (the Poynting theorem) as follows. Let us consider
the classical energy conservation law associated with the 3 Maxwell
equations (\ref{ilag11}): (i) for $\ell $-th EM fields, (ii) for $\ell
^{\prime }$-th EM field and (iii) for their sum. Then subtracting from the
conservation law for the sum of the fields the sum of conservation laws for $%
\ell $-th and $\ell ^{\prime }$-the fields we obtain exactly the
conservation law (\ref{iemt9}).
\end{remark}

\subsection{Elementary EM fields for prescribed elementary currents\label%
{semprecur}}

To study the properties of the elementary EM fields for balanced changes and
the energy-momentum transfer in the space-time it is instructive to consider
a situation of prescribed currents for b-charges similarly to the same for
the case the CEM theory Lagrangian (\ref{flagr7}). In the BEM theory the
corresponding Lagrangian $\mathcal{L}_{\mathrm{EMJ}}$ is based on the
representations (\ref{ilag3}) and is of the form 
\begin{equation}
\mathcal{L}_{\mathrm{BEMJ}}=-\frac{\mathcal{F}^{\mu \nu }\mathcal{F}_{\mu
\nu }}{16\pi }+\sum\limits_{1\leq \ell \leq N}\frac{F^{\ell \mu \nu }F_{\mu
\nu }^{\ell }}{16\pi }-\frac{1}{\mathrm{c}}\sum\limits_{1\leq \ell \leq
N}J_{\mu }^{\ell }A_{\neq }^{\ell \mu }.  \label{lagic1}
\end{equation}%
Based on the representation (\ref{ilag5}) it can be alternatively written as 
\begin{gather}
\mathcal{L}_{\mathrm{BEMJ}}=-\sum\limits_{\left\{ \ell ,\ell ^{\prime
}\right\} :\ell ^{\prime }\neq \ell }\frac{F^{\ell \mu \nu }F_{\mu \nu
}^{\ell ^{\prime }}}{16\pi }-\frac{1}{\mathrm{c}}\sum\limits_{1\leq \ell
\leq N}J_{\mu }^{\ell }A_{\neq }^{\ell \mu }  \label{lagic2} \\
=-\sum\limits_{1\leq \ell \leq N}\frac{F^{\ell \mu \nu }F_{\neq \mu \nu
}^{\ell }}{16\pi }-\frac{1}{\mathrm{c}}\sum\limits_{1\leq \ell \leq
N}J^{\ell \nu }A_{\neq \nu }^{\ell }  \notag
\end{gather}%
The field equations for the Lagrangian $\mathcal{L}_{\mathrm{EMJ}}$ can be
obtained directly from the variational principle. Indeed, taking into
account the antisymmetry of every field tensor $F^{\ell \mu \nu }$, the fact
every pair $\ell ,\ell ^{\prime }$ with $\ell ^{\prime }\neq \ell $ appear
two times in the Lagrangian representation (\ref{lagic2}) and the usual
assumption of the decay of all the fields at infinity we obtain 
\begin{equation}
\delta \int \mathcal{L}_{\mathrm{BEMJ}}\,\mathrm{d}x=\int \sum\limits_{1\leq
\ell \leq N}\left( \frac{1}{4\pi }\partial _{\mu }F^{\ell \mu \nu }-\frac{1}{%
\mathrm{c}}J^{\ell \nu }\right) \delta A_{\neq \nu }^{\ell }\,\mathrm{d}x
\label{lagic3}
\end{equation}%
Since the variables $\delta A_{\neq \nu }^{\ell }$, $1\leq \ell \leq N$ can
vary independently (see relations (\ref{gaum3}) and (\ref{gaum4})) the
requirement for the above variation to vanish yield the field equations
which are exactly the same Maxwell equations for the elementary EM fields as
in (\ref{ilag11}), (\ref{ilag11a}), (\ref{ilag11a}).

Importantly, the system of elementary EM fields with prescribed currents
satisfy the elementary conservation laws (\ref{iemt9}), (\ref{iemt10}) which
can also be directly derived from the Maxwell equations (\ref{ilag11b}).
Indeed using the general vector identity (\ref{vecf7a}) and the Maxwell
equations for the indices $\ell $ and $\ell ^{\prime }$ we obtain 
\begin{gather}
\nabla \cdot \left( \mathbf{E}^{\ell }\times \mathbf{B}^{\ell ^{\prime
}}\right) =\mathbf{B}^{\ell ^{\prime }}\cdot \left( \nabla \times \mathbf{E}%
^{\ell }\right) -\mathbf{E}^{\ell }\cdot \left( \nabla \times \mathbf{B}%
^{\ell ^{\prime }}\right) =  \label{iemt11} \\
=-\mathbf{B}^{\ell ^{\prime }}\cdot \frac{1}{\mathrm{c}}\partial _{t}\mathbf{%
B}^{\ell }-\mathbf{E}^{\ell }\cdot \left( \frac{1}{\mathrm{c}}\partial _{t}%
\mathbf{E}^{\ell ^{\prime }}\mathbf{+}\frac{4\pi }{\mathrm{c}}\mathbf{J}%
^{\ell ^{\prime }}\right) .  \notag
\end{gather}%
Adding then to the identity (\ref{iemt11}) a similar one obtained from it by
swapping indices $\ell $ and $\ell ^{\prime }$ we obtain the elementary
energy conservation (\ref{iemt9}). Similar direct derivation is possible for
the elementary momentum conservation (\ref{iemt10}).

\emph{Observe also that according to the BEM conservation laws (\ref{iemt6})
the energy and the momentum are assigned not to the elementary EM fields by
themselves but rather to their\ interacting pairs. This is a noticeable
change compared to the CEM theory where the EM field has an energy and a
momentum on its own.}

\subsubsection{Dipole elementary currents}

For simplicity sake let us assume all prescribed elementary currents to be
in the form of ideal electric dipoles. Recall that an ideal electric dipole
source concentrated at a point $\mathbf{x}_{0}$, with the charge and current
densities $\mathbf{J}_{\mathrm{d}}$ and $\rho _{\mathrm{d}}$, is defined as
follows, \cite[Section 7.10, (7.151), Appendix 8]{van Bladel}, 
\begin{gather}
\mathbf{J}\left( t,\mathbf{x}\right) =\partial _{t}\left[ \mathbf{p}\left(
t\right) \delta \left( \mathbf{x}-\mathbf{x}_{0}\right) \right] =\mathbf{%
\dot{p}}\left( t\right) \delta \left( \mathbf{x}-\mathbf{x}_{0}\right) ,
\label{dipo1} \\
\rho \left( t,\mathbf{x}\right) =-\nabla \cdot \left[ \mathbf{p}\left(
t\right) \delta \left( \mathbf{x}-\mathbf{x}_{0}\right) \right] =-\mathbf{p}%
\left( t\right) \cdot \nabla \delta \left( \mathbf{x}-\mathbf{x}_{0}\right) 
\notag
\end{gather}%
where $\delta \left( \mathbf{x}-\mathbf{x}_{0}\right) $ is the Dirac
delta-function, and $\mathbf{p}\left( t\right) $ and $\partial _{t}\mathbf{p}%
\left( t\right) $ satisfy 
\begin{equation}
p_{j}\left( t\right) =\int x_{j}^{\prime }\rho \left( t,\mathbf{x}^{\prime
}\right) \,\mathrm{d}\mathbf{x}^{\prime },\quad j=1,2,3,\quad \mathbf{p}%
=\left( p_{1},p_{2},p_{3}\right) ,  \label{dipo2}
\end{equation}%
\begin{equation}
\mathbf{\dot{p}}\left( t\right) =\partial _{t}\mathbf{p}\left( t\right)
=\int \mathbf{J}\left( t,\mathbf{x}^{\prime }\right) \,\mathrm{d}\mathbf{x}%
^{\prime }.  \label{dipo3}
\end{equation}%
It readily follows from (\ref{dipo1}) (\ref{grmax12}) that the potentials of
the ideal electric dipole are 
\begin{equation}
\varphi \left( t,\mathbf{x}\right) =0,\quad \mathbf{A}\left( t,\mathbf{x}%
\right) =\frac{\mathbf{\dot{p}}\left( t_{0}\right) }{\mathrm{c}\left\vert 
\mathbf{R}\right\vert },  \label{dipo3a}
\end{equation}%
where%
\begin{equation}
\mathbf{R}=\mathbf{x}-\mathbf{x}_{0},\quad \mathbf{\hat{R}}=\frac{\mathbf{R}%
}{\left\vert \mathbf{R}\right\vert },\quad t_{0}=t-\frac{\left\vert \mathbf{x%
}-\mathbf{x}_{0}\right\vert }{\mathrm{c}}.  \label{dipo3b}
\end{equation}%
Then applying the Jefimenko and the Panofsky-Phillips formulas (\ref{jef2})
and (\ref{jef3}) for the ideal electric dipole sources $\rho \left( t,%
\mathbf{x}\right) $ and $\mathbf{J}\left( t,\mathbf{x}\right) $ defined by
formulas (\ref{dipo1}) we obtain the following formulas for the EM field%
\begin{gather}
\mathbf{E}\left( t,\mathbf{x}\right) =\frac{3\left( \mathbf{\hat{R}}\cdot 
\mathbf{p}\left( t_{0}\right) \right) \mathbf{\hat{R}}-\mathbf{p}\left(
t_{0}\right) }{\left\vert \mathbf{R}\right\vert ^{3}}+  \label{dipo4} \\
+\frac{3\left( \mathbf{\dot{p}}\left( t_{0}\right) \cdot \mathbf{\hat{R}}%
\right) \mathbf{\hat{R}}-\mathbf{\dot{p}}\left( t_{0}\right) }{\mathrm{c}%
\left\vert \mathbf{R}\right\vert ^{2}}+\frac{\left( \mathbf{\ddot{p}}\left(
t_{0}\right) \times \mathbf{\hat{R}}\right) \times \mathbf{\hat{R}}}{\mathrm{%
c}^{2}\left\vert \mathbf{R}\right\vert },  \notag
\end{gather}%
\begin{equation}
\mathbf{B}\left( t,\mathbf{x}\right) =\left[ \frac{\mathbf{\dot{p}}\left(
t_{0}\right) }{\mathrm{c}R^{2}}+\frac{\mathbf{\ddot{p}}\left( t_{0}\right) }{%
\mathrm{c}^{2}R}\right] \times \mathbf{\hat{R}},\text{ where }\mathbf{\ddot{p%
}}=\partial _{t}^{2}\mathbf{p}.  \label{dipo5}
\end{equation}%
When deriving formula (\ref{dipo4}) we used a vector identity (\ref{vecf3a}).

In a simpler case when the dipole function $\mathbf{p}\left( t\right) $ is
time harmonic and complex valued of the form 
\begin{equation}
\mathbf{p}\left( t\right) =\mathbf{p}_{\omega }\mathrm{e}^{-\mathrm{i}\omega
t},\quad \mathbf{p}\left( t_{0}\right) =\mathbf{p}_{\omega }\mathrm{e}%
^{k\left\vert \mathbf{R}\right\vert -\mathrm{i}\omega t},\quad \text{where }%
k=\frac{\omega }{\mathrm{c}},  \label{dipo7}
\end{equation}%
the general formulas (\ref{dipo4}), (\ref{dipo5}) yield the well known
formulas for the ideal electric dipole fields, \cite[Section 9.2]{Jackson1},%
\begin{gather}
\mathbf{E}\left( t,\mathbf{x}\right) =-\frac{k^{2}\mathrm{e}^{k\left\vert 
\mathbf{R}\right\vert -\mathrm{i}\omega t}}{\left\vert \mathbf{R}\right\vert 
}\times  \label{dipo9} \\
\times \left\{ \left( \mathbf{p}\times \mathbf{\hat{R}}\right) \times 
\mathbf{\hat{R}}-\left[ 3\left( \mathbf{\hat{R}}\cdot \mathbf{p}\right) 
\mathbf{\hat{R}}-\mathbf{p}\right] \left( \frac{1}{k^{2}\left\vert \mathbf{R}%
\right\vert ^{2}}-\frac{\mathrm{i}}{k\left\vert \mathbf{R}\right\vert }%
\right) \right\}  \notag
\end{gather}%
\begin{equation}
\mathbf{B}\left( t,\mathbf{x}\right) =-k^{2}\mathrm{e}^{k\left\vert \mathbf{R%
}\right\vert -\mathrm{i}\omega t}\frac{\mathbf{p}\times \mathbf{\hat{R}}}{%
\left\vert \mathbf{R}\right\vert }\left( 1+\frac{\mathrm{i}}{k\left\vert 
\mathbf{R}\right\vert }\right) ,  \label{dipo10}
\end{equation}%
implying in radiation zone $k\left\vert \mathbf{R}\right\vert \gg 1$ the
following asymptotic formulas 
\begin{equation}
\mathbf{E}\left( t,\mathbf{x}\right) =k^{2}\mathrm{e}^{k\left\vert \mathbf{R}%
\right\vert -\mathrm{i}\omega t}\frac{\mathbf{\hat{R}}\times \left( \mathbf{p%
}_{\omega }\times \mathbf{\hat{R}}\right) }{\left\vert \mathbf{R}\right\vert 
}\left[ 1+O\left( \frac{1}{k\left\vert \mathbf{R}\right\vert }\right) \right]
,  \label{dipo12}
\end{equation}%
\begin{equation}
\mathbf{B}\left( t,\mathbf{x}\right) =k^{2}\mathrm{e}^{k\left\vert \mathbf{R}%
\right\vert -\mathrm{i}\omega t}\frac{\mathbf{\hat{R}}\times \mathbf{p}%
_{\omega }}{\left\vert \mathbf{R}\right\vert }\left[ 1+O\left( \frac{1}{%
k\left\vert \mathbf{R}\right\vert }\right) \right] .  \label{dipo13}
\end{equation}%
For the case of a multi-frequency dipole function $\mathbf{p}\left( t\right) 
$ we can introduce%
\begin{equation}
k_{\min }=\min_{\omega \in \Lambda _{\mathbf{p}}}\left\{ \frac{\omega }{%
\mathrm{c}}\right\} >0.  \label{dipo5a}
\end{equation}
Then radiation components decaying as $\left\vert \mathbf{R}\right\vert
^{-1} $ dominate for $k_{\min }\left\vert \mathbf{R}\right\vert \gg 1$ in
formulas (\ref{dipo4}) (\ref{dipo5}) implying the following asymptotic
expressions for the radiation fields, see also \cite[Section 11.1.4]%
{Griffiths}, 
\begin{gather}
\mathbf{E}\left( t,\mathbf{x}\right) =\frac{\left( \mathbf{\ddot{p}}^{\ell
}\left( t_{0}\right) \times \mathbf{\hat{R}}\right) \times \mathbf{\hat{R}}}{%
\mathrm{c}^{2}\left\vert \mathbf{R}\right\vert }\left[ 1+O\left( \frac{1}{%
k_{\min }\left\vert \mathbf{R}\right\vert }\right) \right] ,  \label{dipo11}
\\
\mathbf{B}\left( t,\mathbf{x}\right) =\frac{\mathbf{\ddot{p}}^{\ell }\left(
t_{0}\right) \times \mathbf{\hat{R}}}{\mathrm{c}^{2}\left\vert \mathbf{R}%
\right\vert }\left[ 1+O\left( \frac{1}{k_{\min }\left\vert \mathbf{R}%
\right\vert }\right) \right] ,\text{ for }k_{\min }\left\vert \mathbf{R}%
\right\vert \gg 1.  \notag
\end{gather}

\paragraph{Energy flux for a system of arbitrary elementary dipoles}

Using expression (\ref{dipo11}) for the EM fields and the vector identities (%
\ref{vecf3b}) and (\ref{vecf3c}) we consequently obtain the following
formulas for the energy flux $\mathbf{S}^{\ell \ell ^{\prime }}$ and the
corresponding total powers $P^{\ell \ell ^{\prime }}$and $P^{\left\{ \ell
,\ell ^{\prime }\right\} }$ radiated through a sphere centered at the dipole
location%
\begin{gather}
\mathbf{S}^{\ell \ell ^{\prime }}=\frac{\mathrm{c}}{4\pi }\mathbf{E}^{\ell
}\times \mathbf{B}^{\ell ^{\prime }}=  \label{dipo11a} \\
=\frac{\mathbf{\ddot{p}}^{\ell }\left( t_{0}\right) \cdot \mathbf{\ddot{p}}%
^{\ell ^{\prime }}\left( t_{0}\right) -\left( \mathbf{\ddot{p}}^{\ell
}\left( t_{0}\right) \cdot \mathbf{\hat{R}}\right) \left( \mathbf{\hat{R}}%
\cdot \mathbf{\ddot{p}}^{\ell ^{\prime }}\left( t_{0}\right) \right) }{4\pi 
\mathrm{c}^{3}\left\vert \mathbf{R}\right\vert ^{2}}\mathbf{\hat{R}}=  \notag
\\
=\frac{\left[ \left( \mathbf{\ddot{p}}^{\ell }\left( t_{0}\right) \times 
\mathbf{\hat{R}}\right) \cdot \left( \mathbf{\ddot{p}}^{\ell ^{\prime
}}\left( t_{0}\right) \times \mathbf{\hat{R}}\right) \right] }{4\pi \mathrm{c%
}^{3}\left\vert \mathbf{R}\right\vert ^{2}}\mathbf{\hat{R}},  \notag
\end{gather}%
\begin{gather}
P^{\ell \ell ^{\prime }}=\int\nolimits_{\left\vert \mathbf{x}\right\vert
=\left\vert \mathbf{R}\right\vert }\mathbf{S}^{\ell \ell ^{\prime }}\,%
\mathrm{d}\sigma =\frac{2}{3\mathrm{c}^{3}}\mathbf{\ddot{p}}^{\ell }\left(
t_{0}\right) \cdot \mathbf{\ddot{p}}^{\ell ^{\prime }}\left( t_{0}\right) ,
\label{dipo11b} \\
P^{\left\{ \ell ,\ell ^{\prime }\right\} }=P^{\ell \ell ^{\prime }}+P^{\ell
^{\prime }\ell }=\frac{4}{3\mathrm{c}^{3}}\mathbf{\ddot{p}}^{\ell }\left(
t_{0}\right) \cdot \mathbf{\ddot{p}}^{\ell ^{\prime }}\left( t_{0}\right) . 
\notag
\end{gather}%
Let us assume now that the dipole functions $\mathbf{p}^{\ell }\left(
t\right) $ depend on time $t$ almost periodically as in Section \ref{saper}.
Then the representation (\ref{dipo11b}) together with relations (\ref{iemt6b}%
), (\ref{dipo3b}) and (\ref{apef8}) consequently imply the following
formulas for the time-averaged radiated powers%
\begin{gather}
\left\langle \mathbf{S}^{\ell \ell ^{\prime }}\right\rangle =\frac{%
\left\langle \mathbf{\ddot{p}}^{\ell }\cdot \mathbf{\ddot{p}}^{\ell ^{\prime
}}\right\rangle -\left\langle \left( \mathbf{\ddot{p}}^{\ell }\cdot \mathbf{%
\hat{R}}\right) \left( \mathbf{\hat{R}}\cdot \mathbf{\ddot{p}}^{\ell
^{\prime }}\right) \right\rangle }{4\pi \mathrm{c}^{3}\left\vert \mathbf{R}%
\right\vert ^{2}}\mathbf{\hat{R}}=  \label{dipo11e} \\
=\frac{\left[ \left\langle \left( \mathbf{\ddot{p}}^{\ell }\times \mathbf{%
\hat{R}}\right) \cdot \left( \mathbf{\ddot{p}}^{\ell ^{\prime }}\times 
\mathbf{\hat{R}}\right) \right\rangle \right] }{4\pi \mathrm{c}%
^{3}\left\vert \mathbf{R}\right\vert ^{2}}\mathbf{\hat{R}=}  \notag \\
\mathbf{=}\frac{\mathbf{\hat{R}}}{8\pi \mathrm{c}^{3}\left\vert \mathbf{R}%
\right\vert ^{2}}\dsum\limits_{\omega \in \Lambda _{\mathbf{p}^{\ell }}\cap
\Lambda _{\mathbf{p}^{\ell ^{\prime }}}}\omega ^{4}\func{Re}\left\{ \left( 
\mathbf{p}_{\omega }^{\ell }\times \mathbf{\hat{R}}\right) \left( \mathbf{p}%
_{\omega }^{\ell ^{\prime }\ast }\times \mathbf{\hat{R}}\right) \right\} , 
\notag
\end{gather}%
\begin{equation}
\left\langle P^{\ell \ell ^{\prime }}\right\rangle =\frac{2\left\langle 
\mathbf{\ddot{p}}^{\ell }\cdot \mathbf{\ddot{p}}^{\ell ^{\prime
}}\right\rangle }{3\mathrm{c}^{3}}=\dsum\limits_{\omega \in \Lambda _{%
\mathbf{p}^{\ell }}\cap \Lambda _{\mathbf{p}^{\ell ^{\prime }}}}\frac{\omega
^{4}}{3\mathrm{c}^{3}}\func{Re}\left\{ \mathbf{p}_{\omega }^{\ell }\cdot 
\mathbf{p}_{\omega }^{\ell ^{\prime }\ast }\right\} ,  \label{dipo11c}
\end{equation}%
\begin{equation}
\left\langle P^{\left\{ \ell ,\ell ^{\prime }\right\} }\right\rangle =\frac{%
4\left\langle \mathbf{\ddot{p}}^{\ell }\cdot \mathbf{\ddot{p}}^{\ell
^{\prime }}\right\rangle }{3\mathrm{c}^{3}}=\dsum\limits_{\omega \in \Lambda
_{\mathbf{p}^{\ell }}\cap \Lambda _{\mathbf{p}^{\ell ^{\prime }}}}\frac{%
2\omega ^{4}}{3\mathrm{c}^{3}}\omega ^{4}\func{Re}\left\{ \mathbf{p}_{\omega
}^{\ell }\cdot \mathbf{p}_{\omega }^{\ell ^{\prime }\ast }\right\} ,
\label{dipo11d}
\end{equation}%
where $\Lambda _{\mathbf{p}^{\ell }}$ and $\Lambda _{\mathbf{p}^{\ell
^{\prime }}}$ are respectively the frequency spectra of $\mathbf{p}^{\ell
}\left( t\right) $ and $\mathbf{p}^{\ell ^{\prime }}\left( t\right) $. It
readily follows from the formula (\ref{dipo11d}) then that \emph{the
time-averaged radiated power }$\left\langle P^{\left\{ \ell ,\ell ^{\prime
}\right\} }\right\rangle $\emph{\ can take any real value: negative, zero or
positive}. Indeed, according to the formulas (\ref{dipo11d}), (\ref{apef8}) $%
\left\langle P^{\left\{ \ell ,\ell ^{\prime }\right\} }\right\rangle $
vanishes if the frequency spectra $\Lambda _{\mathbf{p}^{\ell }}$ and $%
\Lambda _{\mathbf{p}^{\ell ^{\prime }}}$ don't have any common frequencies,
i.e.%
\begin{equation}
\left\langle P^{\left\{ \ell ,\ell ^{\prime }\right\} }\right\rangle =\frac{%
4\left\langle \mathbf{\ddot{p}}^{\ell }\left( t\right) \cdot \mathbf{\ddot{p}%
}^{\ell ^{\prime }}\left( t\right) \right\rangle }{3\mathrm{c}^{3}}=0\text{
if }\Lambda _{\mathbf{p}^{\ell }}\dbigcap \Lambda _{\mathbf{p}^{\ell
}}=\varnothing .  \label{ppp1}
\end{equation}%
\emph{In particular, the relation (\ref{ppp1}) shows that if the both }$\ell 
$\emph{-th and }$\ell ^{\prime }$\emph{-th b-charges are monochromatic of
different frequencies then the time-averaged radiated power is exactly zero
or in other words there is no radiation}. The formulas (\ref{dipo11d}), (\ref%
{apef8}) readily imply%
\begin{equation}
\left\langle P^{\left\{ \ell ,\ell ^{\prime }\right\} }\right\rangle =\frac{%
4\left\langle \mathbf{\ddot{p}}^{\ell }\left( t\right) \cdot \mathbf{\ddot{p}%
}^{\ell }\left( t\right) \right\rangle }{3\mathrm{c}^{3}}=\frac{2}{3\mathrm{c%
}^{3}}\dsum\limits_{\omega \in \Lambda _{\mathbf{p}^{\ell }}}\omega
^{4}\left\vert \mathbf{p}_{\omega }^{\ell }\right\vert ^{2}>0\text{ if }%
\mathbf{p}^{\ell ^{\prime }}\left( t\right) =\mathbf{p}^{\ell }\left(
t\right) ,  \label{ppp2}
\end{equation}%
\begin{equation}
\left\langle P^{\left\{ \ell ,\ell ^{\prime }\right\} }\right\rangle =-\frac{%
4\left\langle \mathbf{\ddot{p}}^{\ell }\left( t\right) \cdot \mathbf{\ddot{p}%
}^{\ell }\left( t\right) \right\rangle }{3\mathrm{c}^{3}}=-\frac{2}{3\mathrm{%
c}^{3}}\dsum\limits_{\omega \in \Lambda _{\mathbf{p}^{\ell }}}\omega
^{4}\left\vert \mathbf{p}_{\omega }^{\ell }\right\vert ^{2}<0\text{ if }%
\mathbf{p}^{\ell ^{\prime }}\left( t\right) =-\mathbf{p}^{\ell }\left(
t\right) .  \label{ppp3}
\end{equation}%
\emph{Evidently the relation (\ref{ppp3}) describes a situation when for a
given pair }$\left\{ \ell ,\ell ^{\prime }\right\} $\emph{\ the
time-averaged radiated power }$\left\langle P^{\left\{ \ell ,\ell ^{\prime
}\right\} }\right\rangle $\emph{\ is negative, that is the radiated energy
propagates with the speed of light toward the source rather than away from it%
}.

\paragraph{Radiated power of a system of identical dipoles}

A basis for simple comparison of radiative properties in the BEM and CEM
theories let us consider a system of $N>1$ b-charges described by one and
the same dipole moment $\mathbf{p}\left( t\right) $. Then in view of (\ref%
{ppp2}) we have for the system the following time-average radiated power%
\begin{eqnarray}
\left\langle P_{\mathrm{BEM}}\right\rangle &=&\sum\limits_{\ell ^{\prime
}\neq \ell }\left\langle P^{\left\{ \ell ,\ell ^{\prime }\right\}
}\right\rangle =  \label{tarp1} \\
N\left( N-1\right) \frac{2\left\langle \mathbf{\ddot{p}}^{\ell }\left(
t\right) \cdot \mathbf{\ddot{p}}^{\ell }\left( t\right) \right\rangle }{3%
\mathrm{c}^{3}} &=&N\left( N-1\right) \frac{1}{3\mathrm{c}^{3}}%
\dsum\limits_{\omega \in \Lambda _{\mathbf{p}}}\omega ^{4}\left\vert \mathbf{%
p}_{\omega }\right\vert ^{2}.  \notag
\end{eqnarray}%
In the CEM theory a similar system of $N$ identical dipoles $\mathbf{p}$ has
an effective dipole moment $N\mathbf{p}$. Substituting this number for $%
\mathbf{p}$ in the expression (\ref{dipo11c}) we get 
\begin{equation}
\left\langle P_{\mathrm{CEM}}\right\rangle =N^{2}\frac{2\left\langle \mathbf{%
\ddot{p}}^{\ell }\left( t\right) \cdot \mathbf{\ddot{p}}^{\ell }\left(
t\right) \right\rangle }{3\mathrm{c}^{3}}=N^{2}\frac{1}{3\mathrm{c}^{3}}%
\dsum\limits_{\omega \in \Lambda _{\mathbf{p}}}\omega ^{4}\left\vert \mathbf{%
p}_{\omega }\right\vert ^{2}.  \label{tarp2}
\end{equation}%
Relating representations (\ref{tarp1}) and (\ref{tarp2}) we readily obtain%
\begin{equation}
\left\langle P_{\mathrm{BEM}}\right\rangle =\left( 1-\frac{1}{N}\right)
\left\langle P_{\mathrm{CEM}}\right\rangle .  \label{tarp3}
\end{equation}

\subsubsection{Elementary currents for point charges}

It is curious to see how the BEM theory describes fields and radiations
phenomena when the elementary currents are caused by a system of point
charges as, for instance, in the Rutherford atom model. In the case of point
charges we use the acceleration fields $\mathbf{E}_{\mathrm{a}}\left( t,%
\mathbf{x}\right) $ and $\mathbf{B}_{\mathrm{a}}\left( t,\mathbf{x}\right) $
defined by formulas (\ref{erad4}), (\ref{erad5}) to find the time-averaged
radiation power. Consequently, for every pair $\left\{ \ell ,\ell ^{\prime
}\right\} $ of b-charges we have%
\begin{equation}
\mathbf{S}^{\ell \ell ^{\prime }}=\frac{\mathrm{c}}{4\pi }\mathbf{E}_{%
\mathrm{a}}^{\ell }\times \mathbf{B}_{\mathrm{a}}^{\ell ^{\prime }}\text{
and }\mathbf{S}^{\ell ^{\prime }\ell }=\frac{\mathrm{c}}{4\pi }\mathbf{E}_{%
\mathrm{a}}^{\ell ^{\prime }}\times \mathbf{B}_{\mathrm{a}}^{\ell }.
\label{ecpc1}
\end{equation}%
Suppose that one of the two b-charges, say $\ell $-th charge, is at rest or
moves uniformly implying that is $\mathbf{\dot{\beta}}^{\ell }=0$. Then it
follows from formulas (\ref{erad4}), (\ref{erad5}) that $\mathbf{E}_{\mathrm{%
a}}^{\ell }=0$ and $\mathbf{B}_{\mathrm{a}}^{\ell }=0$ implying that $%
\mathbf{S}^{\ell \ell ^{\prime }}$ and $\mathbf{S}^{\ell ^{\prime }\ell }$
vanish, namely 
\begin{equation}
\mathbf{S}^{\ell \ell ^{\prime }}=\mathbf{S}^{\ell ^{\prime }\ell }=0\text{
if }\mathbf{\dot{\beta}}^{\ell }=0\text{ or }\mathbf{\dot{\beta}}^{\ell
^{\prime }}=0.  \label{ecpc2}
\end{equation}%
Suppose that the position functions $\mathbf{r}^{\ell }\left( t\right) $ and 
$\mathbf{r}^{\ell ^{\prime }}\left( t\right) $ of the corresponding
b-charges are almost periodic functions as described in Section \ref{saper},
and that $\func{mod}\left( \mathbf{r}^{\ell }\left( t\right) \right) $ and $%
\func{mod}\left( \mathbf{r}^{\ell ^{\prime }}\left( t\right) \right) $ have
no common frequencies. Then in this case the time-averaged flux $%
\left\langle \mathbf{S}^{\ell \ell ^{\prime }}\right\rangle $ is exactly
zero, i.e.%
\begin{equation}
\left\langle \mathbf{S}^{\ell \ell ^{\prime }}\right\rangle =0\text{ if }%
\func{mod}\left( \mathbf{r}^{\ell }\left( t\right) \right) \dbigcap \func{mod%
}\left( \mathbf{r}^{\ell ^{\prime }}\left( t\right) \right) =\varnothing .
\label{ecpc3}
\end{equation}%
Now let us compare this estimate with a similar computation in the framework
of the CEM theory with the total field $\mathbf{E}=\mathbf{E}_{\mathrm{a}%
}^{\ell ^{\prime }}+\mathbf{E}_{\mathrm{a}}^{\ell }$, $\mathbf{B}=\mathbf{B}%
_{\mathrm{a}}^{\ell ^{\prime }}+\mathbf{B}_{\mathrm{a}}^{\ell }\ $or in the
framework of the BEM theory with two prescribed charges and a test charge.
In the framework of the BEM theory if we consider action of the system of
two charges onto a test charge or onto a distant system of test charges
which does not noticebly affect the radiation of two charges, the field of
two charges turns into an external field given by the same formula $\mathbf{E%
}=\mathbf{E}_{\mathrm{a}}^{\ell ^{\prime }}+\mathbf{E}_{\mathrm{a}}^{\ell }$%
, $\mathbf{B}=\mathbf{B}_{\mathrm{a}}^{\ell ^{\prime }}+\mathbf{B}_{\mathrm{a%
}}^{\ell }$. The radiation power at large distances generated by two charges
in accordance with (\ref{erad3}) is expressed by the formula 
\begin{equation}
\mathbf{S}=\frac{\mathrm{c}}{4\pi }\left( \mathbf{E}_{\mathrm{a}}^{\ell
^{\prime }}+\mathbf{E}_{\mathrm{a}}^{\ell }\right) \times \left( \mathbf{B}_{%
\mathrm{a}}^{\ell }+\mathbf{B}_{\mathrm{a}}^{\ell ^{\prime }}\right) .
\label{ecpc4}
\end{equation}%
If $\ell $-th charge is at rest $\mathbf{B}_{\mathrm{a}}^{\ell }=0$ and $%
\mathbf{E}_{\mathrm{a}}^{\ell }=0$ and we obtain the leading part and its
time average%
\begin{equation}
\mathbf{S}=\frac{\mathrm{c}}{4\pi }\mathbf{E}_{\mathrm{a}}^{\ell ^{\prime
}}\times \mathbf{B}_{\mathrm{a}}^{\ell ^{\prime }},\quad \left\langle 
\mathbf{S}\right\rangle =\frac{\mathrm{c}}{4\pi }\left\langle \mathbf{E}_{%
\mathrm{a}}^{\ell ^{\prime }}\times \mathbf{B}_{\mathrm{a}}^{\ell ^{\prime
}}\right\rangle ,  \label{sa}
\end{equation}%
which differs from (\ref{ecpc3}) and (\ref{ecpc2}).

\begin{remark}
Note that formulas (\ref{ecpc3}) and (\ref{ecpc2}) \ describe the radiation
of EM interaction energy between the two charges, and it is not the same as
the EM radiation of the fields which act on test charges. One may ask why we
are interested in the interaction energy? The answer is that this is the
energy which is derived from the Lagrangian and the interaction energy is
conserved when the system is closed. Of course the system where the motion
of sources is prescribed is not closed, and there must be an external source
of energy to provide for the prescribed currents and for the radiation which
acts onto test charges.
\end{remark}

Let us consider in conclusion the case when the both charges move slowly and
hence $\left\vert \mathbf{\beta }^{\ell }\right\vert ,\left\vert \mathbf{%
\beta }^{\ell }\right\vert \ll 1$. Notice that that relations (\ref{erad6})
defining the point charge EM fields are similar to the dipole expressions (%
\ref{dipo11}) and can be obtained from them by a substitution $\mathbf{\ddot{%
p}}^{\ell }\left( t\right) \rightarrow \mathrm{c}q\mathbf{\dot{v}}^{\ell
}\left( t\right) $. Consequently, expressions for the energy flux and the
radiated power for the pair of slowly moving charges can be readily obtained
from the dipole expressions (\ref{dipo11a})-(\ref{dipo11a}) by a
substitution $\mathbf{\ddot{p}}^{\ell }\left( t\right) \rightarrow \mathrm{c}%
q\mathbf{\dot{v}}^{\ell }\left( t\right) $ with an additional correcting
asymptotic factor $\delta ^{\ell }=\left( 1+O\left( \left\vert \mathbf{\beta 
}^{\ell }\right\vert \right) \right) $. In particular, 
\begin{gather*}
\mathbf{S}^{\ell \ell ^{\prime }}=\frac{q^{2}}{4\pi \mathrm{c}}\frac{\mathbf{%
\dot{v}}^{\ell }\left( t\right) \cdot \mathbf{\dot{v}}^{\ell ^{\prime
}}\left( t\right) -\left( \mathbf{\dot{v}}^{\ell }\left( t\right) \cdot 
\mathbf{\hat{R}}\right) \left( \mathbf{\hat{R}}\cdot \mathbf{\dot{v}}^{\ell
^{\prime }}\left( t\right) \right) }{\left\vert \mathbf{R}\right\vert ^{2}}%
\mathbf{\hat{R}}\delta ^{\ell \ell ^{\prime }}= \\
=\frac{q^{2}}{4\pi \mathrm{c}}\frac{\left[ \left( \mathbf{\dot{v}}^{\ell
}\left( t\right) \times \mathbf{\hat{R}}\right) \cdot \left( \mathbf{\dot{v}}%
^{\ell ^{\prime }}\times \mathbf{\hat{R}}\right) \right] }{4\pi \mathrm{c}%
^{3}\left\vert \mathbf{R}\right\vert ^{2}}\mathbf{\hat{R}}\delta ^{\ell \ell
^{\prime }},\text{ where} \\
\delta ^{\ell \ell ^{\prime }}=\left( 1+O\left( \left\vert \mathbf{\beta }%
^{\ell }\right\vert \right) \right) \left( 1+O\left( \left\vert \mathbf{%
\beta }^{\ell ^{\prime }}\right\vert \right) \right) .
\end{gather*}

\subsection{Comparison with the classical EM theory\label{scompcem}}

We would like to show here that the CEM theory can be viewed as a limit case
of the BEM theory. Particularly, we are interested in seeing: (i) how the
single classical EM field is modeled by elementary EM fields in BEM theory;
(ii) how classical EM phenomena\ including the radiation are modeled within
the BEM theory.

As an example of relation between the classical EM field and elementary EM
fields we consider clusters of many tightly bound identical b-charges, every
cluster is labeled by index $\ell .$ Namely, for every $\ell $ we introduce $%
N_{\ell }$ b-charges having identical wave functions and elementary EM
potentials with charges $\ q_{\mathrm{w}}^{\ell }=N_{\ell }q^{\ell }$: 
\begin{equation}
\psi ^{\left( \ell ,s\right) }=\psi ^{\left( \ell ,s\right) }=\psi ^{\ell
},\qquad A^{\left( \ell ,s\right) \mu }=A^{\left( \ell ,s\right) \mu }\quad 
\text{where }1\leq \ell \leq N,\quad 1\leq s\leq N_{\ell },  \label{ibch1}
\end{equation}%
and identical EM fields $F^{\ell \mu \nu }=F^{\left( \ell ,s\right) \mu \nu
} $ and currents \ $J^{\ell ,\nu }=J^{\left( \ell ,s\right) \nu }$. Now we
compare these fields with classical EM fields $\mathcal{F}^{\mu \nu }\ $%
defined by (\ref{ilag1}) with the same currents by comparing the
electromagnetic parts of the Lagrangians in BEM and CEM theories. We have
for classical theory 
\begin{eqnarray*}
\mathcal{L}_{\mathrm{CEM}} &=&-\frac{1}{16\pi }\mathcal{F}^{\mu \nu }%
\mathcal{F}_{\mu \nu }=-\frac{1}{16\pi }\dsum\nolimits_{\ell ^{\prime },\ell
}N_{\ell ^{\prime }}N_{\ell }F^{\ell \mu \nu }F_{\mu \nu }^{\ell ^{\prime }}
\\
&=&-\frac{1}{16\pi }\dsum\nolimits_{\ell ^{\prime }\neq \ell }N_{\ell
^{\prime }}N_{\ell }F^{\ell \mu \nu }F_{\mu \nu }^{\ell ^{\prime }}-\frac{1}{%
16\pi }\dsum\nolimits_{\ell =1}^{N}N_{\ell }^{2}F^{\ell \mu \nu }F_{\mu \nu
}^{\ell }.
\end{eqnarray*}%
For the BEM theory\ (\ref{ilag5}) takes the form 
\begin{eqnarray*}
\mathcal{L}_{\mathrm{BEM}} &=&-\frac{1}{16\pi }\dsum\nolimits_{\left\{
\left( \ell ,s\right) ,\left( \ell ^{\prime },s^{\prime }\right) :\left(
\ell ^{\prime },s^{\prime }\right) \neq \left( \ell ,s\right) \right\}
}F^{\left( \ell ,s\right) \mu \nu }F_{\mu \nu }^{\left( \ell ^{\prime
},s^{\prime }\right) } \\
&=&-\frac{1}{16\pi }\dsum\nolimits_{\ell ^{\prime }\neq \ell }N_{\ell
^{\prime }}N_{\ell }F^{\ell \mu \nu }F_{\mu \nu }^{\ell ^{\prime }}-\frac{1}{%
16\pi }\dsum\nolimits_{\ell }N_{\ell }\left( N_{\ell }-1\right) F^{\ell \mu
\nu }F_{\mu \nu }^{\ell }.
\end{eqnarray*}%
The difference between $\mathcal{L}_{\mathrm{CEM}}$ and $\mathcal{L}_{%
\mathrm{BEM}}$ can be attributed to to interactions inside every cluster. In
particular, we have for the both theories 
\begin{equation*}
\mathcal{L}_{\mathrm{CEM}\ell }=N_{\ell }^{2}F^{\ell \mu \nu }F_{\mu \nu
}^{\ell },\quad \mathcal{L}_{\mathrm{BEM}\ell }=N_{\ell }\left( N_{\ell
}-1\right) F^{\ell \mu \nu }F_{\mu \nu }^{\ell }
\end{equation*}%
\ readily implying the following expression for the relative difference 
\begin{equation}
\mathcal{L}_{\mathrm{BEM}\ell }/\mathcal{L}_{\mathrm{CEM}\ell }-1=1/N_{\ell }
\label{dif2}
\end{equation}%
The difference (\ref{dif2}) evidently becomes small as number $N_{\ell }$ of
particles in the cluster becomes large.

To summarize, the CEM theory with its single EM field can be formally
derived from the BEM theory as a limit obtained by binding together $N_{\ell
}$ identical particles with $N_{\ell }\rightarrow \infty $.

\subsubsection{Lagrangians for clusters of charges}

Here we make a comparison of two Lagrangians for $N$ clusters of charges
with $N_{\ell }$\ particles within every cluster which are formally derived
from WCM and BEM theories respectively. To this end we consider first a
Lagrangian as in WCM theory with $N_{1}+...+N_{N}$ charges and fixed
nonlinearities $G_{\mathrm{w}}^{\ell }$

To this end we consider first a Lagrangian as in WCM theory \ with $%
N_{1}+...+N_{N}$ charges \ and a fixed nonlinearity $G_{\mathrm{w}}^{\ell }$ 
\begin{equation}
\mathcal{L}\left( \{\psi ^{\left( \ell ,s_{\ell }\right) }\},\{\psi _{;\mu
}^{\left( \ell ,s_{\ell }\right) }\}\right) =\sum\limits_{1\leq \ell \leq
N}\sum\limits_{1\leq s_{\ell }\leq N_{\ell }}L^{\ell }\left( \psi ^{\left(
\ell ,s_{\ell }\right) },\psi _{;\mu }^{\left( \ell ,s_{\ell }\right)
}\right) -\frac{\mathcal{F}^{\mu \nu }\mathcal{F}_{\mu \nu }}{16\pi },
\label{ibch21}
\end{equation}%
and to introduce clusters we impose additional constraints \ as in (\ref%
{ibch1}), namely 
\begin{equation}
\psi ^{\ell ,s}=\psi _{\mathrm{w}}^{\ell },\quad q^{\ell ,s}=q^{\ell },\quad
s=1,...,N_{\ell }.  \label{ibch2}
\end{equation}%
\ A cluster in the equilibrium is described by the real valued functions $%
\psi ^{\ell }$ and $\mathring{\varphi}^{\ell ,s}$ which satisfy the
following system of equations:%
\begin{gather}
-\Delta \varphi _{\mathrm{w}}^{\ell }=4\pi \sum_{s=1}^{N_{\ell }}q^{\ell
,s}\left( 1-\frac{q^{\ell ,s}\varphi _{\mathrm{w}}^{\ell }}{m^{\ell ,s}%
\mathrm{c}^{2}}\right) \left( \psi ^{\ell ,s}\right) ^{2},  \label{psfi1a} \\
-\Delta \psi ^{\ell ,s}+\frac{m^{\ell ,s}\varphi _{\mathrm{w}}^{\ell }}{\chi
^{2}}q^{\ell ,s}\left( 2-\frac{q\varphi _{\mathrm{w}}^{\ell }}{m^{\ell ,s}%
\mathrm{c}^{2}}\right) \mathring{\psi}^{\ell ,s}+G_{\mathrm{w}}^{\prime
}\left( |\mathring{\psi}^{\ell ,s}|^{2}\right) \mathring{\psi}^{\ell ,s}=0.
\label{psfi1b}
\end{gather}%
Note that the WCM theory developed in \cite{BabFig1}, \cite{BabFig2}
coincides with the case where every cluster contains exactly one charge. In
this case the charge equilibrium equation has the form 
\begin{gather}
-\Delta \mathring{\varphi}^{\ell }=4\pi q^{\ell }\left( 1-\frac{q^{\ell }%
\mathring{\varphi}^{\ell }}{m\mathrm{c}^{2}}\right) \mathring{\psi}^{\ell 2},
\label{psif01} \\
-\Delta \mathring{\psi}^{\ell }+\frac{m^{\ell }\mathring{\varphi}^{\ell }}{%
\chi ^{2}}q^{\ell }\left( 2-\frac{q^{\ell }\mathring{\varphi}^{\ell }}{%
m^{\ell }\mathrm{c}^{2}}\right) \mathring{\psi}^{\ell }+G_{\mathrm{w}%
}^{\prime }\left( |\mathring{\psi}^{\ell }|^{2}\right) \mathring{\psi}^{\ell
}=0.  \label{psif02}
\end{gather}%
\ Comparing with (\ref{psfi1a}) and (\ref{psfi1b}) and setting%
\begin{gather}
\varphi _{\mathrm{w}}^{\ell }=N_{\ell }\mathring{\varphi}^{\ell },\quad q_{%
\mathrm{w}}^{\ell }=N_{\ell }q^{\ell },\quad m_{\mathrm{w}}^{\ell }=N_{\ell
}^{2}m^{\ell }  \label{ibch3} \\
\psi ^{\ell ,s}=\mathring{\psi}^{\ell }=\psi _{\mathrm{w}}^{\ell },\quad
s=1,...,N_{\ell },  \notag
\end{gather}%
we find then that $\varphi _{\mathrm{w}}^{\ell }$ satisfies \ the following
equation 
\begin{equation}
-\Delta \varphi _{\mathrm{w}}^{\ell }=4\pi q_{\mathrm{w}}^{\ell }\left( 1-%
\frac{q_{\mathrm{w}}^{\ell }\varphi _{\mathrm{w}}^{\ell }}{m_{\mathrm{w}%
}^{\ell }\mathrm{c}^{2}}\right) \mathring{\psi}^{\ell 2},  \label{psif011}
\end{equation}%
which has exactly the same form as (\ref{psif01}) in the WCM theory. The
charge normalization condition is also fulfilled. Equations(\ref{psfi1a})
and (\ref{psif02}) can be rewritten in the form \ 
\begin{equation}
-\Delta \mathring{\psi}^{\ell }+\frac{m_{\mathrm{w}}^{\ell }\varphi _{%
\mathrm{w}}^{\ell }}{N_{\ell }^{4}\chi ^{2}}q_{\mathrm{w}}^{\ell }\left( 2-%
\frac{q_{\mathrm{w}}^{\ell }\varphi _{\mathrm{w}}^{\ell }}{m_{\mathrm{w}%
}^{\ell }\mathrm{c}^{2}}\right) \mathring{\psi}^{\ell }+G_{\mathrm{w}%
}^{\prime }\left( |\mathring{\psi}^{\ell }|^{2}\right) \mathring{\psi}^{\ell
}=0.  \label{psif02a}
\end{equation}%
This equation has the form of (\ref{psfi1b}) \ for $\mathring{\psi}^{\ell }=$
$\mathring{\psi}_{\mathrm{w}}^{\ell }$ if we set%
\begin{equation}
\chi _{\mathrm{w}}=N_{\ell }^{2}\chi .  \label{ibch31}
\end{equation}%
Observe that relations (\ref{ibch1a})- (\ref{ibch3}) readily imply the
following identities for the model constants 
\begin{equation}
\kappa _{\mathrm{w}}^{\ell }=\frac{m_{\mathrm{w}}^{\ell }\mathrm{c}}{\chi _{%
\mathrm{w}}}=\frac{m^{\ell }\mathrm{c}}{\chi }=\kappa ^{\ell },
\label{ibch4}
\end{equation}%
If we introduce Lagrangian $\mathcal{L}_{\mathrm{w}}\left( \{\psi _{\mathrm{w%
}}^{\ell }\},\{\psi _{\mathrm{w};\mu }^{\ell }\}\right) $ with (i) constants
defined by (\ref{ibch3}), (\ref{ibch31}) and (ii) EM potential defined by%
\begin{equation}
A_{\mathrm{w}}^{\mu }=\sum\limits_{\ell ,s}A^{\ell ,s\mu }=\sum\limits_{\ell
}A_{\mathrm{w}}^{\ell \mu },\quad A_{\mathrm{w}}^{\ell \mu }=N_{\ell
}A^{\ell \mu },\quad 1\leq \ell \leq N,  \label{ibch1a}
\end{equation}%
then the Euler-Lagrange Equations for Lagrangian $\mathcal{L}\left( \{\psi
^{\left( \ell ,s_{\ell }\right) }\},\{\psi _{;\mu }^{\left( \ell ,s_{\ell
}\right) }\}\right) $ defined by (\ref{ibch21}) with restrictions (\ref%
{ibch2}) are equivalent to the Euler-Lagrange Equations for $\mathcal{L}_{%
\mathrm{w}}$. Hence, we can conclude that introduction of clusters of
charges is equivalent to a proper rescaling of the constants of the WCM
theory. Note that the quadratic dependence of $m_{\mathrm{w}}^{\ell }$ on
the number $N_{\ell }$ of charges in the cluster is natural in the
relativistic theory since the energy of interactions depends on the number
of particles quadratically, and relativistic mass of the cluster is
proportional to its energy.

Now let us relate the above treatment with the BEM theory. To make the
comparison more transparent we consider a generalized version of the\ BEM
theory, namely the theory where the basic equilibrium object is a cluster of 
$N_{\ell }$\ b-charges, $N_{\ell }\geq 1$. The BEM theory for elementary
b-charges corresponds to the case $N_{\ell }=1$ for all $\ell $ in this
generalized setting. The Lagrangian $\mathcal{L}_{\mathrm{BEMC}}\left(
\{\psi ^{\left( \ell ,s_{\ell }\right) }\},\{\psi _{;\mu }^{\left( \ell
,s_{\ell }\right) }\}\right) $ for clusters is given by an obvious
modification of (\ref{ilag3})\ and the variables are subjected to additional
constraints\ as in (\ref{ibch2}). The equilibrium state for a generalized
cluster of $n=N_{\ell }$ identical b-charges is denoted by $\mathring{\psi}_{%
\mathrm{b,}n}^{\ell ,s},\mathring{\varphi}_{\mathrm{b,}n}^{\ell ,s}$, $%
s=1,...,n$. \ The equilibrium condition takes the following form similar to (%
\ref{psfi1a}) and (\ref{psfi1b}) 
\begin{gather}
-\Delta \mathring{\varphi}_{\mathrm{b},n}^{\ell }=4\pi \sum_{s=2}^{N_{\ell
}}q_{\mathrm{b},n}^{\ell ,s}\left( 1-\frac{q_{\mathrm{b,}n}^{\ell ,s}%
\mathring{\varphi}_{\mathrm{b,}n}^{\ell }}{m_{\mathrm{b,}n}^{\ell ,s}\mathrm{%
c}^{2}}\right) \left( \mathring{\psi}_{\mathrm{b,}n}^{\ell ,s}\right) ^{2},
\label{psifi011} \\
-\Delta \mathring{\psi}_{\mathrm{b,}n}^{\ell ,s}+\frac{m^{\ell ,s}\mathring{%
\varphi}_{\mathrm{b,}n}^{\ell }}{\chi _{\mathrm{b,}n}^{2}}q_{\mathrm{b}%
,n}^{\ell ,s}\left( 2-\frac{q\mathring{\varphi}_{\mathrm{b,}n}^{\ell }}{m_{%
\mathrm{b,}n}^{\ell ,s}\mathrm{c}^{2}}\right) \mathring{\psi}_{\mathrm{b,}%
n}^{\ell ,s}+G_{\mathrm{b,}n}^{\prime }\left( |\mathring{\psi}_{\mathrm{b,}%
n}^{\ell ,s}|^{2}\right) \mathring{\psi}_{\mathrm{b,}n}^{\ell ,s}=0.
\label{psifi012}
\end{gather}%
Note that the only difference between (\ref{psfi1a}) and (\ref{psfi1b})\ is
that the first term with $s=1$ in the sum is omitted. Now $\mathring{\psi}_{%
\mathrm{b,}2}^{\ell },\mathring{\varphi}_{\mathrm{b,}2}^{\ell }$ \ is the
basis for comparison and once again we fix the nonlinearity $G_{\mathrm{b,}%
n}^{\prime }=G_{\mathrm{b,}2}^{\prime }=G_{\mathrm{b}}^{\prime }$ and set \
similarly to (\ref{ibch3}), (\ref{ibch31}): 
\begin{gather}
\mathring{\varphi}_{\mathrm{b,}n}^{\ell }=\left( n-1\right) \mathring{\varphi%
}_{\mathrm{b,}2}^{\ell },\quad q_{\mathrm{b,}n}^{\ell }=\left( n-1\right) q_{%
\mathrm{b,}2}^{\ell },  \label{ibchb} \\
m_{\mathrm{b,}n}^{\ell }=\left( n-1\right) ^{2}m_{\mathrm{b,}2}^{\ell
},\quad \chi _{\mathrm{b,}n}^{\ell }=\left( n-1\right) ^{2}\chi _{\mathrm{b,}%
2}^{\ell }.  \notag
\end{gather}%
Let us introduce Lagrangian $\mathcal{L}_{\mathrm{BEM,}n}\left( \{\psi _{%
\mathrm{b,}n}^{\ell }\}\{\varphi _{\mathrm{b,}n}^{\ell }\}\right) $ with so
defined constants. Observe that the Euler-Lagrange equations for $\mathcal{L}%
_{\mathrm{BEMC}}\left( \{\psi ^{\left( \ell ,s_{\ell }\right) }\},\{\psi
_{;\mu }^{\left( \ell ,s_{\ell }\right) }\}\right) $ with restrictions (\ref%
{ibch2}) are equivalent to the Euler-Lagrange equations for $\mathcal{L}_{%
\mathrm{BEM,}n}\left( \{\psi _{\mathrm{b,}n}^{\ell }\}\{\varphi _{\mathrm{b,}%
n}^{\ell }\}\right) $. Obviously if all $N_{\ell }\geq 2$ then $\mathcal{L}_{%
\mathrm{BEM,}n}$ has the same form as $\mathcal{L}_{\mathrm{w}}$. In
addition to that, if 
\begin{equation*}
G_{\mathrm{b,}2}^{\prime }=G_{\mathrm{w}}^{\prime },\quad q_{\mathrm{b,}%
2}^{\ell }=q^{\ell },\quad m_{\mathrm{b,}2}^{\ell }=m^{\ell },\quad \chi _{%
\mathrm{b,}2}^{\ell }=\chi ,
\end{equation*}%
then the relative difference of coefficients of Lagrangians $\mathcal{L}_{%
\mathrm{BEM,}n}$ and $\mathcal{L}_{\mathrm{w}}$ \ with the same $n\geq 2$ is
of order $\frac{1}{n}$. For example, $\left( q_{\mathrm{w}}^{\ell }-q_{%
\mathrm{b,}n}^{\ell }\right) /q_{\mathrm{w}}^{\ell }=1/n$. Notice that the
case $n=1$ (which is the primary one considered in this article) is special,
and there is a non-vanishing difference between $\mathring{\psi}_{\mathrm{b,}%
2}^{\ell },\mathring{\varphi}_{\mathrm{b,}2}^{\ell }$ and the fundamental
equilibrium $\mathring{\psi}_{\mathrm{b,}1}^{\ell },\mathring{\varphi}_{%
\mathrm{b,}1}^{\ell }=\mathring{\psi}^{\ell },\mathring{\varphi}^{\ell }$
determined by (\ref{nop40}), (\ref{stst}).

Thus, based on the above analysis we conclude that the WCM theory can be
considered as an approximation for the generalized BEM theory if the WCM
charge is identified with a cluster of a large number of b-charges.

\subsection{New EM features of the BEM theory}

In this section we discuss new EM features of BEM theory not presented in
the CEM theory. Looking at the b-charge Lagrangian $\mathcal{L}$ with its
field equations and conserved currents in Section \ref{slagbem} we can
already see new important features of the theory of interacting charges and
EM fields. The first striking feature of the new theory compare to the
classical one is that \emph{the single EM field as independent entity is no
more, instead we have elementary EM potentials }$A^{\ell \mu }$\emph{\ and
fields }$F^{\ell \mu \nu }$\emph{\ defined by the classical formulas and
satisfying individually Maxwell equations (\ref{ilag11})}. Though we still
can naturally define the total potential $\mathcal{A}^{\mu }$ and the
corresponding total EM field $\mathcal{F}^{\mu \nu }$ as the sum of
individual potentials and fields by formulas (\ref{ilag1})-(\ref{ilag2}),
and though the total EM field $\mathcal{F}^{\mu \nu }$ satisfies the Maxwell
equations (\ref{ilag13}) this total field is not an independent entity.

With all that said about the new status of the total EM field its physical
significance as the field sensed by a small test charge remains valid in the
BEM theory. Indeed let us look at how different charges sense each other via
the field equations (\ref{ilag9})-(\ref{ilag11}). It is evident from the
equations and the formulas (\ref{ilag12}) for individual currents that the
entire evolution of $\ell $-th charge is completely determined by a single
quantity, its exterior potential $A_{\neq }^{\ell \mu }$, and, consequently,
charges $\left( \psi ^{\ell },A^{\ell \mu }\right) $ sense each other
exclusively by their potentials $A^{\ell \mu }$. In other words, the state
of every b-charge $\left( \psi ^{\ell },A^{\ell \mu }\right) $ and its
current $J^{\ell \mu }$ are completely determined by the action upon it of
all other charges potentials via the exterior potential $A_{\neq }^{\ell \mu
}$ and every charge $\left( \psi ^{\ell },A^{\ell \mu }\right) $ acts upon
other charges. So in this theory as we can see it is the individual $A_{\neq
}^{\ell \mu }$ potentials as components of respective b-charges $\left( \psi
^{\ell },A^{\ell \mu }\right) $ that facilitate the EM interaction between
charges. This is in contrast to the classical theory as well to our own WCM
theory in \cite{BabFig1}, \cite{BabFig2} in which it is a single and
independent EM field interacting with every charge facilitates all EM
interactions between different charges $\psi ^{\ell }$. In particular, in
the BEM theory by its very set up there is no any EM self-interaction as in
the classical theory and our theory in \cite{BabFig1}, \cite{BabFig2}, where
such interaction exists because there is only a single EM field.

Based on the analysis in Section \ref{scompcem} one may expect noticeable
differences between CEM theory and BEM theory for example when $1/N_{\ell }$
in (\ref{dif2}) is not small. These differences can become more pronounced
when $\mathcal{L}_{\mathrm{e}}$ in (\ref{relsh1}) is comparable with the
classical EM Lagrangian $\mathcal{L}_{\mathrm{CEM}}$ not to mention when if
it is a dominant term in $\mathcal{L}_{\mathrm{BEM}}$.

An important signature of the BEM theory differentiating it from the CEM
theory is a \emph{mechanism of negative radiation for certain prescribed
currents}, i.e. a situation when the EM energy propagates with the speed of
light toward the current source rather than away from it as we have shown in
Section \ref{semprecur}. This mechanism can conceivably work for a limited
time in a system of several bound charges, such as an atom or a molecule,
resulting in effective energy gain coming from matching energy loss of
b-charges outside of this system. Such energy transfer is completely
accounted for by interacting elementary EM fields as the components of the
involved b-charges including external ones which might be represented
effectively by an external EM field. The very possibility of the negative
radiation indicates a significant difference between the EM energy transport
at the elementary atomic scale from the same at the macroscopic scale. For
the later most of the time one would observe well known classical charge
behavior including the EM self-interaction and exclusively normal (positive)
radiation of the EM energy away from the source. In other words the BEM
theory allows for a differentiation between macroscopic and atomic scales at
the level of EM fields alone, and that makes the concept of a b-charge with
its elementary EM field to be truly elementary. This is in noticeable
contrast with some of the classical charge models which are based on the
concept of a single EM field, and we would like to quote here F. Rohrich, 
\cite[Section 6-1]{Rohrlich}: "This was the problem faced by Abraham,
Lorentz, and Poincare in the first few years of this century. The most
obvious model of a charged particle is a sphere carrying a spherically
symmetrical charge wave function. While such a model is meant (and was
indeed proposed) as a picture of a charged elementary particle (an electron,
for example), it is obvious that it is basically a macroscopic charged body,
only much smaller. There is nothing "elementary" about it."

Notice also that as we can see from the BEM expression (\ref{iemt6b}) for
the interaction energy density for a pair of b-charges the interaction
energy density can be negative. The later is analogous to the negative sign
of the electrostatic energy for two classical point charges when one of them
is positive and another one is negative, and the interaction energy is
defined to be zero when charges are separated by infinite distance. This way
to calibrate the interaction energy in the CEM theory is in line with
defining the interaction energy as work done to assemble a system of charges
from the state when they don't interact, that is when they are separated by
infinite distance.

Let us look at the gauge properties of the BEM system Lagrangian $\mathcal{L}
$ defined by (\ref{ilag3}), (\ref{ilag3a}). Recall that gauge \emph{%
transformation of the first or the second kind} (known also as respectively 
\emph{global and local gauge transformation}) are described respectively by
the following formulas, \cite[(17), (23a), (23b)]{Pauli RFT}, \cite[Section
11, (11.4)]{Wentzel} 
\begin{equation}
\psi ^{\ell }\rightarrow \mathrm{e}^{\mathrm{i}\gamma ^{\ell }}\psi ^{\ell
},\ \psi ^{\ell \ast }\rightarrow \mathrm{e}^{-\mathrm{i}\gamma ^{\ell
}}\psi ^{\ell \ast },\text{ where }\gamma ^{\ell }\text{ is any real
constant,}  \label{flagr8}
\end{equation}%
\begin{equation}
\psi ^{\ell }\rightarrow \mathrm{e}^{-\frac{\mathrm{i}q^{\ell }\lambda
\left( x\right) }{\chi \mathrm{c}}}\psi ^{\ell },\ \psi ^{\ell \ast
}\rightarrow \mathrm{e}^{\frac{\mathrm{i}q^{\ell }\lambda \left( x\right) }{%
\chi \mathrm{c}}}\psi ^{\ell \ast },\ A^{\mu }\rightarrow A^{\mu }+\partial
^{\mu }\lambda .  \label{flagr8a}
\end{equation}%
The significance of individual EM fields in the new theory is manifested in
the existence of a new much large group of gauge transformations than
described by (\ref{flagr8})-(\ref{flagr8a}). We have already pointed out
that the system Lagrangian $\mathcal{L}$ is invariant with respect to the
gauge transformation of the first kind (\ref{flagr8}) and used that to
construct an instrumental for the theory conserved individual currents $%
J^{\ell \mu }$ defined in (\ref{ilag11}). We introduce now a new gauge
transformation which we call \emph{elementary gauge transformations} \emph{%
gauge or transformations of the third kind:}%
\begin{gather}
A^{\ell \mu }\rightarrow A^{\ell \mu }+\partial ^{\mu }\lambda ^{\ell
}\left( x\right) ,  \label{gaum1} \\
\psi ^{\ell }\rightarrow \mathrm{e}^{-\frac{\mathrm{i}q^{\ell }\grave{\lambda%
}^{\ell }\left( x\right) }{\chi \mathrm{c}}}\psi ^{\ell },\ \psi ^{\ell \ast
}\rightarrow \mathrm{e}^{\frac{\mathrm{i}q^{\ell }\grave{\lambda}^{\ell
}\left( x\right) }{\chi \mathrm{c}}}\psi ^{\ell \ast },  \label{gaum2}
\end{gather}%
where functions $\lambda ^{\ell }\left( x\right) ,\ 1\leq \ell \leq N$, are
independent real-valued scalar functions of $x$ and 
\begin{equation}
\lambda \left( x\right) =\sum\limits_{1\leq \ell \leq N}\lambda ^{\ell
}\left( x\right) ,\ \grave{\lambda}^{\ell }\left( x\right) =\lambda \left(
x\right) -\lambda ^{\ell }\left( x\right) ,  \label{gaum3}
\end{equation}%
It is readily follows from (\ref{gaum3}) that%
\begin{equation}
\lambda ^{\ell }=\lambda -\grave{\lambda}^{\ell }=\frac{\bar{\lambda}}{N-1}-%
\grave{\lambda}^{\ell },\text{ where }\grave{\lambda}=\sum\limits_{1\leq
\ell \leq N}\grave{\lambda}^{\ell }=\left( N-1\right) \lambda ,
\label{gaum4}
\end{equation}%
implying the independence of functions $\grave{\lambda}^{\ell }\left(
x\right) ,\ 1\leq \ell \leq N$. A straightforward examination shows that for 
$N\geq 2$ the system Lagrangian $\mathcal{L}$ defined by (\ref{ilag5})-(\ref%
{ilag7}) is invariant with respect to the gauge transformation of the third
kind. As we have already pointed out it is due to the gauge invariance with
respect to elementary gauge transformations (\ref{gaum1}), (\ref{gaum2}) the
two-way representation (\ref{ilag8}) holds for the elementary currents that
accounts for an important fact that the conserved Noether's elementary
current is exactly the source current in the corresponding elementary
Maxwell equations (\ref{ilag11}), (\ref{ilag11a}).

Another feature of the BEM theory is the new expressions (\ref{iemt4}), (\ref%
{iemt4b}) and (\ref{iemt4d}) of the total EM energy and the entire EM field
EnMT compare with the classical expressions (\ref{maxw11})-(\ref{maxw13}).
Let us take a look at the representation (\ref{iemt4d}) which relates the EM
EnMT to the classical one. This identity shows that the EnMT can be
effectively obtained by the removal from the classical EnMT for the total EM
field the sum of all classical EnMT's for individual EM fields. So if our
intention was to remove the EM self-interaction in a consistent way keeping
the Lagrangian structure it is perfectly accomplished by the BEM theory.

\section{Non-relativistic dynamics of localized charges\label{snonrbem}}

In this section based on our relativistic model we introduce a
non-relativistic model for the case where charges move slowly compared with
speed of light. First we introduce field equations and the Lagrangian and
briefly describe their derivation from the relativistic model.

Using frequency-shifting substitution (\ref{psioml}) with more general $\psi
=\psi _{\omega }^{\ell }\left( t,\mathbf{x}\right) $ which depends on $%
\left( t,\mathbf{x}\right) $ we observe that the second time derivative in
the Klein-Gordon equation (\ref{KG}) can be written in the form 
\begin{equation}
-\frac{1}{\mathrm{c}^{2}}\tilde{\partial}_{t}^{\ell }\tilde{\partial}%
_{t}^{\ell }\psi _{\omega }^{\ell }=\frac{1}{\mathrm{c}^{2}}\left( \partial
_{t}+\frac{\mathrm{i}q^{\ell }}{\chi }\varphi _{\neq }^{\ell }\right)
^{2}\psi _{\omega }^{\ell }-2i\frac{m^{\ell }}{\chi }\left( \partial _{t}+%
\frac{\mathrm{i}q^{\ell }}{\chi }\varphi _{\neq }^{\ell }\right) \psi
_{\omega }^{\ell }+\kappa _{0}^{2}\psi _{\omega }^{\ell }  \label{dtl2}
\end{equation}%
where%
\begin{equation}
\varphi _{\neq }^{\ell }=\sum_{\ell ^{\prime }\neq \ell }\varphi ^{\ell
^{\prime }}.  \label{fineq}
\end{equation}%
To get a non-relativistic approximation, we neglect in (\ref{dtl2}) the term
with the factor $\frac{1}{\mathrm{c}^{2}}$ and substitute $-2i\frac{m^{\ell }%
}{\chi }\left( \partial _{t}+\frac{\mathrm{i}q^{\ell }}{\chi }\varphi _{\neq
}^{\ell }\right) \psi _{\omega }^{\ell }+\kappa _{0}^{2}\psi _{\omega
}^{\ell }$ instead of the term $-\frac{1}{\mathrm{c}^{2}}\tilde{\partial}%
_{t}^{\ell }\tilde{\partial}_{t}^{\ell }\psi _{\omega }^{\ell }$ in (\ref{KG}%
). As a result, the nonlinear Klein-Gordon equation (\ref{KG}) is
approximated by the following nonlinear Schr\"{o}dinger equation 
\begin{equation}
\chi \mathrm{i}\partial _{t}\psi ^{\ell }+\frac{\chi ^{2}}{2m^{\ell }}\left( 
\tilde{\nabla}^{\ell }\right) ^{2}\psi ^{\ell }-\frac{\chi ^{2}}{2m^{\ell }}%
G^{\ell \prime }\left( \psi ^{\ell \ast }\psi ^{\ell }\right) \psi ^{\ell
}-q^{\ell }\varphi _{\neq }^{\ell }\psi ^{\ell }=0,  \label{eqp1}
\end{equation}%
with $\tilde{\nabla}^{\ell }$ given by (\ref{ilag4}), where for notational
simplicity we write $\psi ^{\ell }$\ in place of $\psi _{\omega }^{\ell }$.
Since magnetic fields generated by moving charges also have coefficient $%
\frac{1}{\mathrm{c}}$ and are small for small velocities, they can be also
neglected, and consequently we preserve only the external magnetic fields
and replace $\tilde{\nabla}^{\ell }$ by the covariant gradient 
\begin{equation}
\tilde{\nabla}_{\mathrm{ex}}^{\ell }=\nabla -\frac{\mathrm{i}q^{\ell }%
\mathbf{A}_{\mathrm{ex}}}{\chi c}.  \label{nop1}
\end{equation}%
\ Notice that in the nonrelativistic case the $\ell $-th b-charge is
described by a pair $\psi ^{\ell }$, $\varphi ^{\ell }$ where the elementary
EM field is represented only by the scalar electric potential $\varphi
^{\ell }$. The \emph{non-relativistic model} has the the Lagrangian 
\begin{gather}
\mathcal{\hat{L}}\left( \varphi ,\left\{ \psi ^{\ell }\right\} _{\ell
=1}^{N},\left\{ \varphi ^{\ell }\right\} _{\ell =1}^{N}\right) =\frac{%
\left\vert \nabla \varphi \right\vert ^{2}}{8\pi }+\sum_{\ell }\hat{L}^{\ell
}\left( \psi ^{\ell },\psi ^{\ell \ast },\varphi \right) ,\text{ where}
\label{Lbet} \\
\hat{L}^{\ell }=\frac{\chi \mathrm{i}}{2}\left[ \psi ^{\ell \ast }\partial
_{t}\psi ^{\ell }-\psi ^{\ell }\partial _{t}\psi ^{\ell \ast }\right] -\frac{%
\chi ^{2}}{2m^{\ell }}\left\{ \left\vert \tilde{\nabla}_{\mathrm{ex}}^{\ell
}\psi ^{\ell }\right\vert ^{2}+G^{\ell }\left( \psi ^{\ell \ast }\psi ^{\ell
}\right) \right\} -  \notag \\
-q^{\ell }\left( \varphi +\varphi _{\mathrm{ex}}-\varphi ^{\ell }\right)
\psi ^{\ell }\psi ^{\ell \ast }-\frac{\left\vert \nabla \varphi ^{\ell
}\right\vert ^{2}}{8\pi },  \notag
\end{gather}%
and 
\begin{equation}
\varphi =\sum_{\ell }\varphi ^{\ell }.  \label{fisuml}
\end{equation}%
Importantly, to total potential $\varphi $ defined by (\ref{fisuml}) is not
an independent entity but just the sum of the elementary potentials $\varphi
^{\ell }$, and $\psi ^{\ell \ast }$ a variable complex conjugate to $\psi
^{\ell }$.

The Euler-Lagrange equations for charge densities $\psi ^{\ell }$ take the
form similar to (\ref{eqp1}), namely%
\begin{equation}
\mathrm{i}\chi \partial _{t}\psi ^{\ell }=-\frac{\chi ^{2}}{2m^{\ell }}%
\left( \tilde{\nabla}_{\mathrm{ex}}^{\ell }\right) ^{2}\psi ^{\ell }+q^{\ell
}\left( \varphi _{\neq }^{\ell }+\varphi _{\mathrm{ex}}\right) \psi ^{\ell }+%
\frac{\chi ^{2}}{2m^{\ell }}\left[ G_{a}^{\ell }\right] ^{\prime }\left(
\left\vert \psi ^{\ell }\right\vert ^{2}\right) \psi ^{\ell },  \label{NLSj0}
\end{equation}%
where $\varphi _{\neq \ell }$ is given by (\ref{fineq}) and every potential $%
\varphi ^{\ell }$ can be determined from the equation 
\begin{equation}
\nabla ^{2}\varphi ^{\ell }=-4\pi q^{\ell }\left\vert \psi ^{\ell
}\right\vert ^{2},\qquad \ell =1,...,N.  \label{delfi}
\end{equation}%
The solution of the above equation (\ref{delfi}) is given by the Green's
formula 
\begin{equation}
\varphi ^{\ell }\left( t,\mathbf{x}\right) =q^{\ell }\dint_{\mathbb{R}^{3}}%
\frac{\left\vert \psi ^{\ell }\right\vert ^{2}\left( t,\mathbf{y}\right) }{%
\left\vert \mathbf{y}-\mathbf{x}\right\vert }\mathrm{d}\mathbf{y.}
\label{jco3}
\end{equation}%
The properties and examples of the nonlinearities $G_{a}^{\ell }$ are
provided in the following \ref{snonlin}. As the result of charge
conservation the norms $\left\Vert \psi ^{\ell }\right\Vert ^{2}$ remain
constant on solutions of (\ref{NLSj0}) and choose those constants to 1. In
other words we impose the following charge normalization condition:%
\begin{equation}
\left\Vert \psi ^{\ell }\right\Vert ^{2}=\int_{\mathbb{R}^{3}}\left\vert
\psi ^{\ell }\right\vert ^{2}{}\mathrm{d}\mathbf{x}=1,\ t\geq 0,\ \ell
=1,...,N.  \label{norm10}
\end{equation}%
A motivation for this particular normalization is based on the formula (\ref%
{jco3}) and requirement that the elementary potential $\varphi ^{\ell }$ is
asymptotically the Coulomb's potential $q^{\ell }/\left\vert \mathbf{x}%
\right\vert $ for large $\left\vert \mathbf{x}\right\vert $.

Recall that in the classical electrodynamics the evolution of a point charge 
$q$ of a mass $m$ and and position vector $\mathbf{r}\left( t\right) $ in an
external electromagnetic (EM) field in the non-relativistic case is governed
by the Newton's equation (\ref{pchar1}). \ In our model a charge is
described by a wave function $\psi ^{\ell }\left( t,\mathbf{x}\right) $ with
dynamics described by a system of nonlinear Schrodinger equations (NLS) (\ref%
{NLSj0}) coupled through corresponding electric potentials. Nevertheless we
show below that in macroscopic regimes we can derive from the field
equations as an approximation the Newton's equations for centers $\mathbf{r}%
_{\ell }\left( t\right) $ of localized waves defined by 
\begin{equation*}
\mathbf{r}_{\ell }\left( t\right) =\int_{\mathbb{R}^{3}}\mathbf{x}\left\vert
\psi _{\ell }\left( t,\mathbf{x}\right) \right\vert ^{2}d\mathbf{x.}
\end{equation*}%
These center adequately describe the positions of wave functions $\psi
^{\ell }\left( t,\mathbf{x}\right) $ when they are localized. To introduce
localized solutions we use the nonlinearity which depends on a size
parameter $a>0$ which, in turn, determines the spatial scale of a the charge
when it is at the rest state. \ We consider below the macroscopic dynamics,
with a macroscopic spatial scale $R\gg a$. We prove in Section \ref{nrapr1}
that in the case of the non-relativistic field equations when $a\rightarrow
0 $ the centers of the interacting charges converge to solutions of the
Newton's equations with the Lorentz forces if $\psi _{\ell }$ remain
localized. We also provide examples of exact solutions of the field
equations in the form of accelerating solitons for which the localization
assumption holds.

\subsection{Single charge}

In the case of a single charge the Lagrangian and the field equations are
obtained by setting $N=1$ in (\ref{Lbet}), (\ref{NLSj0}), (\ref{delfi}) and
this case evidently $\varphi _{\neq }^{\ell }=0$. In particular, for a
single charge without the external field we have $\varphi _{\neq }^{\ell }=0$%
, $\varphi _{\mathrm{ex}}=0$, $\mathbf{A}_{\mathrm{ex}}=0$ and we obtain the
following equations for the rest state $\psi $ with $\partial _{t}\psi =0$: 
\begin{gather}
-\frac{\chi ^{2}}{2m^{\ell }}\nabla ^{2}\psi +q\varphi _{\mathrm{ex}}\psi +%
\frac{\chi ^{2}}{2m^{\ell }}\left[ G_{a}^{\ell }\right] ^{\prime }\left(
\left\vert \psi \right\vert ^{2}\right) \psi =0,  \label{nrstate} \\
\nabla ^{2}\varphi =-4\pi q\left\vert \psi \right\vert ^{2}.  \notag
\end{gather}%
Obviously, the equations coincide with (\ref{stst}), and therefore the rest
solutions of relativistic and non-relativistic equations coincide, as it
should be expected in the case of zero velocity.

Let us consider now a single charge, omitting the index $\ell $, in an
external EM field and set $N=1$ in (\ref{NLSj0}), (\ref{delfi}). For a
special class of external fields we present explicit solutions to the field
equations (\ref{NLSj0}), (\ref{delfi}) for a single charge in the form of
wave-corpuscles (accelerating solitons). We assume here for simplicity a
purely electric external EM field, i.e. when $\mathbf{A}_{\mathrm{ex}}=0$, $%
\mathbf{E}_{\mathrm{ex}}\left( t,\mathbf{x}\right) =-\nabla \varphi _{%
\mathrm{ex}}\left( t,\mathbf{x}\right) $ \ (see \cite{BabFig1}, \cite%
{BabFig2} for a similar exact solution with non-zero magnetic field). For
the purely electric external field the field equations (\ref{NLSj0}), (\ref%
{delfi}) take the form 
\begin{gather}
\mathrm{i}\chi \partial _{t}\psi =-\frac{\chi ^{2}\nabla ^{2}\psi }{2m^{\ell
}}+q\varphi _{\mathrm{ex}}\psi +\frac{\chi ^{2}}{2m}G_{a}^{\prime }\left(
\left\vert \psi \right\vert ^{2}\right) \psi ,  \label{NLSel} \\
\nabla ^{2}\varphi =-4\pi q\left\vert \psi \right\vert ^{2}.  \label{fiel}
\end{gather}%
We define then\emph{\ wave-corpuscle} $\psi ,\varphi $ by the following
formula: 
\begin{gather}
\psi \left( t,\mathbf{x}\right) =\mathrm{e}^{\mathrm{i}S/\chi }\hat{\psi}%
,\qquad S=m\mathbf{v}\cdot \left( \mathbf{x}-\mathbf{r}\right) +s_{\mathrm{p}%
}\left( t\right) ,  \label{psil0} \\
\hat{\psi}=\mathring{\psi}\left( \left\vert \mathbf{x}-\mathbf{r}\right\vert
\right) ,\qquad \varphi =\mathring{\varphi}\left( \left\vert \mathbf{x}-%
\mathbf{r}\right\vert \right) ,\qquad \mathbf{r}=\mathbf{r}\left( t\right) .
\notag
\end{gather}%
In the above formula $\mathring{\psi}$ is the form factor satisfying (\ref%
{stp}), $\mathring{\varphi}$ is a radial function determined from (\ref%
{delfi}), and we refer to the function $\mathbf{r}\left( t\right) $\ as
wave-corpuscle center.\emph{\ }Since\emph{\ }$\hat{\psi}$ is
center-symmetric, this definition agrees with more general definition (\ref%
{peu4}).

Suppose that $\varphi _{\mathrm{ex}}\left( \left( t,\mathbf{x}\right)
\right) $ is a continuous function which is linear with respect to $\mathbf{x%
}$\textbf{.} Then $\psi $ defined by (\ref{psil0}) provide an exact solution
to (\ref{NLSel}), provided that $\mathbf{r}\left( t\right) $ is determined
from the equation 
\begin{equation}
m\frac{\mathrm{d}^{2}\mathbf{r}\left( t\right) }{\mathrm{d}t^{2}}=q\mathbf{E}%
_{\mathrm{ex}}\left( t,\mathbf{r}\right) ,  \label{nlaw}
\end{equation}%
and $\mathbf{v}\left( t\right) ,s_{\mathrm{p}}\left( t\right) $ are
determined by formulas%
\begin{equation}
\mathbf{v}=\frac{\mathrm{d}r}{\mathrm{d}t}\mathbf{,\qquad }s_{\mathrm{p}%
}=\dint_{0}^{t}\left( \frac{m\mathbf{v}^{2}}{2}-q\varphi _{\mathrm{ex}%
}\left( t,\mathbf{r}\left( t\right) \right) \right) \,\mathrm{d}t^{\prime }.
\label{eel}
\end{equation}%
The verification of the fact that (\ref{psil0}), (\ref{nlaw}) and (\ref{eel}%
) determine an exact solution is straightforward and details can be found in 
\cite{BabFig1}, \cite{BabFig2} or \cite{BabFig3}.

Note that in a simpler case when the external fields $\varphi _{\mathrm{ex}}$
and $\mathbf{A}_{\mathrm{ex}}$ vanish, a simpler solution of (\ref{NLSel})-(%
\ref{fiel}) is provided by (\ref{psil0}) with $\mathbf{r}\left( t\right) =%
\mathbf{r}_{0}+\mathbf{v}t$ with constant velocity $\mathbf{v}$. In this
case the wave-corpuscle\emph{\ }solution (\ref{psil0}) of the field
equations (\ref{NLSel})-(\ref{fiel}) can be obtained from the rest solution $%
\mathring{\psi},\mathring{\varphi}$ by certain Galilean-gauge
transformations. Solutions of a similar form are known in the theory of
nonlinear Schr\"{o}dinger equations, see \cite{Sulem} and references
therein. For the particular case of the logarithmic nonlinearity solutions
of the form (\ref{psil0}) were found in \cite{Bialynicki} in the form of
accelerating gaussons. The exponential factor in wave-corpuscle solution of
the form (\ref{psil0}) can be identified with the de Broglie wave, for
details see \cite{BabFig1}, \cite{BabFig2}.

\begin{remark}
\label{R:unia}The construction of the solution (\ref{psil0}) does not depend
on a particular form of the nonlinearity $G^{\prime }=G_{a}^{\prime }$ as
long as (\ref{nop40}) is satisfied. It is uniform with respect to $a>0$, and
the dependence on $a$ in (\ref{psil0}) is only through $\mathring{\psi}%
\left( \left\vert \mathbf{x}-\mathbf{r}\right\vert \right) =a^{-3/2}%
\mathring{\psi}_{1}\left( a^{-1}\left\vert \mathbf{x}-\mathbf{r}\right\vert
\right) $. Obviously, if $\psi \left( t,\mathbf{x}\right) $ is defined by (%
\ref{psil0}) then $\left\vert \psi \left( t,\mathbf{x}\right) \right\vert
^{2}\rightarrow \delta \left( \mathbf{x}-\mathbf{r}\right) $ as $%
a\rightarrow 0$.
\end{remark}

\begin{remark}
In \cite{BabFig1}, \cite{BabFig2} \ we introduced a non-relativistic model
where equations for a steady states coincide with (\ref{nrstate}). We
derived there nonrelativistic equations from the corresponding relativistic
once not only assuming that velocities are vanishingly small, but also
assuming that the Sommerfeld's fine structure constant is vanishingly small.
If we would not neglect there the terms with Sommerfeld's fine structure
constant the steady states \ would satisfy more involved equations, see \cite%
{BabFig1} \ for details. We also would like to note that though $\mathring{%
\psi}$ would be defined differently, equations (\ref{nlaw}) would be the
same. In the present paper the rest state equations (\ref{nrstate}) are
simpler even if the Sommerfeld's fine structure constant is not replaced by
zero.
\end{remark}

\subsection{Determination of Nonlinearity, \label{snonlin}}

As we have already mentioned, the \emph{nonlinear self interaction function }%
$G$\emph{\ is determined from the charge equilibrium\ equation} (\ref{nop40}%
) based on the form factor (ground state) $\mathring{\psi}$. Important
features of our nonlinearity include: (i) the boundedness or slow
subcritical growth of its derivative $G^{\prime }\left( s\right) $ for $%
s\rightarrow \infty $ with consequent boundedness from below of the wave
energy; (ii) slightly singular behavior about $s=0,$ that is for small wave
amplitudes.

In this section we consider construction of the function $G$, study its
properties and provide examples for which the construction of $G$ is carried
out explicitly. Throughout this section we have 
\begin{equation*}
\psi ,\mathring{\psi}\geq 0\text{ and hence }\left\vert \psi \right\vert
=\psi .
\end{equation*}%
We introduce explicitly the dependence of the free ground state $\mathring{%
\psi}$ on the size parameter $a>0$ through the following representation of
the function $\mathring{\psi}\left( r\right) $ 
\begin{equation}
\mathring{\psi}\left( r\right) =\mathring{\psi}_{a}\left( r\right) =a^{-3/2}%
\mathring{\psi}_{1}\left( a^{-1}r\right) ,  \label{nrac5}
\end{equation}%
where $\mathring{\psi}_{1}\left( \mathsf{r}\right) $ is a twice continuously
differentiable function of the dimensionless parameter $\mathsf{r}\geq 0$,
and, as a consequence of (\ref{norm10}), the function $\mathring{\psi}%
_{a}\left( r\right) $ satisfies the charge normalization condition for every 
$a>0$. Obviously, definition (\ref{nrac5}) is consistent with (\ref{nop40})
and (\ref{totgkap}). The size parameter $a$ naturally has the dimension of
length. A properly defined spatial size of $\mathring{\psi}_{a}$, based, for
instance, on the variance, is proportional to $a$\ with a coefficient
depending on $\mathring{\psi}_{1}$. The charge equilibrium\ equation (\ref%
{nop40}) can be written in the following form: 
\begin{equation}
\nabla ^{2}\mathring{\psi}_{a}=G_{a}^{\prime }\left( \mathring{\psi}%
_{a}^{2}\right) \mathring{\psi}_{a}.  \label{stp}
\end{equation}%
The function $\mathring{\psi}_{a}\left( r\right) $ is assumed to be a smooth
positive monotonically decreasing function of $r\geq 0$ which is square
integrable with weight $r^{2}$ and we assume it to satisfy the charge
normalization condition of the form (\ref{norm10}); such a function is
usually called in literature a ground state.

Let us look first at the case $a=1$,\ $\mathring{\psi}_{a}=\mathring{\psi}%
_{1}$,\ $\mathring{\varphi}_{a}=\mathring{\varphi}_{1}$, for which the
equation (\ref{stp}) yields the following representation for $G^{\prime }(%
\mathring{\psi}_{1}^{2})$ from (\ref{stp})%
\begin{equation}
G_{1}^{\prime }\left( \mathring{\psi}_{1}^{2}\left( r\right) \right) =\frac{%
(\nabla ^{2}\mathring{\psi}_{1})\left( r\right) }{\mathring{\psi}_{1}\left(
r\right) }.  \label{gg}
\end{equation}%
Since $\mathring{\psi}_{1}^{2}\left( r\right) $ is a \emph{monotonic}
function,\ we can find its inverse $r=r\left( \psi ^{2}\right) ,$ yielding 
\begin{equation}
G_{1}^{\prime }\left( s\right) =\left[ \frac{\nabla ^{2}\mathring{\psi}_{1}}{%
\mathring{\psi}_{1}}\right] \left( r\left( s\right) \right) ,\ 0=\mathring{%
\psi}_{1}^{2}\left( \infty \right) \leq s\leq \mathring{\psi}_{1}^{2}\left(
0\right) .  \label{intps}
\end{equation}%
Since $\mathring{\psi}_{1}\left( r\right) $ is smooth, $G^{\prime
}(\left\vert \psi \right\vert ^{2})$ is smooth for $0<\left\vert \psi
\right\vert ^{2}\leq \mathring{\psi}_{1}^{2}\left( 0\right) $. If we do not
need $G^{\prime }\left( s\right) $ to be smooth, we extend $G^{\prime
}\left( s\right) $ for $s\geq \mathring{\psi}_{1}^{2}\left( 0\right) $ as a
constant, namely 
\begin{equation}
G_{1}^{\prime }\left( s\right) =G_{1}^{\prime }\left( \mathring{\psi}%
_{1}^{2}\left( 0\right) \right) \text{ if }s\geq \mathring{\psi}%
_{1}^{2}\left( 0\right) .  \label{intps1}
\end{equation}%
First derivative of such an extension at $s=\mathring{\psi}_{1}^{2}\left(
0\right) $ has a discontinuity point. If $\mathring{\psi}_{a}\left( r\right) 
$ is a smooth function of class $C^{n}$, we always can define an extension
of $G^{\prime }\left( s\right) $ for $s\geq \mathring{\psi}_{1}^{2}\left(
0\right) $ as a bounded function of class $C^{n}$ for all $r>0$ and 
\begin{equation}
G_{1}^{\prime }\left( s\right) =G_{1}^{\prime }\left( \mathring{\psi}%
_{1}^{2}\left( 0\right) \right) -1\text{ if }s\geq \mathring{\psi}%
_{1}^{2}\left( 0\right) +1.  \label{intps2}
\end{equation}%
Slowly growing functions $G^{\prime }\left( s\right) $ also can be used.

In the case of arbitrary size parameter $a>0$ we define $G_{a}^{\prime
}\left( s\right) $ by formula (\ref{totgkap}), this definition is consistent
with (\ref{nrac5}) and (\ref{intps}).

Let us take a look at general properties of $G^{\prime }\left( s\right) $ as
they follow from defining them relations (\ref{intps})-(\ref{totgkap}). In
the examples below the $G^{\prime }\left( s\right) $ is not differentiable
at $s=0$. But if $\mathring{\psi}\left( r\right) $ decays exponentially or
by a power law the nonlinearity $g\left( \psi \right) =G^{\prime
}(\left\vert \psi \right\vert ^{2})\psi $ as it enters the field equations
is differentiable for all $\psi $ including zero, and hence it satisfies\
the Lipschitz condition. For a Gaussian $\mathring{\psi}_{1}\left( r\right) $
which decays superexponentially $G^{\prime }(\left\vert \psi \right\vert
^{2})$ is unbounded at zero and $g\left( \psi \right) $ is not
differentiable at zero. Since $\mathring{\psi}\left( \left\vert \mathbf{x}%
\right\vert \right) >0,$ the sign of $G_{1}^{\prime }\left( \left\vert \psi
\right\vert ^{2}\right) $ coincides with the sign of $\nabla ^{2}\mathring{%
\psi}_{1}\left( \left\vert \mathbf{x}\right\vert \right) $. At the origin $%
\mathbf{x}=\mathbf{0}$ the function $\mathring{\psi}_{1}\left( \left\vert 
\mathbf{x}\right\vert \right) $ has its maximum and, consequently, $%
G_{1}^{\prime }\left( s\right) \leq 0$ for $s$ close to $s=\mathring{\psi}%
_{1}^{2}\left( 0\right) $. The Laplacian applied to radial functions $%
\mathring{\psi}_{1}$ takes the form $\frac{1}{r}\frac{\partial ^{2}}{%
\partial r^{2}}\left( r\mathring{\psi}_{1}\left\vert \mathbf{x}\right\vert
\right) $. Consequently, if $r\mathring{\psi}_{1}\left( r\right) $ is convex
at $r=\left\vert \mathbf{x}\right\vert $ we have $\nabla ^{2}\mathring{\psi}%
_{1}\left( \left\vert \mathbf{x}\right\vert \right) \geq 0$. Since $r^{2}%
\mathring{\psi}_{1}\left( r\right) $ is integrable, we naturally assume that 
$\left\vert \mathbf{x}\right\vert \mathring{\psi}_{1}\left( \left\vert 
\mathbf{x}\right\vert \right) \rightarrow 0$ as $\left\vert \mathbf{x}%
\right\vert \rightarrow \infty $. Then if the second derivative of $r%
\mathring{\psi}_{1}\left( r\right) $ has a constant sign near infinity, it
must be non-negative. For an exponentially decaying $\mathring{\psi}%
_{1}\left( r\right) $ the second derivative of $r\mathring{\psi}_{a}\left(
r\right) $ is positive implying $G_{1}^{\prime }\left( s\right) >0$ for $%
s\ll 1.$ In the examples we give below $G_{1}^{\prime }\left( s\right) $ has
exactly one zero on the half-axis.

\textbf{Example 1.} Consider a form factor $\mathring{\psi}_{1}\left(
r\right) $ decaying as a power law, namely 
\begin{equation}
\mathring{\psi}_{1}\left( r\right) =\frac{c_{\mathrm{pw}}}{\left(
1+r^{2}\right) ^{5/4}},\   \label{expsi1}
\end{equation}%
where $c_{\mathrm{pw}}$ is the normalization factor, $c_{\mathrm{pw}%
}=3^{1/2}/\left( 4\pi \right) ^{1/2}$. This function evidently is positive
and monotonically decreasing. Let us find now $G^{\prime }\left( s\right) $
based on the relations (\ref{intps}). An elementary computation of $\nabla
^{2}\mathring{\psi}_{1}$ shows that%
\begin{equation}
G^{\prime }\left( s\right) =\frac{15s^{2/5}}{4c_{\mathrm{pw}}^{4/5}}-\frac{%
45s^{4/5}}{4c_{\mathrm{pw}}^{8/5}},\ G\left( s\right) =\frac{75s^{7/5}}{28c_{%
\mathrm{pw}}^{4/5}}-\frac{25s^{9/5}}{4c_{\mathrm{pw}}^{8/5}},\text{ for }%
0\leq s\leq c_{\mathrm{pw}}^{2}.  \label{exGd1}
\end{equation}%
The extension for $s\geq c_{\mathrm{pw}}^{2}$ can be defined as a constant
or the same formula (\ref{exGd1}) can be used for all $s\geq 0$.

If we explicitly introduce size parameter $a$ into the form factor using (%
\ref{nrac5}), we define $G_{a}^{\prime }\left( s\right) $ by (\ref{totgkap}%
). Notice that the variance of the form factor $\mathring{\psi}%
_{1}^{2}\left( \left\vert \mathbf{x}\right\vert \right) $ decaying as a
power law (\ref{expsi1}) is infinite.

\textbf{Example 2}. Now we consider an exponentially decaying form factor $%
\mathring{\psi}_{1}$ of the form 
\begin{equation}
\mathring{\psi}_{1}\left( r\right) =c_{\mathrm{e}}\mathrm{e}^{-\left(
r^{2}+1\right) ^{1/2}}\text{,\qquad }c_{\mathrm{e}}=\left( 4\pi
\int_{0}^{\infty }r^{2}\mathrm{e}^{-2\left( r^{2}+1\right) ^{1/2}}\,\mathrm{d%
}r\right) ^{-1/2}.  \label{grs}
\end{equation}%
Evidently $\mathring{\psi}_{1}\left( r\right) $ is positive and
monotonically decreasing as required. The dependence $r\left( s\right) $
defined by the relation (\ref{grs})\ is as follows:\ 
\begin{equation}
r=\left[ \ln ^{2}\left( c_{\mathrm{e}}/\sqrt{s}\right) -1\right] ^{1/2},%
\text{ if }\sqrt{s}\leq \mathring{\psi}_{1}\left( 0\right) =c_{\mathrm{e}}%
\mathrm{e}^{-1}.  \label{rofp}
\end{equation}%
An elementary computation shows that 
\begin{equation}
-\frac{\nabla ^{2}\mathring{\psi}_{1}}{\mathring{\psi}_{1}}=\frac{2}{\left(
r^{2}+1\right) ^{\frac{1}{2}}}+\frac{1}{\left( r^{2}+1\right) }+\frac{1}{%
\left( r^{2}+1\right) ^{\frac{3}{2}}}-1.  \label{rofp2}
\end{equation}%
Combining (\ref{rofp}) with (\ref{rofp2}) we readily obtain the following
function for $s\leq c_{\mathrm{e}}^{2}\mathrm{e}^{-2}$ 
\begin{equation}
G_{1}^{\prime }\left( s\right) =\left[ 1-\frac{4}{\ln \left( c_{\mathrm{e}%
}^{2}/s\right) }-\frac{4}{\ln ^{2}\left( c_{\mathrm{e}}^{2}/s\right) }-\frac{%
8}{\ln ^{3}\left( c_{\mathrm{e}}^{2}/s\right) }\right] .\text{ }
\label{ggkap}
\end{equation}%
We can extend it for larger $s$ as follows:%
\begin{equation}
G_{1}^{\prime }\left( s\right) =G_{1}^{\prime }\left( c_{\mathrm{e}}^{2}%
\mathrm{e}^{-2}\right) =-3\text{ if }s\geq c_{\mathrm{e}}^{2}\mathrm{e}^{-2}.
\label{ggkap1}
\end{equation}%
or we can use a smooth extension as in (\ref{intps2}). The function $%
G_{1}^{\prime }\left( s\right) $ is not differentiable at $s=0$. At the same
time if we set $g\left( 0\right) =0$ the function $g\left( \psi \right)
=G_{1}^{\prime }\left( \psi \left( r\right) \right) \psi $ is continuous and 
$g\left( \psi \right) $ is continuously differentiable with respect to $\psi
\ $at zero and $g\left( \psi \right) $ satisfies a Lipschitz condition. The
variance of the exponential form factor $\mathring{\psi}_{1}\left( r\right) $
is obviously finite. To find $G_{a}^{\prime }\left( s\right) $ for arbitrary 
$a$ we use its representation (\ref{totgkap}).

\textbf{Example 3}. Now we define \emph{Gaussian form factor} by the formula 
\begin{equation}
\mathring{\psi}\left( r\right) =C_{g}e^{-r^{2}/2},C_{g}=\frac{1}{\pi ^{3/4}}.
\end{equation}%
Such a ground state is called \emph{gausson} in \cite{Bialynicki}.
Elementary computation shows that 
\begin{equation*}
\frac{\nabla ^{2}\mathring{\psi}\left( r\right) }{\mathring{\psi}\left(
r\right) }=r^{2}-3=-\ln \left( \mathring{\psi}^{2}\left( r\right)
/C_{g}^{2}\right) -3.
\end{equation*}%
Hence, we define the nonlinearity by the formula%
\begin{equation}
G^{\prime }\left( \left\vert \psi \right\vert ^{2}\right) =-\ln \left(
\left\vert \psi \right\vert ^{2}/C_{g}^{2}\right) -3,
\end{equation}%
we call this nonlinearity \emph{logarithmic nonlinearity. }The nonlinear
potential function has the form%
\begin{equation}
G\left( s\right) =\int_{0}^{s}\left( -\ln \left( s^{\prime
}/C_{g}^{2}\right) -3\right) ds^{\prime }=-s\ln s+s\left( \ln \frac{1}{\pi
^{3/2}}-2\right) .  \label{g1gauss}
\end{equation}%
\emph{\ }Dependence on the size parameter $a>0$ is given by the formula 
\begin{equation}
G_{a}^{\prime }\left( \left\vert \psi \right\vert ^{2}\right) =-a^{-2}\ln
\left( a^{3}\left\vert \psi \right\vert ^{2}/C_{g}^{2}\right) -3.
\label{Gpa}
\end{equation}%
Obviously $g\left( \psi \right) =G_{1}^{\prime }\left( |\psi |^{2}\right)
\psi $ is continuous for all $\psi \in \mathbb{C}$ if at zero we set $%
g\left( 0\right) $ $=0$ and is differentiable for every $\psi \neq 0$ but is
not differentiable at $\psi =0$ and does not satisfy Lipschitz condition.

\section{Charges in remote interaction regimes\label{nrapr1}}

The primary focus of this section is to show that if the size parameter $%
a\rightarrow 0$ the dynamics of the centers of localized solutions is
approximated by the Newton's law of motion (the macroscopic limit $%
a\rightarrow 0$ assumes that there is a fixed macroscopic scale $R\gg a$).
This is done in the spirit of the well known in quantum mechanics Ehrenfest\
Theorem\emph{, }\cite[Sections 7, 23]{Schiff}. We also Wave-corpuscle
solutions \ defined by (\ref{psil0}), (\ref{nlaw}) provide an example of
explicit solutions which have such a dynamics.

As we stated, the Lagrangian $\mathcal{\hat{L}}$ in (\ref{relsh}) is gauge
invariant and every $\ell $-th charge has a 4-current $\left( \rho ^{\ell },%
\mathbf{J}^{\ell }\right) $\ defined by 
\begin{equation}
\rho ^{\ell }=q\left\vert \psi ^{\ell }\right\vert ^{2},\ \mathbf{J}^{\ell
}=\left( \frac{\chi q^{\ell }}{m^{\ell }}\func{Im}\frac{\nabla \psi ^{\ell }%
}{\psi ^{\ell }}-\frac{q^{\ell 2}\mathbf{A}_{\mathrm{ex}}}{m^{\ell }\mathrm{c%
}}\right) \left\vert \psi ^{\ell }\right\vert ^{2},  \label{jco1}
\end{equation}%
which satisfies the continuity equations $\partial _{t}\rho ^{\ell }+\nabla
\cdot \mathbf{J}^{\ell }=0$ or explicitly 
\begin{equation}
\partial _{t}\left\vert \psi ^{\ell }\right\vert ^{2}+\nabla \cdot \left( 
\frac{\chi }{m^{\ell }}\func{Im}\frac{\nabla \psi ^{\ell }}{\psi ^{\ell }}%
\left\vert \psi ^{\ell }\right\vert ^{2}-\frac{q^{\ell }}{m^{\ell }\mathrm{c}%
}\mathbf{A}_{\mathrm{ex}}\left\vert \psi ^{\ell }\right\vert ^{2}\right) =0.
\label{jco2}
\end{equation}%
(Note that $\mathbf{J}^{\ell }$ defined by (\ref{jco1}) agrees with the
definition of current (\ref{mplag11}) in Maxwell equations). Equations (\ref%
{jco2}) can be obtained via multiplying (\ref{NLSj0}) by $\psi ^{\ell \ast }$
and taking imaginary part. Integrating the continuity equation we find that $%
\left\Vert \psi ^{\ell }\right\Vert ^{2}=\limfunc{const}$ and we impose the
normalization condition (\ref{norm10}). The momentum density $\mathbf{P}%
^{\ell }$ for the Lagrangian $\mathcal{\hat{L}}_{0}$ in (\ref{Lbet}) is
defined by the formula%
\begin{equation*}
\mathbf{P}^{\ell }=\frac{\mathrm{i}\chi }{2}\left( \psi ^{\ell }\tilde{\nabla%
}^{\ell \ast }\psi ^{\ell \ast }-\psi ^{\ell \ast }\tilde{\nabla}^{\ell
}\psi ^{\ell }\right) .
\end{equation*}%
Note that so defined momentum density $\mathbf{P}^{\ell }$ is related with
the current $\mathbf{J}^{\ell }$ in (\ref{jco1}) by the formula 
\begin{equation}
\mathbf{P}^{\ell }\left( t,\mathbf{x}\right) =\frac{m^{\ell }}{q^{\ell }}%
\mathbf{J}^{\ell }\left( t,\mathbf{x}\right) .  \label{jco5}
\end{equation}%
We introduce the total individual momenta $\mathsf{P}^{\ell }$ for $\ell $%
-th charge by%
\begin{equation}
\mathsf{P}^{\ell }=\int_{\mathbb{R}^{3}}\mathbf{P}^{\ell }\,\mathrm{d}%
\mathbf{x},  \label{peu1}
\end{equation}%
and obtain the following equations for the total individual momenta 
\begin{equation}
\frac{\mathrm{d}\mathsf{P}^{\ell }}{\mathrm{d}t}=q^{\ell }\int_{\mathbb{R}%
^{3}}\left[ \left( \dsum\nolimits_{\ell ^{\prime }\neq \ell }\mathbf{E}%
^{\ell ^{\prime }}+\mathbf{E}_{\mathrm{ex}}\right) \left\vert \psi ^{\ell
}\right\vert ^{2}+\frac{1}{\mathrm{c}}\mathbf{v}^{\ell }\times \mathbf{B}_{%
\mathrm{ex}}\right] \mathrm{d}\mathbf{x},  \label{peu2}
\end{equation}%
where 
\begin{equation}
\mathbf{v}^{\ell }\left( t,\mathbf{x}\right) =\frac{1}{m^{\ell }}\mathbf{P}%
^{\ell }\left( t,\mathbf{x}\right) =\frac{1}{q^{\ell }}\mathbf{J}^{\ell
}\left( t,\mathbf{x}\right) \text{.}  \label{jco6}
\end{equation}%
The external EM fields $\mathbf{E}_{\mathrm{ex}}$,$\ \mathbf{B}_{\mathrm{ex}%
} $ in (\ref{peu2}) corresponding to the potentials $\varphi _{\mathrm{ex}}$%
, $\mathbf{A}_{\mathrm{ex}}$\ are determined by standard formulas (\ref%
{maxw3a}). Derivation of (\ref{peu2}) is\ rather elementary. Indeed, in the
simplest case where $\mathbf{A}_{\mathrm{ex}}=0$ we multiply (\ref{NLSj0})
by $\nabla \psi ^{\ell \ast }$, take the real part and integrate the result
over the entire space using integration by parts. To obtain \ (\ref{peu2})
in more involved general case one can similarly multiply (\ref{NLSj0}) by $%
\tilde{\nabla}^{\ell \ast }\psi ^{\ell \ast }$ and then integrate the result
by parts using some vector algebra manipulation.

Let us show now that if the size parameter $a$ is small compared to the
typical scale of variation of EM fields, the charge evolution can be
described approximately by Newton equations with Lorentz forces similar to (%
\ref{pchar1}). Here we skip technical details. Conditions under which this
kind of derivation is justified are given in \cite{BabFig3}.

We introduce the $\ell $-th charge position $\mathbf{r}^{\ell }\left(
t\right) $ and velocity $\mathsf{v}^{\ell }\left( t\right) $ as the
following spatial averages:%
\begin{equation}
\mathbf{r}^{\ell }\left( t\right) =\mathbf{r}_{a}^{\ell }\left( t\right)
=\int_{\mathbb{R}^{3}}\mathbf{x}\left\vert \psi _{a}^{\ell }\left( t,\mathbf{%
x}\right) \right\vert ^{2}\,\mathrm{d}\mathbf{x},\ \mathsf{v}^{\ell }\left(
t\right) =\frac{1}{q^{\ell }}\int_{\mathbb{R}^{3}}\mathbf{J}^{\ell }\left( t,%
\mathbf{x}\right) \,\mathrm{d}\mathbf{x},  \label{peu4}
\end{equation}%
where the current density $\mathbf{J}^{\ell }$ is defined by (\ref{jco1}).
We show below that a combination of the continuity equation (\ref{jco2})
with the momentum evolution equations (\ref{peu2}) imply the following
remarkable property: the positions $\mathbf{r}^{\ell }\left( t\right) $
satisfy with a high accuracy Newton's equations of motion for the system of $%
N$ point charges if the size parameter $a$ is small.

Multiplying continuity equation (\ref{jco2}) by $\mathbf{x\ }$and
integrating we find the following identities%
\begin{equation}
\frac{\mathrm{d}\mathbf{r}^{\ell }\left( t\right) }{\mathrm{d}t}=\int_{%
\mathbb{R}^{3}}\mathbf{x}\partial _{t}\left\vert \psi ^{\ell }\right\vert
^{2}\,\mathrm{d}\mathbf{x}=\mathbf{-}\frac{1}{q^{\ell }}\int_{\mathbb{R}^{3}}%
\mathbf{x}\nabla \cdot \mathbf{J}^{\ell }\,\mathrm{d}\mathbf{x}=\frac{1}{%
q^{\ell }}\int_{\mathbb{R}^{3}}\mathbf{J}^{\ell }\mathrm{d}\mathbf{x}=%
\mathsf{v}^{\ell }\left( t\right) ,  \label{peu5}
\end{equation}%
showing the positions and velocities defined by formulas (\ref{peu4}) are
related exactly as in the point charge mechanics. Then integrating (\ref%
{jco5}) we obtain the following kinematic representation for the total
momentum%
\begin{equation}
\mathsf{P}^{\ell }\left( t\right) =\frac{m^{\ell }}{q^{\ell }}\int_{\mathbb{R%
}^{3}}\mathbf{J}^{\ell }\left( t,\mathbf{x}\right) \,\mathrm{d}\mathbf{x}%
=m^{\ell }\mathsf{v}^{\ell }\left( t\right) ,  \label{peu6}
\end{equation}%
which also is exactly the same as for point charges mechanics. Relations (%
\ref{peu5}) and (\ref{peu6}) yield 
\begin{equation}
m^{\ell }\frac{\mathrm{d}^{2}\mathbf{r}^{\ell }\left( t\right) }{\mathrm{d}%
^{2}t}=m^{\ell }\frac{\mathrm{d}}{\mathrm{d}t}\mathsf{v}^{\ell }\left(
t\right) =\frac{\mathrm{d}\mathsf{P}^{\ell }}{\mathrm{d}t},  \label{mp1}
\end{equation}%
and we obtain from (\ref{peu2}) the following system of equations of motion
for $N$ charges:%
\begin{equation}
m^{\ell }\frac{\mathrm{d}^{2}\mathbf{r}^{\ell }\left( t\right) }{\mathrm{d}%
^{2}t}=q^{\ell }\int_{\mathbb{R}^{3}}\left[ \left( \dsum\nolimits_{\ell
^{\prime }\neq \ell }\mathbf{E}^{\ell ^{\prime }}+\mathbf{E}_{\mathrm{ex}%
}\right) \left\vert \psi ^{\ell }\right\vert ^{2}+\frac{1}{\mathrm{c}}%
\mathbf{v}^{\ell }\times \mathbf{B}_{\mathrm{ex}}\right] \,\mathrm{d}\mathbf{%
x},\ \ell =1,...,N,  \label{Couf}
\end{equation}%
where $\mathbf{E}^{\ell ^{\prime }}\left( t,\mathbf{x}\right) =-\nabla
\varphi ^{\ell ^{\prime }}\left( t,\mathbf{x}\right) $, $\mathbf{E}_{\mathrm{%
ex}}$ and $\mathbf{B}_{\mathrm{ex}}$\ are defined by (\ref{maxw3a}).

The derivation of the above system is analogous to the well known in quantum
mechanics \emph{Ehrenfest\ Theorem}, \cite[Sections 7, 23]{Schiff} and \cite%
{Bialynicki}. Now we give a formal derivation of the Newton's law of motion
for charge centers.

Let us suppose that for every $\ell $-th charge density $\left\vert \psi
^{\ell }\right\vert ^{2}$ and the corresponding current density $\mathbf{J}%
^{\ell }$ are localized in $a$-vicinity of the position $\mathbf{r}^{\ell
}\left( t\right) $, and that $\left\vert \mathbf{r}^{\ell }\left( t\right) -%
\mathbf{r}^{\ell ^{\prime }}\left( t\right) \right\vert \geq \gamma >0$ with 
$\gamma $ independent on $a$ on time interval $\left[ 0,T\right] $. Then if $%
a\rightarrow 0$ we get 
\begin{equation}
\left\vert \psi ^{\ell }\right\vert ^{2}\left( t,\mathbf{x}\right)
\rightarrow \delta \left( \mathbf{x}-\mathbf{r}^{\ell }\left( t\right)
\right) ,\ \mathbf{v}^{\ell }\left( t,x\right) =\mathbf{J}^{\ell }/q^{\ell
}\rightarrow \mathbf{v}^{\ell }\left( t\right) \delta \left( \mathbf{x}-%
\mathbf{r}^{\ell }\left( t\right) \right) ,  \label{todel}
\end{equation}%
where the coefficients before the Dirac delta-functions are determined by
the charge normalization conditions (\ref{norm10}) and relations (\ref{peu4}%
). Using potential representations (\ref{jco3}) we infer from (\ref{todel})
the convergence of the potentials $\varphi ^{\ell }$ to the corresponding
Coulomb's potentials, namely%
\begin{equation}
\varphi ^{\ell }\left( t,\mathbf{x}\right) \rightarrow \varphi _{0}^{\ell
}\left( t,\mathbf{x}\right) =\frac{q^{\ell }}{\left\vert \mathbf{x}-\mathbf{r%
}^{\ell }\right\vert },\ \nabla _{\mathbf{r}}\varphi ^{\ell }\left( t,%
\mathbf{x}\right) \rightarrow \frac{q^{\ell }\left( \mathbf{x}-\mathbf{r}%
^{\ell }\right) }{\left\vert \mathbf{x}-\mathbf{r}^{\ell }\right\vert ^{3}}%
\text{ as }a\rightarrow 0.  \label{tocou}
\end{equation}%
Hence, if we pass to the limit as $a\rightarrow 0,$ we can recast the
equations of motion (\ref{Couf}) as the system 
\begin{equation}
m^{\ell }\frac{\mathrm{d}^{2}\mathbf{r}^{\ell }}{\mathrm{d}t^{2}}=\mathsf{f}%
^{\ell }+\epsilon _{0},\text{ }  \label{peu8}
\end{equation}%
where%
\begin{equation*}
\mathsf{f}^{\ell }=\sum_{\ell ^{\prime }\neq \ell }q^{\ell }\mathbf{E}%
_{0}^{\ell ^{\prime }}+q^{\ell }\mathbf{E}_{\mathrm{ex}}\left( \mathbf{r}%
^{\ell }\right) +\frac{1}{\mathrm{c}}\mathbf{v}^{\ell }\times \mathbf{B}_{%
\mathrm{ex}}\left( \mathbf{r}^{\ell }\right) ,\ \ell =1,...,N,
\end{equation*}%
and $\epsilon _{0}\rightarrow 0$ as $a\rightarrow 0.$ Notice that the terms $%
\mathsf{f}^{\ell }$ in equations (\ref{peu8}) coincide with the Lorentz
forces and we see that the limit equations of motion obtained from (\ref%
{peu8}) coincide with \emph{Newton's equations of motion} for point charges
interacting via the Coulomb forces and with the external EM field via
corresponding Lorentz forces, namely 
\begin{equation}
m^{\ell }\frac{\mathrm{d}^{2}\mathbf{r}^{\ell }}{\mathrm{d}t^{2}}%
=-\sum_{\ell ^{\prime }\neq \ell }\frac{q^{\ell }q^{\ell ^{\prime }}\left( 
\mathbf{r}^{\ell ^{\prime }}-\mathbf{r}^{\ell }\right) }{\left\vert \mathbf{r%
}^{\ell ^{\prime }}-\mathbf{r}^{\ell }\right\vert ^{3}}+q^{\ell }\mathbf{E}_{%
\mathrm{ex}}\left( \mathbf{r}^{\ell }\right) +\frac{1}{\mathrm{c}}\mathbf{v}%
^{\ell }\times \mathbf{B}_{\mathrm{ex}}\left( \mathbf{r}^{\ell }\right) ,\
\ell =1,\ldots ,N.  \label{Newt}
\end{equation}%
Note that we essentially use the fact that the nonlinearity $G_{a}^{\prime }$%
, which, according to (\ref{totgkap}), singularly depends on $a$ as $%
a\rightarrow 0$, does not enter the system (\ref{Couf}) explicitly. For
mathematical details of the above derivation see \cite{BabFig3}.

\begin{remark}
In special case of one particle in an external EM field which depends only
on time we presented explicit wave-corpuscle solution of (\ref{NLSj0}), (\ref%
{delfi}) given by formula (\ref{psil0}).. Note that if the EM fields have a
general form of spatial dependence, as they do in the case of multiple
particles \ we cannot find write solutions explicitly. Nevertheless, if the
EM fields do not vary fast in space, which is the case of a system of
charges in the regime of remote interaction considered in this section, the
wave-corpuscle solutions of the form (\ref{psil0}) provide approximate
solutions to the field equations with high accuracy. The dynamics of charge
centers $\mathbf{r}^{\ell }$ and their velocities $\mathbf{v}^{\ell }$ in
general case is given by the Newton's equations (\ref{Newt}). Since the
spatial extent of\ $\mathring{\psi}$ is small (of order $a $ ) and we can
linearize potentials of external fields acting on the charge near the center
of the wave-corpuscle, the substitution of wave-corpuscle into field
equations produces relative discrepancy of order 
\begin{equation*}
O\left( \frac{a^{2}}{R^{2}}+\frac{a}{R}\frac{\left\vert \mathbf{v}%
\right\vert }{\mathrm{c}}\right)
\end{equation*}%
where $R$ is the typical spatial scale of variation of external fields near
the charge trajectory. Consequently, we may expect the wave-corpuscle form
of localized solutions to be preserved for long times. We would like to
stress that even for very small $a$ the moving charges are not reduced just
to points and the oscillatory de Broglie wave factors $\mathrm{e}^{\mathrm{i}%
S^{\ell }/\chi }$ in (\ref{psil0}) remain to be significant. More detailed
studies of approximate wave-corpuscle solutions in similar situation in the
framework of the WCM can be found in \cite{BabFig1}, \cite{BabFig2}.
\end{remark}

\section{Hydrogen atom model\label{sbemha}}

In this section we provide a detailed sketch of our hydrogen atom (HA) model
with an intention to write a separate detailed paper on this subject. In
this model $\psi _{\ell }$ \ are spinless, therefore one cannot expect spin
related effects to be modeled here (though it is quite clear that extensions
of this model to multi-component $\psi _{\ell }$ which can possess spin are
possible, this is a subject of future research).

To model the hydrogen atom (HA) we set $N=2$ in the non-relativistic system (%
\ref{NLSj0}), (\ref{delfi}) where the indices $\ell $\ take two values $\ell
=1$, for electron, and $\ell =2$, for proton, and the charges values $%
q_{1}=-q=q_{2}$. The electric fields in the resting hydrogen atom have to be
time-independent, \ hence $\left\vert \psi _{\ell }\right\vert ^{2}$ in (\ref%
{delfi}) must be time-independent too. Therefore we assume that only phase
factors depend on time and consider the multi-harmonic solutions of this
system, namely solutions of the form 
\begin{equation}
\psi ^{\ell }\left( t,\mathbf{x}\right) =\mathrm{e}^{-\mathrm{i}\omega
_{\ell }t}\psi _{\ell }\left( \mathbf{x}\right) ,\qquad \varphi ^{\ell
}\left( t,\mathbf{x}\right) =\varphi _{\ell }\left( \mathbf{x}\right) ,\quad
\ell =1,2.  \label{psiup}
\end{equation}%
Plugging the expressions (\ref{psiup}) in equations (\ref{NLSj0}), (\ref%
{delfi}) we find that the functions $\psi _{\ell }\left( \mathbf{x}\right) $
satisfy the following nonlinear eigenvalue problem 
\begin{equation}
\chi \omega _{\ell }\psi _{\ell }+\frac{\chi ^{2}}{2m_{\ell }}\nabla
^{2}\psi _{\ell }-q^{\ell }\varphi _{\neq \ell }\psi _{\ell }=\frac{\chi ^{2}%
}{2m_{\ell }}G_{_{\ell }}^{\prime }\left( \left\vert \psi _{\ell
}\right\vert ^{2}\right) \psi _{\ell },  \label{nep1}
\end{equation}%
where, in accordance with (\ref{fineq}), 
\begin{equation}
\varphi _{\neq 1}=\varphi _{2},\quad \varphi _{\neq 2}=\varphi _{1}\qquad 
\frac{1}{4\pi }\nabla ^{2}\varphi _{\ell }=-q^{\ell }\left\vert \psi _{\ell
}\right\vert ^{2}.  \label{nrs2}
\end{equation}%
Let us introduce 
\begin{equation}
\Phi _{\ell }=\frac{\varphi _{\ell }}{q_{\ell }},\qquad a_{\ell }=\frac{\chi
^{2}}{q^{2}m_{\ell }},\quad \ell =1,2,  \label{fifi}
\end{equation}%
where the quantity $a_{1}$ turns into \emph{the Bohr radius} if $\chi $
equals to the Planck constant $\hbar $, and $m_{1},q$ are the electron mass
and charge respectively. Using (\ref{fifi}) we rewrite the system (\ref{nep1}%
), (\ref{nrs2}) as the following \emph{nonlinear eigenvalue problem} 
\begin{equation}
\frac{\chi }{q^{2}}\omega _{1}\psi _{1}+\frac{a_{1}}{2}\nabla ^{2}\psi
_{1}+\Phi _{2}\psi _{1}=\frac{a_{1}}{2}G_{1}^{\prime }\left( \left\vert \psi
_{1}\right\vert ^{2}\right) \psi _{1},  \label{H1}
\end{equation}%
\begin{equation}
\frac{\chi }{q^{2}}\omega _{2}\psi _{2}+\frac{a_{2}}{2}\nabla ^{2}\psi
_{2}+\Phi _{1}\psi _{2}=\frac{a_{2}}{2}G_{2}^{\prime }\left( \left\vert \psi
_{2}\right\vert ^{2}\right) \psi _{2},  \label{H2}
\end{equation}%
\begin{equation}
\nabla ^{2}\Phi _{1}=-4\pi \left\vert \psi _{1}\right\vert ^{2},\qquad
\nabla ^{2}\Phi _{2}=-4\pi \left\vert \psi _{2}\right\vert ^{2}.
\label{v1v2}
\end{equation}%
where $\psi _{1}$ and $\psi _{2}$ are respectively the wave functions for
the electron and the proton with%
\begin{equation}
\left\Vert \psi _{1}\right\Vert =1,\qquad \left\Vert \psi _{2}\right\Vert =1,
\label{v1v2a}
\end{equation}%
according the charge normalization condition (\ref{norm10}). The
nonlinearities $G_{1}^{\prime },$ $G_{2}^{\prime }$ are assumed to be
logarithmic as defined by (\ref{Gpa}) where the size parameter $a=a^{\ell }$
is different for electron and proton. One can similarly consider other
nonlinearities but in this paper we stay with the logarithmic ones.

Let us introduce an energy functional $\mathcal{E}\left( \psi _{1},\psi
_{2}\right) $ associated with the energy derived from the Lagrangian (\ref%
{Lbet}) by the following formula 
\begin{gather}
\mathcal{E}\left( \psi _{1},\psi _{2}\right) =\mathcal{E}_{1}\left( \psi
_{1},\psi _{2}\right) +\mathcal{E}_{2}\left( \psi _{1},\psi _{2}\right) ,%
\text{ where}  \label{Epe} \\
\mathcal{E}_{1}=\frac{q^{2}a_{1}}{2}\int \left[ \left\vert \nabla \psi
_{1}\right\vert ^{2}+G_{1}\left( \left\vert \psi _{1}\right\vert ^{2}\right) %
\right] \,\mathrm{d}\mathbf{x}-2\pi q^{2}\int \left[ \left( -\nabla
^{2}\right) ^{-1}\left\vert \psi _{2}\right\vert ^{2}\right] \left\vert \psi
_{1}\right\vert ^{2}\mathrm{d}\mathbf{x},  \notag \\
\mathcal{E}_{2}=\frac{q^{2}a_{2}}{2}\int \left[ \left\vert \nabla \psi
_{2}\right\vert ^{2}+G_{2}\left( \left\vert \psi _{2}\right\vert ^{2}\right) %
\right] \,\mathrm{d}\mathbf{x}-2\pi q^{2}\int \left[ \left( -\nabla
^{2}\right) ^{-1}\left\vert \psi _{1}\right\vert ^{2}\right] \left\vert \psi
_{2}\right\vert ^{2}\mathrm{d}\mathbf{x},  \notag
\end{gather}%
and $\left( -\nabla ^{2}\right) ^{-1}\left\vert \psi _{\ell }\right\vert
^{2} $ is defined by the Green's formula (\ref{jco3}). Observe then that
equations (\ref{H1}), (\ref{H2}), (\ref{v1v2}) are the Euler equations for
critical points of the functional $\mathcal{E}$\ under the normalization
constraint (\ref{v1v2a}), if we set in (\ref{v1v2})%
\begin{equation*}
\Phi _{1}=4\pi \left( \left( -\nabla ^{2}\right) ^{-1}\psi _{1}^{2}\right)
,\qquad \Phi _{2}=4\pi \left[ \left( -\nabla ^{2}\right) ^{-1}\psi _{2}^{2}%
\right]
\end{equation*}%
and the frequencies $\omega _{1},\omega _{2}$ are the corresponding Lagrange
multipliers. Importantly, it turns out that for the logarithmic as in (\ref%
{g1gauss}) nonlinearities $G_{1},G_{2}$ \emph{the frequencies }$\omega
_{1},\omega _{2}$\emph{\ and critical values of }$E_{1},E_{2}$\emph{\
satisfy Planck-Einstein formula} $E=\hbar \omega $ (see \cite{BabFig3} for
details and also \cite{Bialynicki} where a relation between the
Planck-Einstein formula and the logarithmic nonlinearity was discovered in a
different setting).

The problem of finding critical values for the energy functional $\mathcal{E}
$ defined by (\ref{Epe}) can be reduced approximately to a simpler problem
for a single wave function in a way similar to the Born-Oppenheimer
approximation in the quantum mechanics. To that we introduce a change of
variables 
\begin{equation}
\mathbf{x}=a_{\ell }\mathbf{y}_{\ell },\qquad \ell =1,2,  \label{not121}
\end{equation}%
where $a_{\ell }$ are defined by (\ref{fifi}), and rescale the fields as
follows:%
\begin{equation}
\Phi _{\ell }\left( \mathbf{x}\right) =\frac{\phi _{\ell }\left( \mathbf{y}%
_{\ell }\right) }{a_{\ell }},\qquad \psi _{\ell }\left( \mathbf{x}\right) =%
\frac{\Psi _{\ell }\left( \mathbf{y}_{\ell }\right) }{a_{\ell }^{3/2}},\quad
\ell =1,2.  \label{fipsi}
\end{equation}%
Then the equations (\ref{H1}), (\ref{H2}) turn into the following system%
\begin{equation}
\frac{\chi }{q^{2}}\omega _{1}\Psi _{1}+\frac{1}{2a_{1}}\nabla _{\mathbf{y}%
_{1}}^{2}\Psi _{1}+\frac{1}{a_{2}}\phi _{2}\left( \frac{a_{1}}{a_{2}}\mathbf{%
y}_{1}\right) \Psi _{1}=\frac{1}{2a_{1}}G_{1}^{\prime }\left( \left\vert
\psi _{1}\right\vert ^{2}\right) \Psi _{1},  \label{nh1}
\end{equation}%
\begin{equation}
\frac{\chi }{q^{2}}\omega _{2}\Psi _{2}+\frac{1}{2a_{2}}\nabla _{\mathbf{y}%
_{2}}^{2}\Psi _{2}+\frac{1}{a_{1}}\phi _{1}\left( \frac{a_{2}}{a_{1}}\mathbf{%
y}_{2}\right) \Psi _{2}=\frac{1}{2a_{2}}G_{2}^{\prime }\left( \left\vert
\Psi _{2}\right\vert ^{2}\right) \Psi _{2},  \label{nh2}
\end{equation}%
\begin{equation}
\nabla _{\mathbf{y}_{1}}^{2}\phi _{1}=-4\pi \left\vert \Psi _{1}\right\vert
^{2},\qquad \nabla _{\mathbf{y}_{2}}^{2}\phi _{2}=-4\pi \left\vert \Psi
_{2}\right\vert ^{2},  \label{vv2}
\end{equation}%
where not to complicate notations we use the same letter $G$ in rescaled
variables. Recall now that the electron/proton mass ratio is small, that is%
\begin{equation}
b=\frac{m_{1}}{m_{2}}=\frac{a_{2}}{a_{1}}\simeq \frac{1}{1800}\ll 1.
\label{baa}
\end{equation}%
Then we recast (\ref{nh1}), (\ref{nh2}) in the following system 
\begin{equation}
\frac{\chi a_{1}}{q^{2}}\omega _{1}\Psi _{1}+\frac{1}{2}\nabla ^{2}\Psi _{1}+%
\frac{1}{b}\phi _{2}\left( \frac{1}{b}\mathbf{y}\right) \Psi _{1}=\frac{1}{2}%
G_{1}^{\prime }\left( \left\vert \psi _{1}\right\vert ^{2}\right) \Psi _{1},
\label{eqh1}
\end{equation}%
\begin{equation}
\frac{\chi a_{2}}{q^{2}}\omega _{2}\Psi _{2}+\frac{1}{2}\nabla ^{2}\Psi
_{2}+b\phi _{1}\left( b\mathbf{y}\right) \Psi _{2}=\frac{1}{2}G_{2}^{\prime
}\left( \left\vert \Psi _{2}\right\vert ^{2}\right) \Psi _{2}.  \label{eqh2}
\end{equation}%
Notice that equation (\ref{eqh2}) for the proton wave function $\Psi _{2}$
depends on $\Psi _{1}$ \ through the potential $b\phi _{1}\left( b\mathbf{y}%
\right) $ which is small since $b$ \ is small. On the other hand, the
electron wave function$\Psi _{1}$ depends on $\Psi _{2}$ through the
potential $\frac{1}{b}\phi _{2}\left( \frac{1}{b}\mathbf{y}\right) $, and if
we look for radial solutions we can use the formula 
\begin{equation}
\frac{1}{b}\phi _{2}\left( \frac{1}{b}r\right) =\frac{1}{r}-\frac{4\pi }{r}%
\int_{r/b}^{\infty }\left( r_{1}-r/b\right) r_{1}\left\vert \Psi _{2}\left(
r_{1}\right) \right\vert ^{2}\,\mathrm{d}r_{1}.  \label{ficr1}
\end{equation}%
Restricting ourselves to lower energy levels for $\mathcal{E}\left( \psi
_{1},\psi _{2}\right) $ we can conclude using (\ref{ficr1}) that $\frac{1}{b}%
\phi _{2}\left( \frac{1}{b}r\right) $ can be replaced by the Coulomb
potential $\frac{1}{r}$ with an error of order $b^{2}$ (see \cite{BabFig3} \
for details). With that in mind we introduce the following energy functional
with the Coulomb potential for a single wave function: 
\begin{equation}
\mathcal{E}_{\mathrm{Cb}}\left( \Psi _{1}\right) =\frac{q^{2}}{a_{1}}\int_{%
\mathbb{R}^{3}}\left[ \frac{1}{2}\left\vert \nabla \Psi _{1}\right\vert ^{2}+%
\frac{1}{2}G_{1}\left( \left\vert \Psi _{1}\right\vert ^{2}\right) -\frac{1}{%
\left\vert \mathbf{y}\right\vert }\left\vert \Psi _{1}\right\vert ^{2}\right]
\,\mathrm{d}\mathbf{y}.  \label{E10}
\end{equation}%
Then with a small and controlled error we substitute the original problem of
finding frequencies $\omega _{1}$ based on critical points of $\mathcal{E}%
\left( \psi _{1},\psi _{2}\right) $ for lower energy levels with a simpler
problem of finding critical points, lower critical levels and corresponding
frequencies $\omega _{1}$ for the energy functional $\mathcal{E}_{\mathrm{Cb}%
}\left( \Psi _{1}\right) $ subjected to the constraint $\left\Vert \Psi
_{1}\right\Vert =1$. The corresponding nonlinear eigenvalue problem for the
electron wave function$\ \Psi _{1}$ and dimensionless spectral parameter $%
\omega =\frac{\chi a_{1}}{q^{2}}\omega _{1}$ is%
\begin{equation}
\omega \Psi _{1}+\frac{1}{2}\nabla ^{2}\Psi _{1}+\frac{1}{\left\vert \mathbf{%
y}\right\vert }\Psi _{1}=\frac{1}{2}G_{1}^{\prime }\left( \left\vert \psi
_{1}\right\vert ^{2}\right) \Psi _{1}.  \label{nlh}
\end{equation}%
As we have already mentioned a similar reduction to a single Schrodinger
equation with the Coulomb potential is made in the quantum mechanics via the
Born-Oppenheimer approximation.

Let us exploit the dependence of the nonlinearity $G_{1}^{\prime
}=G_{1a}^{\prime }$ on the small parameter $\kappa =\frac{a_{1}}{a}$ which
is the the ratio of the electron Bohr radius $a_{1}$ $\ $to the size
parameter $a$. If $\kappa $ is small then the nonlinearity $G_{1a}^{\prime
}\left( s\right) =\kappa ^{2}G_{1}^{\prime }\left( \kappa ^{-3}s\right) $ is
small and plays a role of a small perturbation in the eigenvalue problem (%
\ref{nlh}). A detailed analysis shows that lower energy levels of the
functional $\mathcal{E}_{\mathrm{Cb}}\left( \Psi _{1}\right) $ \ are
arbitrary close to the energy levels of the Schrodinger operator for HA \
provided that $\kappa =\frac{a_{1}}{a}$ is sufficiently small. Consequently,
based on estimates obtained in \cite{BabFig3} we can conclude that $n$-th
lower frequency $\omega _{1n}$ for solution of (\ref{H1}), (\ref{H2}), (\ref%
{v1v2}) are given by the following approximate formula 
\begin{equation}
\chi \omega _{1n}=-\frac{1}{n^{2}}\frac{q^{2}}{2a_{1}}\left[ 1+O\left(
b^{2}+\left( \frac{a_{1}}{a}\right) ^{2}\left\vert \ln \left( \frac{a_{1}}{a}%
\right) \right\vert \right) \right] ,\quad n=1,2,...  \label{nlh1}
\end{equation}%
The correction term $O\left( b^{2}+\left( \frac{a_{1}}{a}\right)
^{2}\left\vert \ln \left( \frac{a_{1}}{a}\right) \right\vert \right) $ in (%
\ref{nlh1}) is small if $b$ given by (\ref{baa}) and $\frac{a_{1}}{a}$ are
small. Observe that differences of energy levels of the nonlinear eigenvalue
problem are very close for the same in the Rydberg formula with relative
error of order $10^{-4}$ if $\frac{a_{1}}{a}$ \ is of order $10^{-2}$.
Hence, if we assume that the size $a$ of a free electron is 100 larger than
the Bohr radius, \ then the introduced here hydrogen atom model is a good
quantitative agreement with the hydrogen spectroscopic data. We think that
it is quite reasonable to assume that a free electron has much larger size
then an electron bound in a hydrogen atom where it is naturally contracted
by the electric force of the positively charged proton.

\begin{remark}
If we use the provisional HA model from \cite{BabFig2} we still obtain
discrete energy levels, but as $\frac{a_{1}}{a}\rightarrow 0$ the limiting
linear eigenvalue problem involves a potential $\frac{-q^{2}}{\left\vert 
\mathbf{y}\right\vert }+q\phi \left( \mathbf{y}\right) _{\ }$where in
addition to the Coulomb potential there is a term $q\phi \left( \mathbf{y}%
\right) $ due the electron EM self-interaction. So, if there is EM
self-interaction the limiting eigenvalue problem as $\frac{a_{1}}{a}%
\rightarrow 0$ does not turn into the same for the linear Schrodinger
operator for the HA and consequently the energy levels do not converge to
the known expressions for HA as $\frac{a_{1}}{a}\rightarrow 0$.
\end{remark}

Now we briefly compare the above non-relativistic treatment of the HA with
the treatment in the framework of the full relativistic version of our
model. We start directly from the relativistic system in (\ref{fiash})-(\ref%
{KG}) and look for time-harmonic solutions with $\mathbf{A}^{\ell }=0$ and
time-independent $\varphi ^{\ell }$ using substitutions (\ref{psioml}) and (%
\ref{psiup}) as follows: 
\begin{equation}
\psi ^{\ell }\left( t,\mathbf{x}\right) =\mathrm{e}^{-\mathrm{i}\left(
\omega _{\ell }+\omega _{0}\right) t}\psi _{\ell }\left( \mathbf{x}\right)
,\qquad \varphi ^{\ell }\left( t,\mathbf{x}\right) =\varphi _{\ell }\left( 
\mathbf{x}\right) ,
\end{equation}%
where $\ell =1,2,$ $\omega _{0}=\frac{m_{\ell }\mathrm{c}^{2}}{\chi }=%
\mathrm{c}\kappa _{0\ell }$. We arrive then at a system similar to (\ref%
{nep1}), (\ref{nrs2})%
\begin{equation}
\frac{1}{\mathrm{c}^{2}}\left( \frac{m_{\ell }\mathrm{c}^{2}}{\chi }+\omega
_{\ell }-\frac{q^{\ell }\varphi _{\neq \ell }}{\chi }\right) ^{2}\psi ^{\ell
}+\nabla ^{2}\psi ^{\ell }-G^{\ell \prime }\left( \psi ^{\ell \ast }\psi
^{\ell }\right) \psi ^{\ell }-\frac{1}{\mathrm{c}^{2}}\left( \frac{m_{\ell }%
\mathrm{c}^{2}}{\chi }\right) ^{2}\psi ^{\ell }=0.
\end{equation}%
Based on smallness of electron/proton mass ratio we similarly to the
non-relativistic case arrive to the following eigenvalue problem for
electron density which is a relativistic version of \ (\ref{nlh}):%
\begin{equation}
\left( m_{1}\mathrm{c}^{2}+\chi \omega _{1}+\frac{q^{2}}{\left\vert \mathbf{y%
}\right\vert }\right) ^{2}\psi _{1}+\mathrm{c}^{2}\chi ^{2}\nabla ^{2}\psi
_{1}-\mathrm{c}^{2}\chi ^{2}G^{\prime }\left( \left\vert \psi
_{1}\right\vert ^{2}\right) \psi _{1}-m_{1}^{2}\mathrm{c}^{4}\psi _{1}=0.
\label{relNLS}
\end{equation}%
If the ratio $\kappa =\frac{a_{1}}{a}$ is small, the nonlinearity $G^{\prime
}\left( s\right) =\kappa ^{2}G^{\prime }\left( \kappa ^{-3}s\right) $ can be
treated as a small perturbation, and the linear part of (\ref{relNLS})
essentially determines the lower energy levels. Note that if we set $\chi
=\hbar $ the linear part of (\ref{relNLS}) coincides with \emph{relativistic
Schrodinger equation} (see \cite{Schiff} p.309 ), which has the form 
\begin{equation*}
\left( -\hbar ^{2}c^{2}\nabla ^{2}+m^{2}c^{4}\right) u=\left( E+\frac{q}{%
\left\vert \mathbf{x}\right\vert }\right) ^{2}u
\end{equation*}%
with energy level $E=m\mathrm{c}^{2}+\chi \omega _{1}.$ According to \cite%
{Schiff} the energy levels of the linear relativistic Schrodinger equation
in a contrast to the non-relativistic hydrogen Schrodinger equation have a
fine structure, the fine structure energy levels are given by Sommerfeld's
formula and the relative scale of the fine structure is controlled by $%
\alpha ^{2}$ where $\alpha $ is Sommerfeld's fine structure constant, $%
\alpha =\frac{q^{2}\ }{\hbar \mathrm{c}\ }\simeq \frac{1}{137}$. This shows
that at atomic scales in our relativistic model relativistic effects are
present even in the case of zero velocities if the square of Sommerfeld's
fine structure constant is not assumed to be negligible.

\begin{remark}
In our treatment of charges in Section \ref{nrapr1} at macroscopic scales we
assume the electron size $a\ $to be very small and in this section we assume 
$\kappa =a_{1}/a$ to be very small. Since there is huge gap of scales
between the macroscopic and atomic scales there is no contradiction if we
take into account the small value of the Bohr radius $a_{1}\sim 5.3\times
10^{-11}$ $m$ compared with the scale of variation of EM fields. Note that
the error $\epsilon _{0}$ of approximation by Newtonian trajectory in (\ref%
{peu8}) is of order $a^{2}/R_{\mathrm{macr}}^{2}\ll 1\ $where $R_{\mathrm{%
macr}}$ \ is the scale of spatial variation of EM fields which act on a
charge. In the treatment of HA in this section we assume $\kappa
^{2}=a_{1}^{2}/a^{2}\ll 1$. Taking $a\sim 10^{2}a_{1}$ \ we arrive at the
restriction $R_{\mathrm{macr}}\gg 5.3\times 10^{-9}m$\ which \ is an
estimate of the scale of spatial variation of EM fields for which the
Newton's equations with Lorentz force holds with a good accuracy.
\end{remark}

\section{Appendix}

\subsection{Classical electrodynamics}

We consider the Maxwell equations for the EM fields and their covariant form
following to \cite[Section 11.9]{Jackson}, \cite[Sections 23, 30]{LandauLif
EM}, \cite[Sections 7.4, 11.2]{Griffiths}, in CGS units%
\index{Maxwell equations}%
\begin{equation}
\nabla \cdot \mathbf{E}=4\pi \varrho ,\qquad \nabla \cdot \mathbf{B}=0,
\label{maxw1}
\end{equation}%
\begin{equation}
\nabla \times \mathbf{E}+%
\frac{1}{\mathrm{c}}\partial _{t}\mathbf{B}=0,\qquad \nabla \times \mathbf{B}%
-\frac{1}{\mathrm{c}}\partial _{t}\mathbf{E}=\frac{4\pi }{\mathrm{c}}\mathbf{%
J}.  \label{maxw2}
\end{equation}%
To represent Maxwell equations in a manifestly Lorentz invariant form it is
common to introduce a four-vector potential $A^{\mu }$ and a four-vector
current density $J^{\nu }$:%
\index{current density!four-vector}%
\index{electromagnetic!four-vector potential} 
\begin{gather}
A^{\mu }=\left( \varphi ,\mathbf{A}\right) ,\ J^{\mu }=\left( \mathrm{c}%
\varrho ,\mathbf{J}\right) ,  \label{maxw2a} \\
\partial _{\mu }=%
\frac{\partial }{\partial x^{\mu }}=\left( \frac{1}{\mathrm{c}}\partial
_{t},\nabla \right) ,\ \partial ^{\mu }=\frac{\partial }{\partial x_{\mu }}%
=\left( \frac{1}{\mathrm{c}}\partial _{t},-\nabla \right) ,  \notag
\end{gather}%
and, then, an antisymmetric second-rank tensor, the "field strength tensor,%
\index{electromagnetic!field tensor}%
\begin{equation}
F^{\mu \nu }=\partial ^{\mu }A^{\nu }-\partial ^{\nu }A^{\mu },
\label{maxw2b}
\end{equation}%
so that%
\begin{equation}
F^{\mu \nu }=\left[ 
\begin{array}{cccc}
0 & -E_{1} & -E_{2} & -E_{3} \\ 
E_{1} & 0 & -B_{3} & B_{2} \\ 
E_{2} & B_{3} & 0 & -B_{1} \\ 
E_{3} & -B_{2} & B_{1} & 0%
\end{array}%
\right] ,\ F_{\mu \nu }=\left[ 
\begin{array}{cccc}
0 & E_{1} & E_{2} & E_{3} \\ 
-E_{1} & 0 & -B_{3} & B_{2} \\ 
-E_{2} & B_{3} & 0 & -B_{1} \\ 
-E_{3} & -B_{2} & B_{1} & 0%
\end{array}%
\right] ,  \label{maxw3}
\end{equation}%
and%
\begin{equation}
\mathbf{E}=-\nabla \varphi -%
\frac{1}{\mathrm{c}}\partial _{t}\mathbf{A},\ \mathbf{B}=\nabla \times 
\mathbf{A}.  \label{maxw3a}
\end{equation}%
Then the two inhomogeneous equations and the two homogeneous equations from
the four Maxwell equations (\ref{maxw1}) take respectively the form%
\index{Maxwell equations!covariant form}%
\begin{equation}
\partial _{\mu }F^{\mu \nu }=%
\frac{4\pi }{\mathrm{c}}J^{\nu },  \label{maxw4}
\end{equation}%
\begin{equation}
\partial _{\alpha }F_{\beta \gamma }+\partial _{\beta }F_{\gamma \alpha
}+\partial _{\gamma }F_{\alpha \beta }=0,\ \alpha ,\beta ,\gamma =0,1,2,3.
\label{maxw4a}
\end{equation}%
It follows from the asymmetry of $F^{\mu \nu }$, the Maxwell equation (\ref%
{maxw4}) and (\ref{maxw2a})-(\ref{maxw2b}) that the four-vector current $%
J^{\mu }$ must satisfy the \emph{continuity equation} 
\begin{equation}
\partial _{\mu }J^{\mu }=0\text{ or }\partial _{t}\varrho +\nabla \cdot 
\mathbf{J}=0.  \label{maxw5}
\end{equation}%
The Maxwell equations (\ref{maxw4}) turn into the following equations for
the four-vector potential $A^{\mu }$ 
\begin{equation}
\square A^{\nu }-\partial ^{\nu }\partial _{\mu }A^{\mu }=\frac{4\pi }{%
\mathrm{c}}J^{\nu },  \label{maxw6}
\end{equation}%
where%
\index{d'Alembertian operator} 
\begin{equation}
\square =\partial _{\mu }\partial ^{\mu }=%
\frac{1}{\mathrm{c}^{2}}\partial _{t}^{2}-\nabla ^{2}\text{ (d'Alembertian
operator).}  \label{maxw7}
\end{equation}

The EM field Lagrangian is, \cite[Section 12.7]{Jackson}, \cite[Section IV.1]%
{Barut}%
\index{Lagrangian!electromagnetic field} 
\begin{equation}
L_{\mathrm{em}}\left( A^{\mu }\right) =-%
\frac{1}{16\pi }F_{\mu \nu }F^{\mu \nu }-\frac{1}{\mathrm{c}}J_{\mu }A^{\mu
},  \label{flagr7}
\end{equation}%
where $J_{\mu }$ is an external (prescribed) current. Using (\ref{maxw3}), (%
\ref{maxw3a}) and (\ref{maxw2a}) we can recast (\ref{flagr7}) as%
\begin{gather}
L_{\mathrm{em}}\left( A^{\mu }\right) =\frac{1}{8\pi }\left( \mathbf{E}^{2}-%
\mathbf{B}^{2}\right) -\rho \varphi +\frac{1}{\mathrm{c}}\mathbf{J}\cdot 
\mathbf{A}  \label{flagr7a} \\
=\frac{1}{8\pi }\left[ \left( \nabla \varphi +\frac{1}{\mathrm{c}}\partial
_{t}\mathbf{A}\right) ^{2}-\left( \nabla \times \mathbf{A}\right) ^{2}\right]
-\rho \varphi +\frac{1}{\mathrm{c}}\mathbf{J}\cdot \mathbf{A.}  \notag
\end{gather}%
In particular, if there are no sources the above Lagrangians turn into%
\begin{gather}
L_{\mathrm{em}}\left( A^{\mu }\right) =-\frac{1}{16\pi }F_{\mu \nu }F^{\mu
\nu }=\frac{1}{8\pi }\left( \mathbf{E}^{2}-\mathbf{B}^{2}\right) =
\label{flagr7aa} \\
=\frac{1}{8\pi }\left[ \left( \nabla \varphi +\frac{1}{\mathrm{c}}\partial
_{t}\mathbf{A}\right) ^{2}-\left( \nabla \times \mathbf{A}\right) ^{2}\right]
.  \notag
\end{gather}

The canonical energy-momentum (stress-, power-momentum) tensor $\mathring{%
\Theta}^{\mu \nu }$ for the EM field is as follows, \cite[(12.104)]{Jackson}%
, \cite[Section III.4.D]{Barut}%
\index{energy-momentum tensor!electromagnetic field!canonical} 
\begin{equation}
\mathring{\Theta}^{\mu \nu }=-\frac{F^{\mu \gamma }\partial ^{\nu }A_{\gamma
}}{4\pi }+g^{\mu \nu }\frac{F^{\xi \gamma }F_{\xi \gamma }}{16\pi },
\label{flagr7b}
\end{equation}%
or, in particular, for $i,j=1,2,3$ 
\begin{gather}
\mathring{\Theta}^{00}=-\frac{\mathbf{E}^{2}-\mathbf{B}^{2}}{8\pi }+\rho
\varphi -\frac{1}{\mathrm{c}}\mathbf{J}\cdot \mathbf{A}-\frac{\partial _{0}%
\mathbf{A}\cdot \mathbf{E}}{4\pi },  \label{flagr7c} \\
\mathring{\Theta}^{0i}=-\frac{\partial _{i}\mathbf{A}\cdot \mathbf{E}}{4\pi }%
,\ \mathring{\Theta}^{i0}=-\frac{E_{i}\partial _{0}\varphi }{4\pi }+\frac{%
\left( \mathbf{B}\times \partial _{0}\mathbf{A}\right) _{i}}{4\pi },  \notag
\\
\mathring{\Theta}^{ij}=-\frac{E_{i}\partial _{j}\varphi }{4\pi }+\frac{%
\left( \mathbf{B}\times \partial _{j}\mathbf{A}\right) _{i}}{4\pi }+\frac{%
\mathbf{E}^{2}-\mathbf{B}^{2}}{8\pi }-\rho \varphi +\frac{1}{\mathrm{c}}%
\mathbf{J}\cdot \mathbf{A},  \notag
\end{gather}%
whereas the symmetric one $\Theta ^{\alpha \beta }$ for the EM field is, 
\cite[Section 12.10, (12.113)]{Jackson}, \cite[Section III.3]{Barut}%
\index{energy-momentum tensor!electromagnetic field!symmetric}%
\begin{equation}
\Theta ^{\alpha \beta }=%
\frac{1}{4\pi }\left( g^{\alpha \mu }F_{\mu \nu }F^{\nu \beta }+\frac{1}{4}%
g^{\alpha \beta }F_{\mu \nu }F^{\mu \nu }\right) ,  \label{maxw11}
\end{equation}%
implying the following formulas for the field energy density $w$, the
momentum density $\mathbf{g}$ and the Maxwell stress tensor $\tau _{ij}$: 
\begin{equation}
w=\Theta ^{00}=\frac{\mathbf{E}^{2}+\mathbf{B}^{2}}{8\pi },\ \mathrm{c}%
g_{i}=\Theta ^{0i}=\Theta ^{i0}=\frac{\mathbf{E}\times \mathbf{B}}{4\pi },
\label{maxw12}
\end{equation}%
\begin{equation}
\Theta ^{ij}=-\frac{1}{4\pi }\left[ E_{i}E_{j}+B_{i}B_{j}-\frac{1}{2}\delta
_{ij}\left( \mathbf{E}^{2}+\mathbf{B}^{2}\right) \right] ,  \label{maxw13}
\end{equation}%
\begin{eqnarray}
\Theta ^{\alpha \beta } &=&%
\begin{bmatrix}
w & \mathrm{c}\mathbf{g} \\ 
\mathrm{c}\mathbf{g} & -\tau _{ij}%
\end{bmatrix}%
,\ \Theta _{\alpha \beta }=%
\begin{bmatrix}
w & -\mathrm{c}\mathbf{g} \\ 
-\mathrm{c}\mathbf{g} & -\tau _{ij}%
\end{bmatrix}%
,  \label{maxw14} \\
\Theta _{\ \beta }^{\alpha } &=&%
\begin{bmatrix}
w & -\mathrm{c}\mathbf{g} \\ 
\mathrm{c}\mathbf{g} & -\tau _{ij}%
\end{bmatrix}%
,\ \Theta _{\alpha }^{\ \beta }=%
\begin{bmatrix}
w & \mathrm{c}\mathbf{g} \\ 
-\mathrm{c}\mathbf{g} & -\tau _{ij}%
\end{bmatrix}%
.  \notag
\end{eqnarray}%
Note that in the special case when the vector potential $\mathbf{A}$
vanishes and the scalar potential $\varphi $ does not depend on time using
the expressions (\ref{maxw3a}) we get the following representation for the
canonical energy density defined by (\ref{flagr7c})%
\begin{gather}
\mathring{\Theta}^{00}=-\frac{\left( \nabla \varphi \right) ^{2}}{8\pi }%
+\rho \varphi \text{ for }\mathbf{A}=\mathbf{0}\text{ and }\partial
_{0}\varphi =0,  \label{maxw14a} \\
\text{whereas }\Theta ^{00}=\frac{\left( \nabla \varphi \right) ^{2}}{8\pi }.
\notag
\end{gather}%
It is instructive to observe a substantial difference between the above
expressions $\mathring{\Theta}^{00}$, which is the Hamiltonian density of
the EM field, and the energy density $\Theta ^{00}$ defined by (\ref{maxw12}%
).

If there no external currents the with differential \emph{conservation laws}
takes the form%
\index{conservation laws!energy-momentum!EM field}%
\begin{equation}
\partial _{\alpha }\Theta ^{\alpha \beta }=0,  \label{maxw15}
\end{equation}%
and, in particular, the energy conservation law%
\begin{gather}
0=\partial _{\alpha }\Theta ^{\alpha \beta }=%
\frac{1}{\mathrm{c}}\left( \frac{\partial w}{\partial t}+\nabla \cdot 
\mathbf{S}\right) ,\text{ where }w\text{ is the energy density, and}
\label{maxw16} \\
\mathbf{S}=\mathrm{c}^{2}\mathbf{g}=\frac{\mathrm{c}}{4\pi }\mathbf{E}\times 
\mathbf{B}\text{ is the Poynting vector.}  \notag
\end{gather}%
In the presence of external currents the conservation laws take the form, 
\cite[Section 12.10]{Jackson}%
\index{conservation laws!energy-momentum!EM field}%
\begin{equation}
\partial _{\alpha }\Theta ^{\alpha \beta }=-f^{\beta },\qquad f^{\beta }=%
\frac{1}{\mathrm{c}}F^{\beta \nu }J_{\nu },  \label{maxw17}
\end{equation}%
and the time and space components of the equations (\ref{maxw17}) are the
conservation of energy $w$ and momentum $\mathbf{g}$ which can be recast 
\begin{equation}
\frac{1}{\mathrm{c}}\left( \frac{\partial w}{\partial t}+\nabla \cdot 
\mathbf{S}\right) =-\frac{1}{\mathrm{c}}\mathbf{J}\cdot \mathbf{E},
\label{maxw18}
\end{equation}%
\begin{equation}
\frac{\partial g_{i}}{\partial t}-\sum\limits_{j=1}^{3}\frac{\partial }{%
\partial x^{j}}\tau _{ij}=-\left[ \rho E_{i}+\frac{1}{\mathrm{c}}\left( 
\mathbf{J}\times \mathbf{B}\right) _{i}\right] .  \label{maxw19}
\end{equation}
The energy conservation law (\ref{maxw18}) is often called the Poynting's
theorem, \cite[Section 6.7]{Jackson}%
\index{conservation laws!Poynting's theorem}. The 4-vector $f^{\beta }$ in
the conservation law (\ref{maxw17}) is known as the \emph{Lorentz force
density}%
\index{Lorentz force!density}%
\begin{equation}
f^{\beta }=%
\frac{1}{\mathrm{c}}F^{\beta \nu }J_{\nu }=\left( \frac{1}{\mathrm{c}}%
\mathbf{J}\cdot \mathbf{E},\rho \mathbf{E}+\frac{1}{\mathrm{c}}\mathbf{J}%
\times \mathbf{B}\right) .  \label{maxw20}
\end{equation}

\subsection{Potentials and fields for prescribed currents\label{sGreenMax}}

In this section we describe EM fields $F^{\mu \nu }$ arising from prescribed
(external) currents $J^{\nu }$ following mostly to \cite[Section 6.4, 6.5
and 12.11]{Jackson}. Namely, the EM fields $F^{\mu \nu }$ satisfy the
inhomogeneous Maxwell equation%
\index{Maxwell equations}%
\begin{equation}
\partial _{\mu }F^{\mu \nu }=%
\frac{4\pi }{\mathrm{c}}J^{\nu },\ F^{\mu \nu }=\partial ^{\mu }A^{\nu
}-\partial ^{\nu }A^{\mu },  \label{grmax1}
\end{equation}%
which take the following form for the potentials $A^{\nu }$%
\begin{equation}
\square A^{\nu }-\partial ^{\nu }\partial _{\mu }A^{\mu }=\frac{4\pi }{%
\mathrm{c}}J^{\nu }.  \label{grmax2}
\end{equation}%
If the potentials satisfy the Lorentz condition, $\partial _{\mu }A^{\mu }=0$%
, they are then solutions of the four-dimensional wave equation,%
\begin{equation}
\square A^{\nu }=\frac{4\pi }{\mathrm{c}}J^{\nu }  \label{grmax3}
\end{equation}%
The solution of the inhomogeneous wave equation (\ref{grmax3}) is
accomplished by finding a Green function $G\left( x,x^{\prime }\right) $ for
the equation%
\begin{equation}
\square G\left( z\right) =\delta ^{\left( 4\right) }\left( z\right) ,\
G\left( x,x^{\prime }\right) =G\left( x-x^{\prime }\right) ,  \label{grmax4}
\end{equation}%
where $\delta ^{\left( 4\right) }\left( z\right) =\delta \left( z_{0}\right)
\delta \left( \mathbf{z}\right) $ is a four-dimensional delta function. One
can introduce then the so-called \emph{retarded or causal Green function}
solving the above equation (\ref{grmax4}), namely%
\index{Green function!retarded, causal}%
\begin{equation}
G^{\left( +\right) }\left( x-x^{\prime }\right) =%
\frac{\theta \left( x_{0}-x_{0}^{\prime }\right) \delta \left(
x_{0}-x_{0}^{\prime }-R\right) }{4\pi R},\ R=\left\vert \mathbf{x}-\mathbf{x}%
^{\prime }\right\vert ,  \label{grmax5}
\end{equation}%
where $\theta \left( x_{0}\right) $ is the Heaviside step function. The name
causal or retarded is justified by the fact that the source-point time $%
x_{0}^{\prime }$ is always earlier then the observation-point time $x_{0}$.
Similarly one can introduce the \emph{advanced Green function}%
\index{Green function!advanced} 
\begin{equation}
G^{\left( -\right) }\left( x-x^{\prime }\right) =%
\frac{\theta \left[ -\left( x_{0}-x_{0}^{\prime }\right) \right] \delta
\left( x_{0}-x_{0}^{\prime }+R\right) }{4\pi R},\ R=\left\vert \mathbf{x}-%
\mathbf{x}^{\prime }\right\vert .  \label{grmax6}
\end{equation}%
These Green functions can be written in the following covariant form%
\begin{eqnarray}
G^{\left( +\right) }\left( x-x^{\prime }\right) &=&\frac{1}{2\pi }\theta
\left( x_{0}-x_{0}^{\prime }\right) \delta \left[ \left( x-x^{\prime
}\right) ^{2}\right] ,  \label{grmax7} \\
G^{\left( -\right) }\left( x-x^{\prime }\right) &=&\frac{1}{2\pi }\theta
\left( x_{0}^{\prime }-x_{0}\right) \delta \left[ \left( x-x^{\prime
}\right) ^{2}\right] ,\   \notag
\end{eqnarray}%
where%
\begin{equation*}
\left( x-x^{\prime }\right) ^{2}=\left( x_{0}-x_{0}^{\prime }\right)
^{2}-\left\vert \mathbf{x}-\mathbf{x}^{\prime }\right\vert ^{2},
\end{equation*}%
\begin{equation}
\delta \left[ \left( x-x^{\prime }\right) ^{2}\right] =\frac{1}{2R}\left[
\delta \left( x_{0}-x_{0}^{\prime }-R\right) +\delta \left(
x_{0}-x_{0}^{\prime }+R\right) \right] .  \label{grmax8}
\end{equation}%
The more explicit form of the Green functions $G^{\left( \pm \right) }$ in
terms of time space variable is%
\begin{equation}
G^{\left( \pm \right) }\left( \tau ,R\right) =\frac{1}{R}\delta \left( \tau
\mp \frac{R}{\mathrm{c}}\right)  \label{grmax8a}
\end{equation}%
where%
\begin{equation}
R=\left\vert \mathbf{x}-\mathbf{x}^{\prime }\right\vert ,\quad \tau
=t-t^{\prime },  \label{grmax8b}
\end{equation}%
or%
\begin{equation}
G^{\left( \pm \right) }\left( t,\mathbf{x};t^{\prime },\mathbf{x}^{\prime
}\right) =\frac{1}{\left\vert \mathbf{x}-\mathbf{x}^{\prime }\right\vert }%
\delta \left( t-\left[ t^{\prime }\mp \frac{\left\vert \mathbf{x}-\mathbf{x}%
^{\prime }\right\vert }{\mathrm{c}}\right] \right) .  \label{grmax8c}
\end{equation}

The solution to the wave equation (\ref{grmax3}) can be written in terms of
the Green functions%
\begin{equation}
A^{\nu }\left( x\right) =A_{\mathrm{in}}^{\nu }\left( x\right) +\frac{4\pi }{%
\mathrm{c}}\dint G^{\left( +\right) }\left( x-x^{\prime }\right) J^{\nu
}\left( x^{\prime }\right) \,\mathrm{d}x^{\prime }  \label{grmax9}
\end{equation}%
or%
\begin{equation}
A^{\nu }\left( x\right) =A_{\mathrm{out}}^{\nu }\left( x\right) +\frac{4\pi 
}{\mathrm{c}}\dint G^{\left( -\right) }\left( x-x^{\prime }\right) J^{\nu
}\left( x^{\prime }\right) \,\mathrm{d}x^{\prime }  \label{grmax10}
\end{equation}%
where $A_{\mathrm{in}}^{\nu }\left( x\right) $ and $A_{\mathrm{out}}^{\nu
}\left( x\right) $ are solutions to the homogeneous wave equation. In (\ref%
{grmax9}) the retarded Green function is used. In the limit $%
x_{0}\rightarrow -\infty $, the integral over the sources vanishes, assuming
the sources are localized in space and time, because of the retarded nature
of the Green function, and $A_{\mathrm{in}}^{\nu }\left( x\right) $ can be
interpreted as "\emph{incident}" or "\emph{incoming}" potential, specified
at $x_{0}\rightarrow -\infty $. Similarly, in (\ref{grmax10}) with the
advanced Green function, the homogeneous solution $A_{\mathrm{out}}^{\nu
}\left( x\right) $ is the asymptotic "\emph{outgoing}" potential, specified
at $x_{0}\rightarrow +\infty $. \emph{The radiation fields are defined as
the difference between the "outgoing" and "incoming" fields}, and their $4$%
-vector potential is, \cite{Dirac CE}%
\begin{gather}
A_{\mathrm{rad}}^{\nu }\left( x\right) =A_{\mathrm{out}}^{\nu }\left(
x\right) -A_{\mathrm{in}}^{\nu }\left( x\right) =\frac{4\pi }{\mathrm{c}}%
\dint G\left( x-x^{\prime }\right) J^{\nu }\left( x^{\prime }\right) \,%
\mathrm{d}x^{\prime },\text{ where}  \label{grmax11} \\
G\left( x-x^{\prime }\right) =G^{\left( +\right) }\left( x-x^{\prime
}\right) -G^{\left( -\right) }\left( x-x^{\prime }\right) .  \notag
\end{gather}

More explicit form of the potential $A^{\nu }\left( x\right) $ solving the
inhomogeneous wave equation (\ref{grmax3}) based on the retarded Green
function $G^{\left( +\right) }$ and with $A_{\mathrm{in}}^{\nu }\left(
x\right) =0$ is%
\begin{equation}
\varphi \left( t,\mathbf{x}\right) =\dint \frac{\left[ \rho \left( t^{\prime
},\mathbf{x}^{\prime }\right) \right] _{\mathrm{ret}}}{R}\,\mathrm{d}\mathbf{%
x}^{\prime },\ \mathbf{A}\left( t,\mathbf{x}\right) =\frac{1}{\mathrm{c}}%
\dint \frac{\left[ \mathbf{J}\left( t^{\prime },\mathbf{x}^{\prime }\right) %
\right] _{\mathrm{ret}}}{R}\,\mathrm{d}\mathbf{x}^{\prime },  \label{grmax12}
\end{equation}%
where%
\begin{equation}
\mathbf{R}=\mathbf{x}-\mathbf{x}^{\prime },\ R=\left\vert \mathbf{x}-\mathbf{%
x}^{\prime }\right\vert ,  \label{grmax13}
\end{equation}%
and the symbol $\left[ \cdot \right] _{\mathrm{ret}}$ means that the
quantity in the square brackets is to be evaluated at the retarded time%
\begin{equation}
t^{\prime }=t_{\mathrm{ret}}=t-\frac{R}{c}=t-\frac{\left\vert \mathbf{x}-%
\mathbf{x}^{\prime }\right\vert }{c}.  \label{grmax14}
\end{equation}%
The corresponding to the potentials \ref{grmax12}) EM fields can be
represented by \emph{Jefimenko formulas}%
\index{Jefimenko formulas}, \cite[Section 15.7]{Jefimenko}, \cite[Section 6.5%
]{Jackson}%
\begin{gather}
\mathbf{E}\left( t,\mathbf{x}\right) =\dint 
\frac{\left[ \rho \left( t^{\prime },\mathbf{x}^{\prime }\right) \right] _{%
\mathrm{ret}}\mathbf{\hat{R}}}{R^{2}}\,\mathrm{d}\mathbf{x}^{\prime }+\dint 
\frac{\left[ \partial _{t^{\prime }}\rho \left( t^{\prime },\mathbf{x}%
^{\prime }\right) \right] _{\mathrm{ret}}\mathbf{\hat{R}}}{\mathrm{c}R}\,%
\mathrm{d}\mathbf{x}^{\prime }-  \label{jef1} \\
-\dint \frac{\left[ \partial _{t^{\prime }}\mathbf{J}\left( t^{\prime },%
\mathbf{x}^{\prime }\right) \right] _{\mathrm{ret}}}{\mathrm{c}^{2}R}\,%
\mathrm{d}\mathbf{x}^{\prime },  \notag
\end{gather}%
\begin{equation}
\mathbf{B}\left( t,\mathbf{x}\right) =\dint \left\{ \frac{\left[ \mathbf{J}%
\left( t^{\prime },\mathbf{x}^{\prime }\right) \right] _{\mathrm{ret}}}{%
\mathrm{c}R^{2}}+\frac{\left[ \partial _{t^{\prime }}\mathbf{J}\left(
t^{\prime },\mathbf{x}^{\prime }\right) \right] _{\mathrm{ret}}}{\mathrm{c}%
^{2}R}\right\} \times \mathbf{\hat{R}}\,\mathrm{d}\mathbf{x}^{\prime },
\label{jef2}
\end{equation}%
where $\mathbf{\hat{R}=}\frac{\mathbf{R}}{\left\vert \mathbf{R}\right\vert }$%
. An essentially equivalent form of the Jefimenko equation (\ref{jef1}) for
the electric field $\mathbf{E}$ is due Panofsky and Phillips, \cite[Section
14.3]{Panofsky Phillips}%
\begin{gather}
\mathbf{E}\left( t,\mathbf{x}\right) =\dint \frac{\left[ \rho \left(
t^{\prime },\mathbf{x}^{\prime }\right) \right] _{\mathrm{ret}}\mathbf{\hat{R%
}}}{R^{2}}\,\mathrm{d}\mathbf{x}^{\prime }+  \label{jef3} \\
+\dint \frac{\left( \left[ \mathbf{J}\left( t^{\prime },\mathbf{x}^{\prime
}\right) \right] _{\mathrm{ret}}\cdot \mathbf{\hat{R}}\right) \mathbf{\hat{R}%
}+\left( \left[ \mathbf{J}\left( t^{\prime },\mathbf{x}^{\prime }\right) %
\right] _{\mathrm{ret}}\times \mathbf{\hat{R}}\right) \times \mathbf{\hat{R}}%
}{\mathrm{c}R^{2}}\,\mathrm{d}\mathbf{x}^{\prime }+  \notag \\
+\dint \frac{\left( \left[ \partial _{t^{\prime }}\mathbf{J}\left( t^{\prime
},\mathbf{x}^{\prime }\right) \right] _{\mathrm{ret}}\times \mathbf{\hat{R}}%
\right) \times \mathbf{\hat{R}}}{\mathrm{c}^{2}R}\,\mathrm{d}\mathbf{x}%
^{\prime }.  \notag
\end{gather}%
It was pointed out by McDonald in \cite{McDonald} that the combination of
equations (\ref{jef2}) and (\ref{jef3}) has a certain advantage since it
"manifestly displays the mutually transverse character of the radiation
fields (those that vary as $1/R$)". Since the radiation fields $\mathbf{E}_{%
\mathrm{rad}}$ and $\mathbf{B}_{\mathrm{rad}}$ decay as $1/R$ for large $R$
we can extract them from the expressions (\ref{jef3}), (\ref{jef2}) for the
entire EM fields obtaining%
\begin{equation}
\mathbf{E}_{\mathrm{rad}}\left( t,\mathbf{x}\right) =\dint \frac{\left( %
\left[ \partial _{t^{\prime }}\mathbf{J}\left( t^{\prime },\mathbf{x}%
^{\prime }\right) \right] _{\mathrm{ret}}\times \mathbf{\hat{R}}\right)
\times \mathbf{\hat{R}}}{\mathrm{c}^{2}R}\,\mathrm{d}\mathbf{x}^{\prime },
\label{erad1}
\end{equation}%
\begin{equation}
\mathbf{B}_{\mathrm{rad}}\left( t,\mathbf{x}\right) =\dint \left[ \partial
_{t^{\prime }}\mathbf{J}\left( t^{\prime },\mathbf{x}^{\prime }\right) %
\right] _{\mathrm{ret}}\times \frac{\mathbf{\hat{R}}}{\mathrm{c}^{2}R}\,%
\mathrm{d}\mathbf{x}^{\prime }  \label{erad2}
\end{equation}

\subsection{Point charge and the Li\'{e}nard-Wiechert Potential}

If the particle is a point charge $q$ whose position and velocity in an
inertial frame are respectively $\mathbf{r}\left( t\right) $ and $\mathbf{v}%
\left( t\right) =\partial _{t}\mathbf{r}\left( t\right) $ the corresponding
charge and current densities in that frame are, \cite[Section 12.11, (12.138)%
]{Jackson}%
\begin{equation}
\rho \left( t,\mathbf{x}\right) =q\delta \left( \mathbf{x}-\mathbf{r}\left(
t\right) \right) \text{ and }\mathbf{J}\left( t,\mathbf{x}\right) =q\mathbf{v%
}\left( t\right) \delta \left( \mathbf{x}-\mathbf{r}\left( t\right) \right) .
\label{liwi1}
\end{equation}%
Using the formulas (\ref{grmax12}) for the charge density and the current as
in the equations (\ref{liwi1}) we obtain the Li\'{e}nard-Wiechert Potential, 
\cite[Section 14.1]{Jackson}%
\begin{equation}
\varphi \left( t,\mathbf{x}\right) =\left[ \frac{q}{\left( 1-\mathbf{\beta }%
\cdot \mathbf{\hat{R}}\right) }\right] _{\mathrm{ret}},\quad \mathbf{A}%
\left( t,\mathbf{x}\right) =\left[ \frac{q\mathbf{\beta }}{\left( 1-\mathbf{%
\beta }\cdot \mathbf{\hat{R}}\right) }\right] _{\mathrm{ret}}  \label{liwi2}
\end{equation}%
where%
\begin{equation}
\mathbf{R}=\mathbf{x}-\mathbf{r}\left( t_{\mathrm{r}}\right) ,\quad
R=\left\vert \mathbf{x}-\mathbf{r}\left( t_{\mathrm{r}}\right) \right\vert
,\quad \mathbf{\hat{R}}=\frac{\mathbf{R}}{\left\vert \mathbf{R}\right\vert }%
,\quad \mathbf{\beta }=\frac{\mathbf{v}\left( t_{\mathrm{r}}\right) }{%
\mathrm{c}},  \label{liwi3}
\end{equation}%
and the retarded time $t_{\mathrm{r}}=t_{\mathrm{r}}\left( \mathbf{x}%
,t\right) $ is defined implicitly by the following equation%
\begin{equation}
t_{\mathrm{r}}=t-\frac{\left\vert \mathbf{x}-\mathbf{r}\left( t_{\mathrm{r}%
}\right) \right\vert }{\mathrm{c}}.  \label{liwi4}
\end{equation}%
Then with the help of the Jefimenko formulas (\ref{jef1}), (\ref{jef2})
applied for the point charge density and the current (\ref{liwi1}) one can
derive the Heaviside-Feynman formulas%
\index{Heaviside-Feynman formula} (first discovered by Heaviside (1902) and
rediscovered by Feynman (1950)), \cite{Heaviside1}, \cite[Subsection 510]%
{Heaviside2}, \cite[Vol. I, Section 28; Vol II, Section 21]{Feynman}, \cite[%
Section 6.5]{Jackson}, \cite{Janah}, \cite{Monaghan}: 
\begin{equation}
\mathbf{E}\left( t,\mathbf{x}\right) =q\left\{ \left[ 
\frac{\mathbf{\hat{R}}}{R^{2}}\right] _{\mathrm{ret}}+\frac{\left[ R\right]
_{\mathrm{ret}}}{\mathrm{c}}\partial _{t}\left[ \frac{\mathbf{\hat{R}}}{R^{2}%
}\right] _{\mathrm{ret}}+\frac{1}{\mathrm{c}^{2}}\partial _{t}^{2}\left[ 
\mathbf{\hat{R}}\right] _{\mathrm{ret}}\right\} ,  \label{liwi5}
\end{equation}%
\begin{gather}
\mathbf{B}\left( t,\mathbf{x}\right) =\frac{q}{\mathrm{c}}\left\{ \left[ 
\frac{\mathbf{v}\times \mathbf{\hat{R}}}{\varkappa ^{2}R^{2}}\right] _{%
\mathrm{ret}}+\frac{1}{\mathrm{c}\left[ R\right] _{\mathrm{ret}}}\partial
_{t}\left[ \frac{\mathbf{v}\times \mathbf{\hat{R}}}{\varkappa }\right] _{%
\mathrm{ret}}\right\} =  \label{liwi6} \\
=\frac{q}{\mathrm{c}}\left\{ \left[ \frac{\mathbf{\hat{R}}}{R}\right] _{%
\mathrm{ret}}\times \partial _{t}\left[ \mathbf{\hat{R}}\right] _{\mathrm{ret%
}}+\frac{\left[ \mathbf{\hat{R}}\right] _{\mathrm{ret}}}{\mathrm{c}}\times
\partial _{t}^{2}\left[ \mathbf{\hat{R}}\right] _{\mathrm{ret}}\right\} = 
\notag \\
=\mathbf{E}\left( t,\mathbf{x}\right) \times \left[ \mathbf{\hat{R}}\right]
_{\mathrm{ret}},  \notag
\end{gather}%
where%
\begin{equation}
\varkappa =1-\frac{\mathbf{v}\cdot \mathbf{\hat{R}}}{\mathrm{c}}.
\label{liwi7}
\end{equation}%
In view of the implicit relations (\ref{liwi4}) between the retarded time $%
t_{\mathrm{r}}$ and the time-space variables $\left( \mathbf{x},t\right) $
it is important to keep in mind that there is evidently a difference between 
$\partial _{t}\left[ \left( \cdot \right) \right] _{\mathrm{ret}}$ and $%
\left[ \partial _{t}\left( \cdot \right) \right] _{\mathrm{ret}}$.

The Heaviside-Feynman formulas (\ref{liwi5}), (\ref{liwi6}) imply the
following formulas for the radiation fields of the moving point charge%
\begin{equation}
\mathbf{E}_{\mathrm{rad}}\left( t,\mathbf{x}\right) =\frac{q}{\mathrm{c}^{2}}%
\partial _{t}^{2}\left[ \mathbf{\hat{R}}\right] _{\mathrm{ret}},\quad 
\mathbf{B}_{\mathrm{rad}}\left( t,\mathbf{x}\right) =\frac{q}{\mathrm{c}^{2}%
\left[ R\right] _{\mathrm{ret}}}\partial _{t}\left[ \frac{\mathbf{v\times 
\hat{R}}}{\varkappa }\right] _{\mathrm{ret}}.  \label{erad3}
\end{equation}%
Another important representation of the EM fields of an arbitrary moving
charge is their decomposition into the \emph{velocity and acceleration
fields, }\cite[Section 14.1]{Jackson1}%
\begin{gather}
\mathbf{E}\left( t,\mathbf{x}\right) =\mathbf{E}_{\mathrm{v}}\left( t,%
\mathbf{x}\right) +\mathbf{E}_{\mathrm{a}}\left( t,\mathbf{x}\right) ,\text{
where for }\mathbf{\beta }=\frac{\mathbf{v}}{\mathrm{c}}  \label{erad4} \\
\mathbf{E}_{\mathrm{v}}\left( t,\mathbf{x}\right) =q\left[ \frac{\left( 
\mathbf{\hat{R}}-\mathbf{\beta }\right) \left( 1-\beta ^{2}\right) }{%
\varkappa ^{3}R^{2}}\right] _{\mathrm{ret}},  \notag \\
\mathbf{E}_{\mathrm{a}}\left( t,\mathbf{x}\right) =\mathbf{E}_{\mathrm{rad}%
}\left( t,\mathbf{x}\right) =\frac{q}{\mathrm{c}}\left[ \frac{\mathbf{\hat{R}%
}}{\varkappa ^{3}R}\times \left\{ \left( \mathbf{\hat{R}}-\mathbf{\beta }%
\right) \times \mathbf{\dot{\beta}}\right\} \right] _{\mathrm{ret}},  \notag
\end{gather}%
\begin{gather}
\mathbf{B}\left( t,\mathbf{x}\right) =\mathbf{\hat{R}}\times \mathbf{E}%
\left( t,\mathbf{x}\right) =\mathbf{B}_{\mathrm{v}}\left( t,\mathbf{x}%
\right) +\mathbf{B}_{\mathrm{a}}\left( t,\mathbf{x}\right) ,  \label{erad5}
\\
\mathbf{B}_{\mathrm{v}}\left( t,\mathbf{x}\right) =\mathbf{\hat{R}}\times 
\mathbf{E}_{\mathrm{v}}\left( t,\mathbf{x}\right) ,  \notag \\
\mathbf{B}_{\mathrm{a}}\left( t,\mathbf{x}\right) =\mathbf{B}_{\mathrm{rad}%
}\left( t,\mathbf{x}\right) =\mathbf{\hat{R}}\times \mathbf{E}_{\mathrm{a}%
}\left( t,\mathbf{x}\right) .  \notag
\end{gather}%
The \emph{velocity fields are essentially static fields} falling off as $%
R^{-2}$, whereas the \emph{acceleration fields are typical radiation fields}%
, both $\mathbf{E}_{\mathrm{a}}$ and $\mathbf{B}_{\mathrm{a}}$ being
transverse to the radius vector $\mathbf{R}$ and varying as $R^{-1}$. For
low velocities the formulas (\ref{erad4}), (\ref{erad5}) turn into the
following simpler asymptotic expressions 
\begin{gather}
\mathbf{E}_{\mathrm{rad}}\left( t,\mathbf{x}\right) =\frac{q}{\mathrm{c}}%
\left[ \frac{\mathbf{\hat{R}\times }\left( \mathbf{\hat{R}\times \dot{v}}%
\right) }{\left\vert \mathbf{R}\right\vert }\right] _{\mathrm{ret}}\left(
1+O\left( \left\vert \mathbf{\beta }\right\vert \right) \right) ,
\label{erad6} \\
\mathbf{B}_{\mathrm{rad}}\left( t,\mathbf{x}\right) =\frac{q}{\mathrm{c}}%
\left[ \frac{\mathbf{\dot{v}\times \hat{R}}}{\left\vert \mathbf{R}%
\right\vert }\right] _{\mathrm{ret}}\left( 1+O\left( \left\vert \mathbf{%
\beta }\right\vert \right) \right) ,\quad \left\vert \mathbf{\beta }%
\right\vert \ll 1.  \notag
\end{gather}

\subsection{Almost periodic functions and their time-averages\label{saper}}

We provide here very basic information on real and complex valued almost
periodic (a.p.) functions following \cite[Section 3.1]{Corduneanu}. In fact,
we consider for simplicity sake a class $\mathcal{A}_{1}\left( \mathbb{R},%
\mathbb{C}\right) $ of a.p. function of the form 
\begin{equation}
f\left( t\right) =\dsum\limits_{s=1}^{\infty }f_{s}\mathrm{e}^{-\mathrm{i}%
\omega _{s}t},\qquad -\infty <t<\infty ,  \label{apef1}
\end{equation}%
where the exponents $\omega _{s}$ are real valued numbers and amplitudes $%
f_{s}$ are complex valued numbers and such that%
\begin{equation}
\left\vert f\right\vert =\dsum\limits_{s=1}^{\infty }\left\vert
f_{s}\right\vert <\infty .  \label{apef2}
\end{equation}%
We refer to the set of exponents $\Lambda _{f}=\left\{ \omega _{s}\right\} $
as the Fourier spectrum of the function $f$ and to the numbers $f_{s}$ as
its Fourier coefficients. The class $\mathcal{A}_{1}\left( \mathbb{R},%
\mathbb{C}\right) $ is a Banach algebra, that is (i) it is linear space and
(ii) for any $f$ and $g$ in $\mathcal{A}_{1}\left( \mathbb{R},\mathbb{C}%
\right) $ the product $fg$ is in $\mathcal{A}_{1}\left( \mathbb{R},\mathbb{C}%
\right) $ as well and $\left\vert fg\right\vert \leq \left\vert f\right\vert
\left\vert g\right\vert $. The derivatives of a.p. $f\left( t\right) $
satisfy%
\begin{equation}
\partial _{t}^{r}f\left( t\right) =\dsum\limits_{s=1}^{\infty }\left( -%
\mathrm{i}\omega _{s}\right) ^{r}f_{s}\mathrm{e}^{-\mathrm{i}\omega _{s}t},
\label{apef2a}
\end{equation}%
provided%
\begin{equation}
\dsum\limits_{s=1}^{\infty }\left\vert \omega _{s}\right\vert ^{r}\left\vert
f_{s}\right\vert <\infty .  \label{apef2b}
\end{equation}%
Every a.p. function $f$ in $\mathcal{A}_{1}\left( \mathbb{R},\mathbb{C}%
\right) $ is assigned the \emph{time average (mean)} value $\left\langle
f\right\rangle $ defined by 
\begin{equation}
\left\langle f\right\rangle =\lim_{T\rightarrow \infty }\frac{1}{2T}%
\dint\nolimits_{-T}^{T}f\left( t\right) \,\mathrm{d}t.  \label{apef3}
\end{equation}%
The mean value has the fundamental property%
\begin{equation}
\left\langle f\left( t\right) \mathrm{e}^{\mathrm{i}\theta t}\right\rangle
=\left\{ 
\begin{tabular}{ll}
$f_{s}$ & if $\theta =\omega _{s}$ \\ 
$0$ & otherwise%
\end{tabular}%
\right. .  \label{apef4}
\end{equation}

In some applications it is convenient to view a real a.p. function $f\left(
t\right) $ as the real part of a complex valued function $\mathsf{f}\left(
t\right) $ written in the following form%
\begin{gather}
\mathsf{f}\left( t\right) =\dsum\limits_{\omega \in \Lambda _{f}}f_{\omega }%
\mathrm{e}^{-\mathrm{i}\omega t},  \label{apef5} \\
\text{where }f_{\omega }=\left\langle f\left( t\right) \mathrm{e}^{\mathrm{i}%
\omega t}\right\rangle =\left\langle f\cos \left( \omega t\right)
\right\rangle +\mathrm{i}\left\langle f\sin \left( \omega t\right)
\right\rangle .  \notag
\end{gather}%
The set $\Lambda _{f}$ in (\ref{apef5}) is at most countable set of
non-negative frequencies $\omega \geq 0$, and we refer to it as the
frequency spectrum of $f$. Then the corresponding real valued a.p. function
has the following representation%
\begin{equation}
f\left( t\right) =\func{Re}\left\{ \mathsf{f}\left( t\right) \right\} =\frac{%
1}{2}\dsum\limits_{\omega \in \Lambda _{f}}\left( f_{\omega }\mathrm{e}^{-%
\mathrm{i}\omega t}+f_{\omega }^{\ast }\mathrm{e}^{\mathrm{i}\omega
t}\right) .  \label{apef6}
\end{equation}%
Evidently, any real a.p. function can be represented in the form (\ref{apef5}%
), (\ref{apef6}). Observe that for any a.p. complex valued function $\mathsf{%
f}$ and $\mathsf{g}$ of the form (\ref{apef5}) we have 
\begin{equation}
\left\langle \func{Re}\left\{ \mathsf{f}\left( t\right) \right\} \func{Re}%
\left\{ \mathsf{g}\left( t\right) \right\} \right\rangle =\left\{ 
\begin{tabular}{ll}
$\frac{1}{2}\dsum\limits_{\omega \in \Lambda _{f}\cap \Lambda _{g}}\func{Re}%
\left\{ f_{\omega }g_{\omega }^{\ast }\right\} $ & if $\Lambda _{f}\cap
\Lambda _{g}\neq \varnothing ,$ \\ 
$0$ & if $\Lambda _{f}\cap \Lambda _{g}=\varnothing $%
\end{tabular}%
\right. .  \label{apef7}
\end{equation}%
In particular, if $f$ and $g$ are real valued the relation (\ref{apef7})
imply%
\begin{equation}
\left\langle f\left( t\right) g\left( t\right) \right\rangle =\left\{ 
\begin{tabular}{ll}
$\frac{1}{2}\dsum\limits_{\omega \in \Lambda _{f}\cap \Lambda _{g}}\func{Re}%
\left\{ f_{\omega }g_{\omega }^{\ast }\right\} $ & if $\Lambda _{f}\cap
\Lambda _{g}\neq \varnothing ,$ \\ 
$0$ & if $\Lambda _{f}\cap \Lambda _{g}=\varnothing $%
\end{tabular}%
\right. .  \label{apef8}
\end{equation}%
The identity (\ref{apef7}) readily follows from formula (\ref{apef5}). It
reads that if the frequency spectra $\Lambda _{f}^{+}$ and $\Lambda _{g}^{+}$
have no common frequencies the time-average $\left\langle \func{Re}\left\{
f\right\} \func{Re}\left\{ g\right\} \right\rangle $ is identically zero. A
particular case of the formula (\ref{apef7}) when the frequency spectra of $%
f $ and $g$ consist of a the same single frequency $\omega $ is customary
used in electrodynamics for time harmonic fields, \cite[Section 11.2]%
{Panofsky Phillips}, \cite[Section 2.20]{Stratton}.

For any a.p. function $f$ with the frequency spectrum $\Lambda _{f}$ one can
introduce the smallest additive group in the set of all real numbers that
contains all frequencies $\omega $ from $\Lambda _{f}$. Such a smallest
group is called the module of $f$ and is denoted by $\func{mod}\left(
f\right) $, \cite[Section 4.6]{Corduneanu}. It is easy to see that $\func{mod%
}\left( f\right) $ consists of all real numbers of the form%
\begin{equation*}
\dsum\limits_{j=1}^{s}m_{j}\omega _{j}\text{ where }m_{j}\text{ are
integers, }\omega _{j}\in \Lambda _{f}\text{ and }s\text{ is a natural
number.}
\end{equation*}

\subsection{Vector Identities}

Here is the fist set of commonly used vector identities%
\begin{equation}
\mathbf{a}\cdot \left( \mathbf{b}\times \mathbf{c}\right) =\mathbf{b}\cdot
\left( \mathbf{c}\times \mathbf{a}\right) =\mathbf{c}\cdot \left( \mathbf{a}%
\times \mathbf{b}\right) ,  \label{vecf1}
\end{equation}%
\begin{equation}
\mathbf{a}\times \left( \mathbf{b}\times \mathbf{c}\right) =\left( \mathbf{a}%
\cdot \mathbf{c}\right) \mathbf{b}-\left( \mathbf{a}\cdot \mathbf{b}\right) 
\mathbf{c},  \label{vecf2}
\end{equation}%
\begin{equation}
\left( \mathbf{a}\times \mathbf{b}\right) \cdot \left( \mathbf{c}\times 
\mathbf{d}\right) =\left( \mathbf{a}\cdot \mathbf{c}\right) \left( \mathbf{b}%
\cdot \mathbf{d}\right) -\left( \mathbf{a}\cdot \mathbf{d}\right) \left( 
\mathbf{b}\cdot \mathbf{c}\right) .  \label{vecf3}
\end{equation}%
Using the above identities we readily obtain for any vectors $\mathbf{a}$, $%
\mathbf{b}$ and any unit vector $\mathbf{u}$%
\begin{equation}
\mathbf{u}\times \left( \mathbf{a}\times \mathbf{u}\right) =\mathbf{a}%
-\left( \mathbf{a}\cdot \mathbf{u}\right) \mathbf{u},\quad \left\vert 
\mathbf{u}\right\vert =1,  \label{vecf3a}
\end{equation}%
\begin{gather}
\left[ \mathbf{u}\times \left( \mathbf{a}\times \mathbf{u}\right) \right]
\times \left( \mathbf{u}\times \mathbf{b}\right) =\left[ \left( \mathbf{a}%
\times \mathbf{u}\right) \times \mathbf{u}\right] \times \left( \mathbf{b}%
\times \mathbf{u}\right) =  \label{vecf3b} \\
=\left[ \mathbf{a}\cdot \mathbf{b}-\left( \mathbf{a}\cdot \mathbf{u}\right)
\left( \mathbf{u}\cdot \mathbf{b}\right) \right] \mathbf{u=}  \notag \\
\mathbf{=}\left[ \left( \mathbf{a}\times \mathbf{u}\right) \cdot \left( 
\mathbf{b}\times \mathbf{u}\right) \right] \mathbf{u},\quad \left\vert 
\mathbf{u}\right\vert =1.  \notag
\end{gather}%
\begin{equation}
\frac{1}{4\pi }\int\nolimits_{\left\vert \mathbf{x}\right\vert =1}\left[ 
\mathbf{a}\cdot \mathbf{b}-\left( \mathbf{a}\cdot \mathbf{\hat{x}}\right)
\left( \mathbf{\hat{x}}\cdot \mathbf{b}\right) \right] \,\mathrm{d}\sigma =%
\frac{2}{3}\mathbf{a}\cdot \mathbf{b},\quad \mathbf{\hat{x}}=\frac{\mathbf{x}%
}{\left\vert \mathbf{x}\right\vert }.  \label{vecf3c}
\end{equation}%
\begin{equation}
\nabla \cdot \left( \mathbf{a}\times \mathbf{b}\right) =\mathbf{b}\cdot
\left( \nabla \times \mathbf{a}\right) -\mathbf{a}\cdot \left( \nabla \times 
\mathbf{b}\right) .  \label{vecf7a}
\end{equation}

\textbf{Acknowledgment.} The research was supported through Dr. A. Nachman
of the U.S. Air Force Office of Scientific Research (AFOSR), under grant
number FA9550-04-1-0359.

\end{document}